\documentclass[twocolumn, amsmath,amssymb,aps]{revtex4-2}
\usepackage{graphicx}
\usepackage{dcolumn}
\usepackage{bm}
\usepackage{appendix}
\usepackage{subcaption}
\usepackage{booktabs}
\usepackage{natbib,har2nat}
\usepackage{placeins}

\usepackage{url}
\usepackage{hyperref}
\begin{document}
\preprint{APS/123-QED}
\title{Seasonal Variation of Polar Ice: Implications for Ultrahigh Energy Neutrino Detectors}
\author{A. Kyriacou}
\email{alexander.luke.kyriacou@ulb.be}
\altaffiliation[Present address: ]{Interuniversity Institute for High Energies,
Universit\'e Libre de Bruxelles, 1050 Brussels, Belgium}
\author{S. Prohira}
\author{D. Besson}
\affiliation{Dept. of Physics and Astronomy,
University of Kansas, Lawrence, KS 66045, USA}
\begin{abstract}
The upper $100 \, \mathrm{m}$ to $150 \, \mathrm{m}$ of the polar ice sheet, called the firn, has a time-dependent density due to seasonal variations in the surface temperature, snow accumulation and surface melt events. We present RF simulations of an in-ice neutrino-induced radio source that show that these density anomalies create variations in the amplitude and propagation times of radio signals propagating through the polar firn at Summit, Greenland. The received power from signals generated in the ice that refract within the upper ${\sim} 15 \, \mathrm{m}$ firn are subject to a seasonal variation on the order of 10\%. These variations result in an inherent and geometry dependent background uncertainty on the reconstructed neutrino energy and arrival direction for detectors using ice as a detection medium.
\end{abstract}
\maketitle
\section{Introduction}
For decades the ultrahigh energy neutrino (UHEN) flux, with energies exceeding $10 \, \mathrm{PeV}$, has been predicted to exist as a by-product of ultrahigh energy cosmic rays (UHECRs) interacting with ambient photons \cite{Beresinsky_Zatsepin, Ahlers_2012}.
The UHEN flux is likely composed of two populations associated with distinct production mechanisms. At lower energies, an \textit{astrophysical} flux arises from cosmic ray-photon interactions within the sources themselves, such as active galactic nuclei (AGNs), supernova remnants, and other compact objects \cite{Kotera_2011}. At higher energies, cosmic rays escape their sources and propagate through intergalactic space, where they scatter off relic photons of the cosmic microwave background (CMB). For UHECR protons with energies above the `Greisen-Zatsepin-Kuzmin' (GZK) limit of $E > 5 \cdot 10^{19} \, \mathrm{eV}$, this interaction produces neutral and charged pions that subsequently decay into photons and neutrinos respectively, the latter of which comprise the \textit{cosmogenic} neutrino flux \cite{Zatsepin_Kuzmin_GZK, Greisen_GZK}. The detection by the KM3NET collaboration of a muon with an estimated energy of $E_{\mu} = 120^{+110}_{-60} \, \mathrm{PeV}$ could constitute the first detection of an UHE neutrino \cite{KM3NeT:2025npi}.
\\~\\
\begin{figure}[t]
    \centering
    \includegraphics[width=\linewidth]{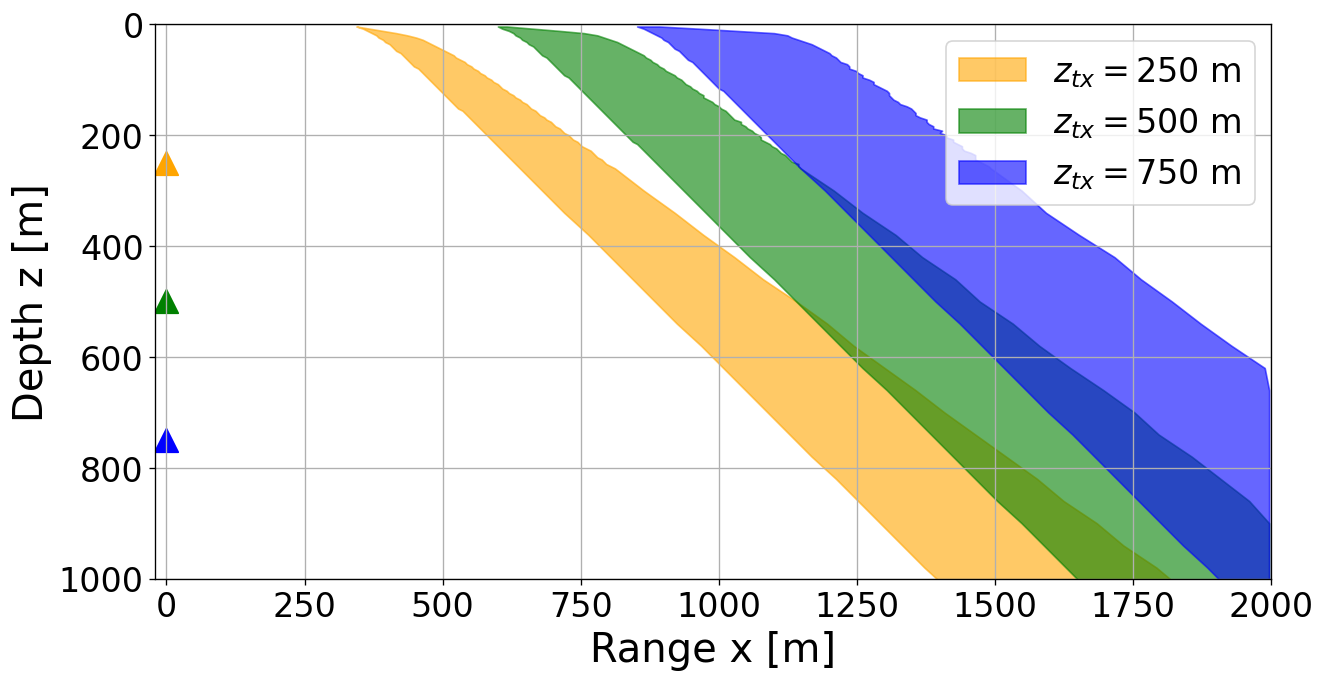}
    \caption{Radio signals that traverse the upper $15 \, \mathrm{m}$ of the ice exhibit a seasonal fluctuation of $\mathcal{O}(0.1)$ in received fluence $\phi^{E}$. Shaded regions highlight the portion of the detection region subject to this seasonal effect for the associated neutrino vertex (shown as triangles at $x=0$, 3 different example vertices shown).}
    \label{fig:shaded-regions}
\end{figure}
Because they experience negligible attenuation over cosmic distances and are undeflected by magnetic fields, UHENs can act as direct tracers of UHECR sources and can also potentially serve as probes of physics beyond the Standard Model~\cite{UHE_cross_section, Muzio_2025}. To date, the UHE neutrino flux $\Phi_{\nu}$ above $10 \, \mathrm{PeV}$ has been constrained to an upper limit of $E_{\nu}^{2}\Phi_{\nu} \lesssim 10^{-8} \, \mathrm{GeV cm^{-2}s^{-1} sr^{-1}}$ with the possible exception of the recent KM3NET event \cite{IceCube_upper_limit_PhysRevLett.135.031001}. This corresponds to a detection rate of less than one event detected per cubic-kilometer per decade \cite{Kotera_2010, Aartsen_2018}. Consequently, fiducial volumes in excess of $\mathcal{O}(10 \, \mathrm{km^{3}})$ are required to achieve useful detection rates. Such detector scales are prohibitively expensive for water- and ice-based optical Cherenkov arrays such as IceCube, ANTARES, KM3NeT, or Baikal-GVD, necessitating alternative detection paradigms.
\\~\\
One approach is to search for Askaryan radiation: coherent radio emission produced at the Cherenkov angle due to the build-up of excess negative charge in a particle cascade~\cite{Askaryan:1961pfb}. Askaryan radiation appears to the observer as a broadband radio-frequency pulse with a time width of $\mathcal{O}(\mathrm{ns})$. The radio attenuation length $L_{\alpha}$ in polar ice sheets is $L_{\alpha} \sim \mathcal{O}(1 \, \mathrm{km})$ for frequencies in the 100~MHz to 1~GHz regime~\cite{aguilar_RF_attenuation, Barwick_Besson_Gorham_Saltzberg_2005, Aguilar_broad_band_atten}, allowing an array of receiving antennas embedded in the ice to monitor an volume of $\mathcal{O}(10 \, \mathrm{km^{3}})$. Multiple such arrays can collectively monitor an area more than an order of magnitude larger than that of existing optical Cherenkov observatories at a reduced cost. The in-ice Askaryan detection method has been field-tested by the RICE \cite{rice_status}, ARIANNA \cite{arianna}, ARA \cite{allison2011designinitialperformanceaskaryan}, and RNO-G \cite{rno_g_design_sensitvity} experiments. Alternatively, airborne observatories can monitor the ice sheet for Askaryan radiation from high altitudes, achieving greater detection volumes during limited-duration flights. This method was first utilized by the ANITA~\cite{Deaconu_2021_ANITA} experiment and the recently-completed PUEO flight will also search for UHENs through this method \cite{Abarr_2021_PUEO}. Although these experiments have not yet detected neutrinos they have refined radio detection techniques, identified sources of background, and established upper limits on the UHE neutrino flux \cite{arianna_upper, KRAVCHENKO2003195_upper, Allison_2020_upper}. Additionally the ANITA, ARIANNA and RNO-G experiments have observed radio showers induced by cosmic rays in the air and in the surface snow \cite{ANITA_detection, Barwick_2019_arianna_CR, RNO_G_CR_detection_shallow}.
\\~\\
A more recently proposed in-ice strategy is the \textit{radar echo method}, in which an array of antennas broadcasts a radar signal into a volume of ice. The signal is effectively scattered by the ionization trail left in the wake of a relativistic particle cascade, and the resulting echo can be measured by receiver antennas. This approach also benefits from the radio transparency and density of glacial ice by allowing monitoring of much larger volumes than possible with optical methods. Radar echo detection of particle cascades was first demonstrated in experiment T576 at SLAC, where a high-energy electron beam was directed into a block of high-density polyethylene, simulating a cascade of similar energy to that produced by an $E_{\nu} \geq 10^{18} \, \mathrm{eV}$ neutrino interaction. A radar reflection signature correlating with the cascade was subsequently identified in an analysis \cite{Prohira_2020}. The Radar Echo Telescope for Cosmic Rays (RET-CR), which operated during the northern summer seasons of 2023 and 2024, aims to validate this method in natural settings by detecting radar echoes coincident with high-energy cosmic rays recorded by surface scintillator and radio detectors \cite{Prohira_2021, allison2024initialperformanceradarecho}. RET-CR has, at the time of this writing, completed a season of data-taking, and analysis is underway to identify radar echo events.
\\~\\
For both approaches, accurate reconstruction of UHE neutrino events require precise modeling of radio wave propagation between the neutrino vertex and the receiving antennas. Beneath depths of up to $\mathcal{O}(100 \, \mathrm{m})$ \cite{springer_firn, Firn_East_Antarctica}, the Greenlandic and Antarctic ice sheets exhibit nearly homogeneous density $\rho_i = {\sim}917 \, \mathrm{kg/m^{3}}$ \cite{springer_firn}, and the refractive index has been measured as $n_{i,\mathrm{gr}} = 1.778 \pm 0.006$ in deep Greenlandic ice \cite{aguilar2023precisionmeasurementindexrefraction}, resulting in straight-line radio propagation. At shallower depths however the ice is characterized by the \textit{firn layer}, a region where snow gradually is compacted into ice. The density increases with depth, causing a velocity gradient in the planar wavefront, which has the effect of \textit{curving} the ray paths of radio signals. Glaciological models and measurements indicate that the firn layer exhibits time-varying fluctuations in both temperature and density \cite{densification_polar_firn_horhold, Jay_Zwally_Jun_2002, Howat_2022}. This implies that radio signals traversing the firn may experience time-dependent variations in path and amplitude.
\\~\\
The aim of this study is to characterize and quantify, to first order, the magnitude of seasonal variation in firn-traversing radio signals and examine the implications for UHE neutrino reconstruction. We simulate the propagation of a broadband radio pulse from a hypothetical neutrino interaction vertex within the ice to a receiver array located between the surface and 160 m depth. This simulation is repeated for a series of refractive index profiles derived from firn density measurements at Summit Station, Greenland ($72^{\circ}36^{\prime}\,\mathrm{N},\;38^{\circ}25^{\prime}\,\mathrm{W};\;{\sim}3200~\mathrm{m~a.s.l.}$) \cite{NOAA_Summit_altitude}, predicted by glaciological models spanning January 1980 to December 2021. The received waveforms are compared across all scenarios to evaluate variations in signal fluence and propagation time. Section~\ref{firn-sim} describes the data-driven modeling of firn profiles using the Community Firn Model. In Section~\ref{radio-sim}, we detail the radio simulation framework. Results are presented in Section~\ref{results-section}, and implications for neutrino detection are discussed in Section~\ref{discussion}.
\section{Firn Evolution at Summit, Greenland}\label{firn-sim}
Firn is the transitional stage between fallen snow and glacial ice, with the densification process primarily driven by gravity via the overburden pressure of shallower layers of firn. The rate of densification $k$ is also a function of temperature, and the densification is governed by multiple physical processes \cite{springer_firn}. Two processes of particular importance are the initial phase of settling and grain growth that occurs until a threshold density of $\rho_{550} = 550 \, \mathrm{kg/m^{3}}$, and a later process known as sintering which becomes dominant beyond this threshold. Sintering is the gradual fusion of distinct ice grains into a continuous bulk as the areas of contact expand due to sublimation and vapor transport of water molecules \cite{Blackford_2007_sintering, springer_firn}. The former process is faster and exhibits greater sensitivity to temperature than the latter. At greater depths, past a threshold density of $\rho \approx 800 \, -\, 840 \, \mathrm{kg/m^{3}}$, the densification is dominated by the compression of the remaining pores by the overburden pressure. Generally, the firn density profile $\rho(z)$ in the polar regions can be approximated using a asymptotic exponential decay function of depth $z$ (where $z>0$ is below the surface and $z<0$ is above the surface) approaching the density of ice $\rho_{i}$:
\begin{equation}\label{firn_exponential}
        \rho(z) = 
        \begin{cases}
                \rho_{i} + (\rho_{0} - \rho_{i}) e^{-k_{0} z} & \text{if } \rho \leq \rho_{550} \\
                \rho_{i} + (\rho_{550} - \rho_{i}) e^{-k_{1} (z-z_{550})} & \text{if } \rho > \rho_{550},
        \end{cases}
\end{equation}
\begin{figure*}[t]
    \centering
    \includegraphics[width=\linewidth]{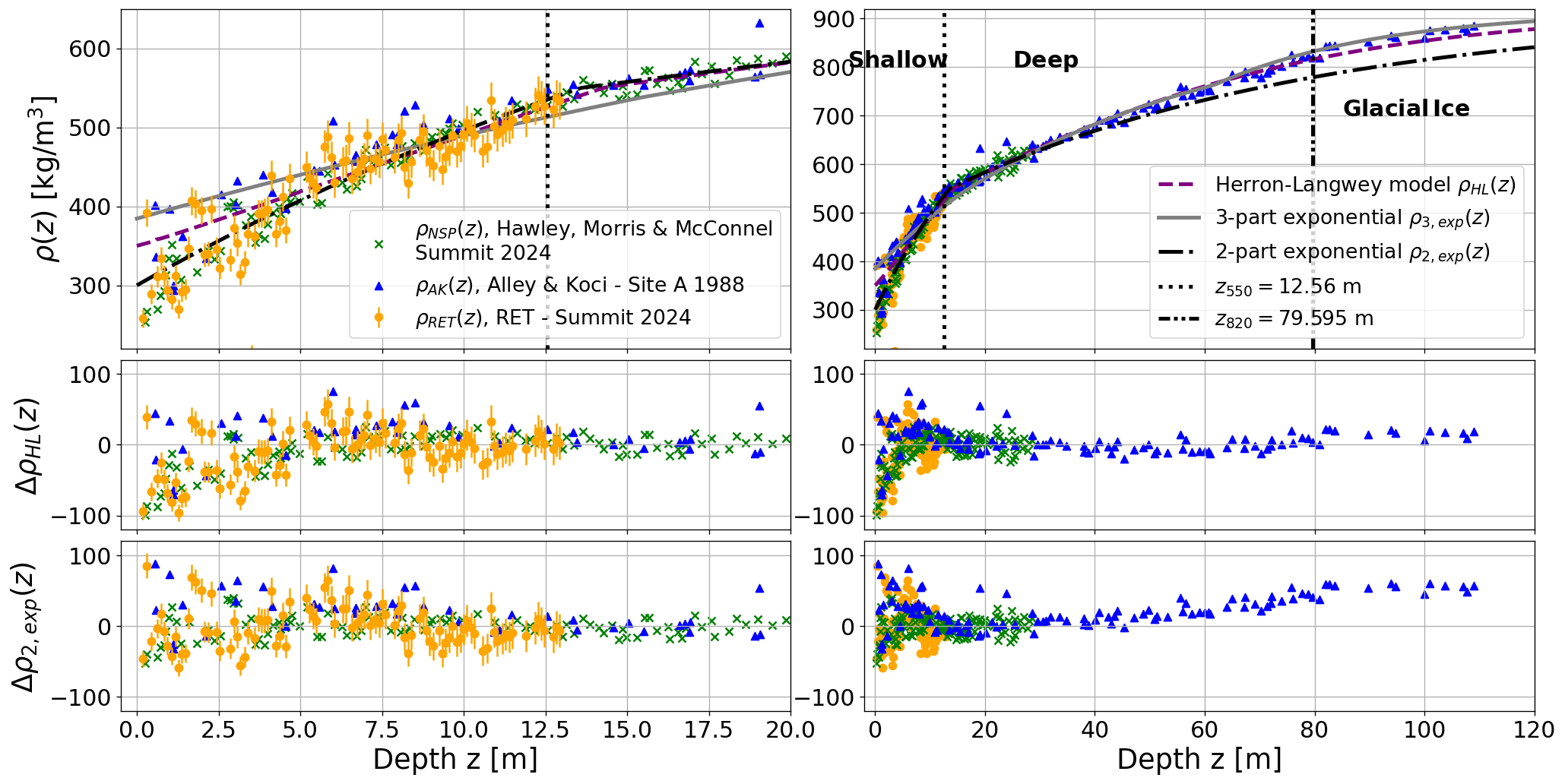}
    \caption{The upper row of plots show (with the left side displaying the upper 20 m, the right down to 120 m) two measured density profiles at Summit; $\rho_{RET}(z)$ \cite{Kyriacou:2025tj} (shown with uncertainties) \& $\rho_{NSP}(z)$ \cite{Hawley_Morris_McConnell_2008}, alongside a density profile $\rho_{AK}(z)$ made at Site A \cite{Alley_Koci_1988}, located 222 km south of Summit station. A 2-part asymptotic exponential function $\rho_{2,exp}(z)$ is fit to the NSP (Neutron Scattering Probe) and a 3-part exponential model $\rho_{3,exp}(z)$ is shown with the parameters taken from a previous work \cite{ali2024modelingrefractiveindexprofile}. The output of the Herron-Langway $\rho_{HL}(z)$ model is also shown. The shallow firn $z < 15 \, \mathrm{m}$ is where the fluctuation in density is most pronounced. The lower two plots show the residuals $\Delta \rho_{HL}(z)$ \& $\Delta \rho_{2,exp}(z)$ of the measured density profiles to HL model and 2-part exponential fit respectively.}
    \label{fig:rho_profile_summit}
\end{figure*}
With $\rho_{0} = \rho(z=0)$ being the surface density, $k_{0}$ and $k_{1}$ corresponding respectively to the densification rate above and below the critical depth $z_{550}$ where the threshold density $\rho_{550}$ is reached. Hereafter the region between the surface and $z_{550}$ is defined as the \textit{shallow firn} and the region between $z_{550}$ and $z_{820}$ (the depth at which $\rho = 820 \, \mathrm{kg/m^{3}}$) is defined as the \textit{deep firn}, with the depth below defined as the \textit{glacial ice}. It is also possible to model the firn's density profile with a three-part asymptotic exponential $\rho_{3,exp}(z)$, with an additional densification rate $k_{2}$ beyond the pore close off threshold depth $z_{820}$ \cite{ali2024modelingrefractiveindexprofile}, where as the former will be designated the two-part exponential model $\rho_{2,exp}(z)$. Moreover one can derive an asymptotic exponential dependence of density on depth using the Herron-Langway (HL) model, with an assumption of constant mass accumulation $A$ and temperature $T$ at 10 m depth \cite{herron_langway_1980}. Various physical models are used to estimate firn density evolution with time-variant temperature and accumulation rates, but for the most part are extensions upon the HL model \cite{cfm_v1}. For Summit, we derive a density profile $\rho_{HL}(z)$ using the HL model calculated from an average firn temperature at 10 m of $T_{10m} = -31.4^{\circ}$, an accumulation rate $A = 0.23 \mathrm{m\,yr^{-1}}$ (ice equivalent), and an assumed surface density of $\rho_{0} = 350 \, \mathrm{kg/m^{3}}$ \cite{herron_langway_1980}. 
\\~\\
In Fig.\,\ref{fig:rho_profile_summit} we show two density measurements of the firn in the vicinity of the Summit site, alongside a separate measurement of the firn at another location on the central Greenlandic ice shelf known as Site A at a similar elevation. Additionally we show the residuals of these data relative to the aforementioned HL model and a 2-part exponential model obtained from a best fit to one of the datasets. This first dataset $\rho_{NSP}(z)$ was measured on June 8th, 2004 to a depth of 30 m approximately 1 km from the main camp at Summit station, utilizing a neutron scatter probe (NSP) lowered into a borehole, which reconstructed the density from the return scatter of neutrons emitted by a radioactive source \cite{Giese_Hawley_2015, Hawley_Morris_McConnell_2008}. The measurement uncertainty was not explicitly defined, but the authors stated that an error of $\delta \rho / \rho \approx 6\% $ is expected when the probe is centered inside a borehole of a diameter of $12 \, \mathrm{cm}$ \cite{Hawley_Morris_McConnell_2008}, whereas the borehole used had a diameter of $\approx 14.1 \, \mathrm{cm}$ \cite{HawleyPhD2004}. Also shown is a direct density measurement of ice cores $\rho_{AK}(z)$ made in 1988 to a depth of 120 m at Site A $70^{\circ}45^{\prime}\,\mathrm{N},\;35^{\circ}57.5^{\prime}\,\mathrm{W};\;3145~\mathrm{m~a.s.l.}$) located approximately 222 km south east of Summit \cite{Alley_Koci_1988}. Finally a direct measurement of the density $\rho_{RET}$ to a depth of 13 m at the RET-CR deployment site on May 16th \& 17th of 2024 \cite{Kyriacou:2025tj}. Cores were taken with a coring drill, trimmed to ensure consistent density, and then sliced to a uniform thickness of 10 cm, and therefore constant volume and then weighed to a precision of $1 \, \mathrm{g}$. Snow chips produced during cutting (corresponding to $\sim 2.2\%$ of the core volume) were included in the mass measurement, though some loss due to wind is expected. Accounting for chip loss and possible core misalignment, we conservatively estimate a relative density uncertainty of $\delta\rho/\rho \lesssim 4.4\%$, superseding a larger estimate ($\delta \rho / \rho >7\%$) reported in earlier proceedings \cite{Kyriacou:2025tj}.
\\~\\
In addition to HL parameterization $\rho_{HL}(z)$, we show the 2-part exponential model $\rho_{2,exp}(z)$ obtained via best fit to the NSP data, and a 3-part exponential model $\rho_{3,exp}(z)$ derived for the Summit site in a previous work \cite{ali2024modelingrefractiveindexprofile}. The best fit parameters to $\rho_{2,exp}(z)$ are: $\rho_{s} = 300 \pm (5 \cdot10^{-3}) \, \mathrm{kg/m^{3}}$, $k_{0} = (38.2 \pm 13) \cdot 10^{-3} \, \mathrm{m^{-1}}$, $k_{1} = (14.8 \pm 2.2) \cdot 10^{-3} \, \mathrm{m^{-1}}$. From this fit we expect the critical depth at $z_{550} = 13.56 \pm 0.69 \, \mathrm{m}$, whereas the HL model predicts this to be slightly deeper at $z_{550} = 14.9 \, \mathrm{m}$. The agreement between the datasets and the models is defined by finding the RMS of the residual of the data to the model $\sigma_{\Delta \rho} = \frac{1}{N} \sqrt{\Sigma^{N}_{i} (\rho_{data,i}-\rho_{model,i})^{2}}$, as shown in the table \ref{tab:rho_model_and_data}.
\begin{table}[t]
    \centering
    \renewcommand{\arraystretch}{1.2}
    \begin{tabular*}{\columnwidth}{@{\extracolsep{\fill}}|c|c|c|c|}
        \hline
         & 2,exp &  3,exp &  HL \\
        \hline
        RET & 3.273 & 5.168 & 3.877 \\
        \hline
        NSP &  1.408 & 3.401 & 2.162 \\
        \hline
        AK & 3.201 & 2.292 & 2.086 \\
        \hline
    \end{tabular*}
    \caption{RMS residual $\sigma_{\Delta \rho}\, [\mathrm{kg/m^{3}}]$ between the density data and models shown in Fig \ref{fig:rho_profile_summit}.}
    \label{tab:rho_model_and_data}
\end{table}
\\~\\
All 3 parameterizations describe the overall densification of the firn as seen in the datasets, particularly in the region of $5 \, \mathrm{m} < z < 60 \, \mathrm{m}$. Unsurprisingly the 2-part exponential shows the best agreement with the NSP dataset, while the HL parameterization has a lower $\sigma_{\Delta \rho}$ value than the 3-part exponential model for all 3 datasets, showing it to be a competitive model compared to the fits. The Site-A density begins to rise relative to the HL model at depths $z > 60 \mathrm{m}$, as seen in Fig. \ref{fig:rho_profile_summit}, while showing better agreement than as compared to the 2-part exponential fit. Density residuals are the most prominent within the shallow firn, with the density residuals contained within an interval of $-100 \, \mathrm{kg/m^{3}} < \rho-\rho_{HL} < 80 \, \mathrm{kg\,m^{3}}$. Negative residuals compared the HL model are noticeable in the RET and NSP datasets for depths below 5 m, while positive residuals are present through the shallow firn, and in the case of the AK dataset, positive residuals density can be seen at depths deeper than 20 m. The consistency of these residuals between the datasets suggests that they reflect real variation in the firn instead of being a result of measurement uncertainties.
\subsection{Modeling Firn Evolution}
To investigate the effect of temporally evolving firn density at Summit Station, Greenland, we used glaciological simulation software known as the `Community Firn Model' (CFM) \cite{cfm_v1}. CFM is a modular code that models the physical and chemical properties of a volume of firn, including, but not limited to: density, temperature, concentration of isotopes and meltwater \cite{cfm_v1}. CFM can model changes in these properties in response to changing air temperature, surface accumulation, changing surface density, melt events, and other factors. CFM uses a Lagrangian grid with a fixed number of volumes, each corresponding to a layer of firn with identical properties. At each iteration $i$, a new layer is added to the top, corresponding to newly accumulated snow over the time interval $\Delta t_{i}$, and the deepest simulated layer is removed from the bottom. The density profile is evolved from $\rho(z, t_{i})$ to $\rho(z, t_{i+1})$ using the specified densification equation, with changes in temperature, melt-water concentration, and overburden pressure being forced from the surface.
\\~\\
For this study, the firn simulation was initiated with an initial firn model, with the starting density profile $\rho_{init}(z)$ estimated using the HL model $\rho_{init} = \rho_{HL}$ with the aforementioned temperature, accumulation and surface density inputs. The density profile for each year $y$ and month $m$ is designated as $\rho_{y,m}(z)$. The model domain extends down to $z = 120 \, \mathrm{m}$, with a depth binning of $\Delta z = 0.01 \, \mathrm{m}$. Below $120 \, \mathrm{m}$ the depths are estimated using a best-fit asymptotic exponential function as shown in Equation \ref{firn_exponential}. The time interval $\Delta t_{i}$ is one month, with the first iteration at January 1980 $\rho_{1980,1}(z)$ and the final at December 2021 $\rho_{2021,12}(z)$. The surface density was held constant at a value of $\rho_{0} = 350 \, \mathrm{kg/m^{3}}$. For each iteration, the monthly averaged accumulation rate $A_{i}$, skin temperature $T_{i}$ for Summit is obtained from the online database \textit{Modern-Era Retrospective analysis for Research and Application version 2} (MERRA-2) \cite{MERRA_2}, with more details about this data explained in Appendix \ref{CFM_explained}. We utilized the Kuipers-Munneke (KM) densification model to define the densification rate for each time step following the initialization \cite{kuipers2015elevation}. The KM model has been widely validated and used for Summit Station, Greenland, and provides a physically based description of dry-snow compaction suitable for modeling long-term firn evolution under cold, low-melt conditions \cite{cfm_v1}. From the output density models $\rho_{y,m}(z)$, a set of refractive index profiles $n_{y,m}(z)$ was calculated using the empirically derived linear relationship between $n$ and firn density $\rho$\cite{kovacs1995situ}:
\begin{equation}\label{kovacs-eq}
    n(z) = 1 + 0.845 \frac{\rho(z)}{\mathrm{g/cm^{3}}}.
\end{equation}
\subsection{Results for Density and Temperature}\label{CFM_results}
\begin{figure}[t]
    \centering
    \begin{subfigure}[t]{\columnwidth}
        \centering
        \includegraphics[width=0.96\linewidth]{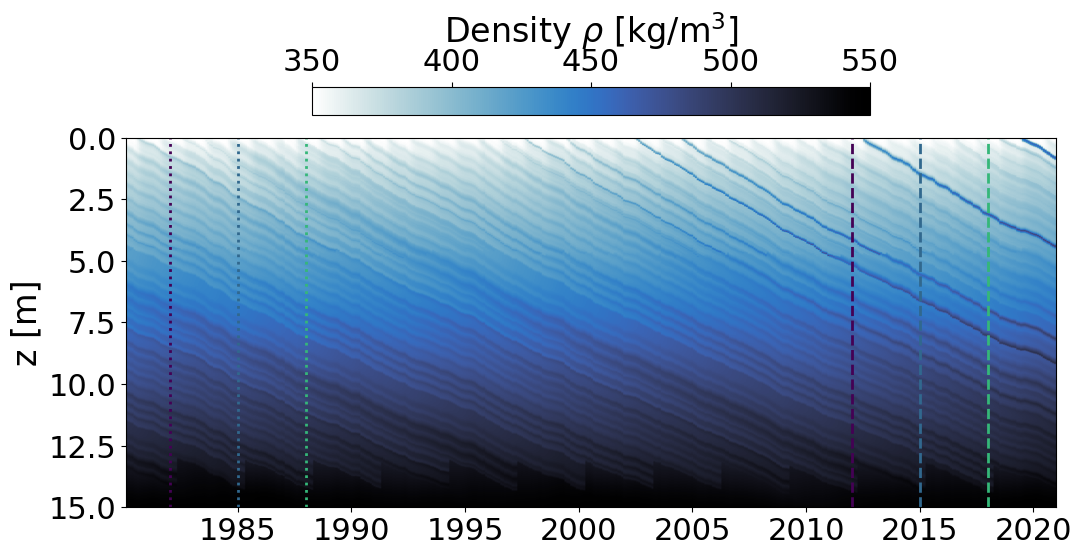}
        \caption{Density evolution map $\rho_{CFM}(z,t)$}
        \label{fig:cfm_map}
    \end{subfigure}
    \begin{subfigure}[t]{\columnwidth}
        \centering
        \includegraphics[width=0.96\linewidth]{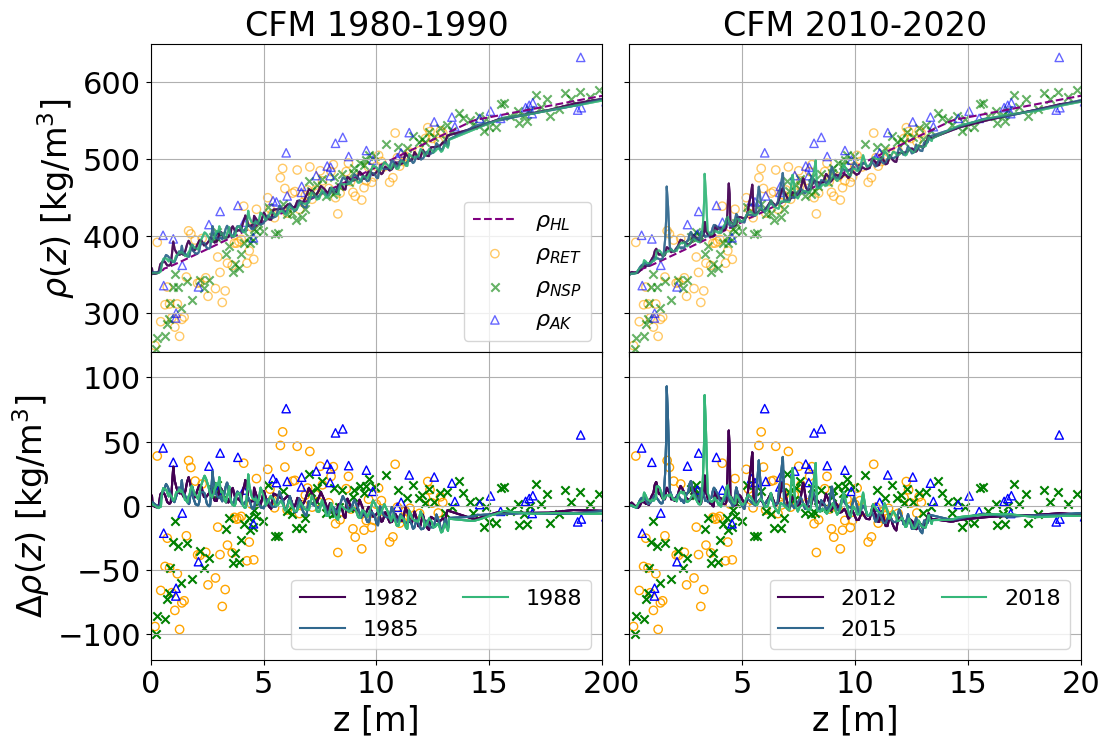}
        \caption{The shallow firn density \& residuals (left: 1980-1990 \& right: 2010-2020)}
        \label{fig:rho_prof_cfm_both}
    \end{subfigure}
    \begin{subfigure}[t]{\columnwidth}
        \centering
        \includegraphics[width=0.96\linewidth]{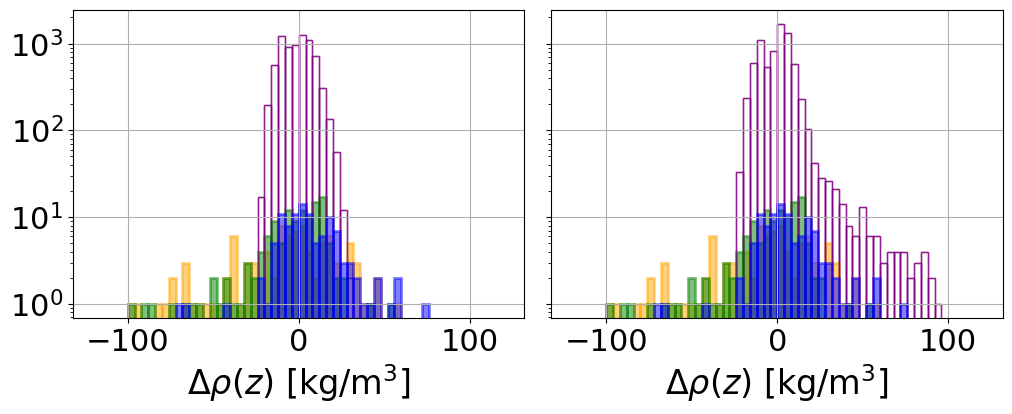}
        \caption{Histogram of density residuals.(\ref{fig:rho_prof_cfm_both})}
        \label{fig:rho_prof_cfm_hist}
    \end{subfigure}
    \caption{Above: A color-map of the time-dependent density profile of the shallow firn from 1980 through to 2020. The colored lines indicate the CFM refractive index model outputs $n_{x,7}(z)$ for month 7 of 2012, 2015 \& 2018, which are some of those used to quantify the RF signal variation. Below: shows the discrete density profiles of the aforementioned dates.}
    \label{fig:cfm_model}
\end{figure}

Using the MERRA-2 data inputs, the firn temperature, melt-water concentration, water run-off, and isotope concentrations were computed as a function of depth for each simulated epoch. The temperature for shallow depths of $z < 10 \, \mathrm{m}$ undergoes a seasonal oscillation, with a maximum surface temperature of $-5^{\circ} \, \mathrm{C}$ to $-10^{\circ} \, \mathrm{C}$ in July to a minimum of approximately $ -50^{\circ} \, \mathrm{C}$ in February. The temperature oscillation propagates deeper into the ice while undergoing dampening, with an isothermal depth at ${\approx}10 \, \mathrm{m}$ (shown in Appendix \ref{atten_appendix}), a result consistent with field measurements of the firn temperature at Summit \cite{Giese_Hawley_2015}. To quantify the density fluctuation, we define density residuals $\Delta \rho_{y,m}(z)$, which are defined with respect to the starting profile $\Delta \rho_{y,m}(z) = \rho_{y,m}(z) - \rho_{init}(z)$. 
\\~\\
For the shallow firn ($z < 15 \, \mathrm{m}$), one can classify `small scale' density fluctuations with $|\Delta \rho| < 25 \, \mathrm{kg/m^{3}}$, that are present across all the dates, and `large scale' density fluctuations $\Delta \rho > 25 \, \mathrm{kg/m^{3}}$ that are prominent from the year 2000 onward. The large scale density fluctuations correspond to surface melt events, which we discuss in more detail below. The residuals of the small-scale fluctuations obey a Gaussian-like distribution, with a least-squares fit yielding a standard deviation $\sigma_{\Delta \rho,CFM}(z<15 \, \mathrm{m}) = 7.24 \, \mathrm{kg/m^{3}}$ and a mean $\mu_{\Delta \rho, CFM}(z<15\, \mathrm{m}) = 0.078 \, \mathrm{kg/m^{3}}$ for the decade of 1980-1990, as shown in Fig.\,\ref{fig:rho_prof_cfm_hist}. The small-scale fluctuations are a result of the seasonal temperature oscillation in the upper $10 \, \mathrm{m}$ of the firn. The large-scale fluctuations act as outliers from this distribution, with $\Delta \rho$ reaching values of $\Delta \rho = 60 - 100 \, \mathrm{kg/m^{3}}$ shown in Fig.\,\ref{fig:rho_prof_cfm_hist}. For the deep firn, the density residuals are smaller, with a standard deviation of $\sigma_{\Delta \rho,CFM}(z>15 \, \mathrm{m}) \approx 1.1 \, \mathrm{kg/m^{3}}$. The situation for the decade of 1990-2000 resembles that of 1980-1990, with an absence of large melt events, whereas the subsequent decade of 2000-2010 more resembles that of 2010-2020, with melting events occurring in 2002 and 2004. 
\\~\\
In the CFM modeling, large-scale fluctuations form within $0.4 \, \mathrm{m}$ of the surface in correlation with surface-melt events, as the melt water percolates down into the snow and refreezes once the surrounding temperature drops below freezing. With typical densities of $400 - 550 \, \mathrm{kg/m^{3}}$ these refrozen layers persist inside the shallow firn at relatively constant densities at progressively deeper depths as snow accumulates above them, acting as a record of past melt events until they arrive at a depth where the surrounding firn's density matches that of the layer. The formation and deepening of the ice layers are visible in Fig \ref{fig:cfm_map}, where a color-map of the time-dependent density profile $\rho(z,t)$ is displayed. For example, a `refrozen' layer forms at $z = 0.3 \, \mathrm{m}$ in the June 2002 profile, and by June 2020 is visible at a depth of $z= 9.1 \, \mathrm{m}$. The four largest surface-melt events in the dataset by melt volume, occurring in the summer months of 2002, 2004, 2012, and 2019 (the latter featuring a rain event at Summit), correspond to the densest refrozen layers.
\\~\\
For depths greater than $z>6\,\mathrm{m}$, the CFM profiles show good agreement with the available density measurements and reproduce the broad densification trend represented by the HL initial profile. In the shallow firn ($z<15\,\mathrm{m}$), the comparison is more complex. The dense layers generated in the 2010--2020 CFM profiles have magnitudes comparable to the largest density excursions observed in the $\rho_{AK}$ and $\rho_{RET}$ datasets. However, in the upper 6 m, the CFM models tends to overestimate the mean density and does not fully reproduce the stochastic variability observed in the measurements. This is partly attributable to the prescribed fixed surface density, $\rho_s=350\,\mathrm{kg/m^{3}}$, and to the absence of unresolved near-surface variability in the forcing mechanisms driving the densification.
\\~\\
As a sensitivity check, we also considered CFM configurations with variable surface density. These produce a broader distribution of shallow densities while preserving the formation of dense layers associated with melt events. The corresponding results for density, and for radio-propagation (investigated using the same procedure as outlined in section \ref{radio-sim}), are discussed in Appendix \ref{appendix:rho_s_var}. We therefore interpret the baseline CFM profiles used in this work as physically motivated realizations of time-dependent firn evolution, rather than as unique reconstructions of the true Summit density profile. The RF simulations presented in the following section quantify the propagation effects associated with such evolving firn structure. Alternative initial conditions, finer temporal forcing, and improved constraints on surface-density variability may further improve agreement with the shallow density measurements.
\section{Simulating RF Propagation through the Firn}\label{radio-sim}
Having modeled the seasonal variation in the refractive index, we produced a set of refractive index profiles, $n_{y,m}(z)$, which were then used to generate simulated radio propagation data using the linear relation between density and refractive index described by Eq. \ref{kovacs-eq} \cite{kovacs1995situ}.
\subsection{RF Simulation Methods}
To examine the resulting variation in radio signal properties, we solved Maxwell’s equations numerically across the entire simulation domain and sampled the signal at discrete receiver positions. These simulations were performed using the open-source code \texttt{MEEP} \cite{MEEP}, a versatile Finite Difference Time-Domain (FDTD) solver. FDTD operates by discretizing the simulation geometry into a grid of cells, within which Maxwell’s equations are solved in discrete time steps. Boundary conditions were implemented using a radiating dipole source initialized at $t_{0}= 50 \, \mathrm{ns}$ and perfectly matched layers (PML) at the edges of the domain to prevent artificial reflections. Given the high level of accuracy of FDTD methods, our conclusions are largely based on the \texttt{MEEP} simulation outputs.
\\~\\
We additionally utilized the parabolic wave equation (PE) method, solved using the simulation code \texttt{paraPropPython} or simply \texttt{paraProp}. The PE method is based on the parabolic approximation of Maxwell's equations, assuming cylindrical symmetry in the medium and a vertically polarized and monochromatic electric field $E_{z}(r,z,\omega)$ propagating from the source outwards in the radial direction and assuming a $e^{i\omega t}$ time dependence \cite{paraprop_prohira}. Using a step-wise spatial solver, the field $E_{z}(r,z,\omega)$ is solved throughout the entire geometry. Time-domain solutions can be obtained by repeating this process across the frequency domain of the source signal. The PE method, as implemented in \texttt{paraProp}, provides accurate radio solutions within a paraxial angle $\theta_{paraxial} = 45^{\circ}$ relative to the radial direction of the domain and has been verified against FDTD codes within geometries comparable to that used in this study \cite{paraprop_prohira}. The greater simulation efficiency of PEs compared to FDTD allows solutions to be computed on time scales that are orders of magnitude smaller, allowing for the possibility of Monte-Carlo-type analyses and the simulation of larger simulation domains than possible for FDTD methods.
\\~\\
Finally, numerical ray tracing (RT) was used to approximate the path taken by RF signals from source to receiver, allowing us to discern useful information such as the arrival angle of different signals at receivers, independently check if the signal has undergone reflection or refraction and the portion of the signal's travel path spent in different zones of the ice. The \texttt{SignalProp} module of the \texttt{NuRadioMC} code was used to generate ray tracing solutions for given pairs of sources and receivers, utilizing the \texttt{radiopropa} solver \cite{NuRadioMC, RadioPropa}. The RT solutions used throughout Sections \ref{results-section} and \ref{discussion} were solved using the best fit of the double exponential model (described in Eq. \ref{firn_exponential}) to the CFM profile corresponding to January 2010, with the parameters and the profile described in the Appendix \ref{vertex-appendix}.
\subsection{Geometry and Source Modeling}\label{geometry-and-source-modelling}
\begin{figure}[t]
    \centering
    \includegraphics[width=0.95\linewidth]{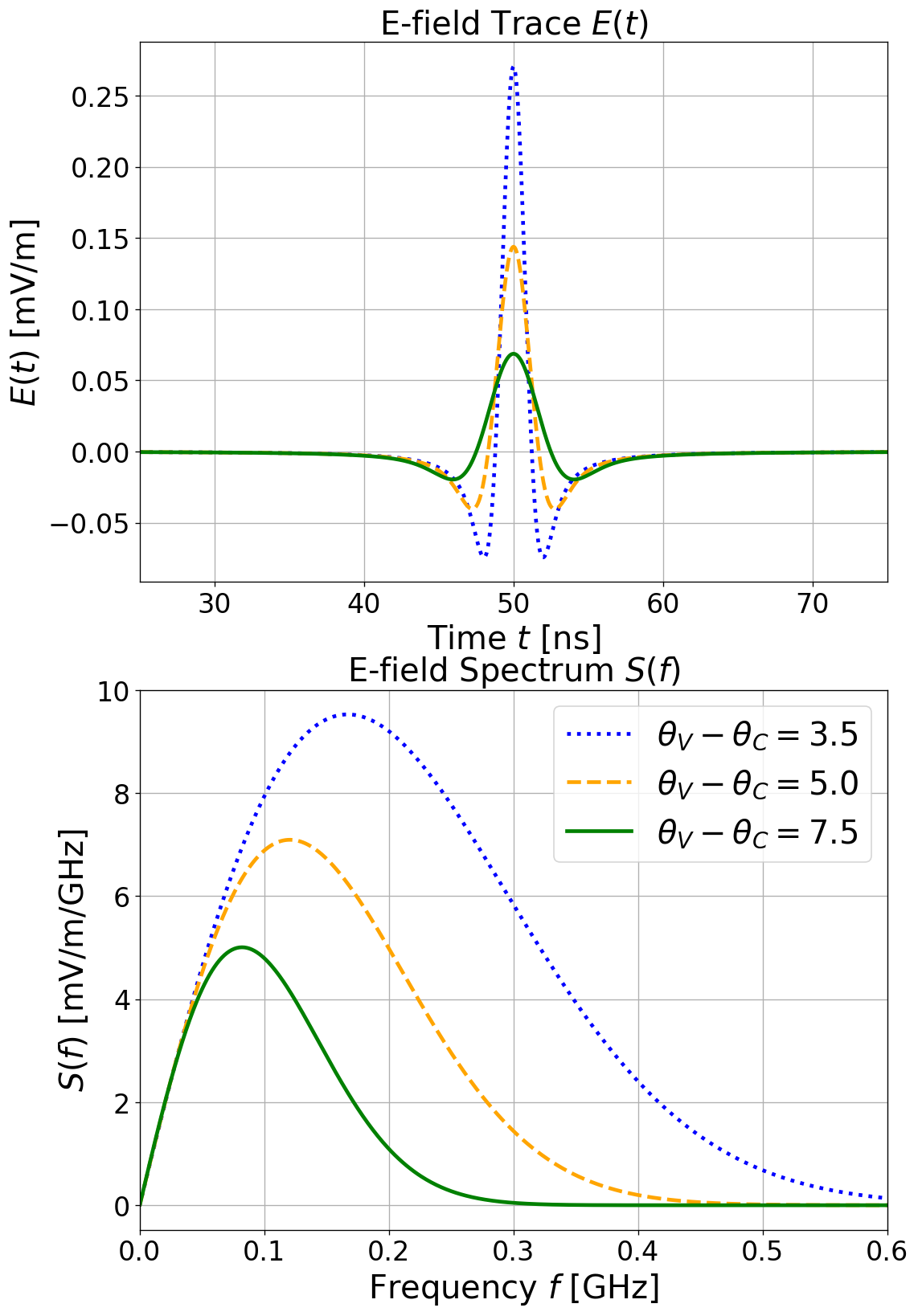}
    \caption{The test pulses (top) \& spectra (bottom) injected into the source in the RF simulations, with the former initialized such that the amplitude maximum occurs at $t_{0} = 50 \, \mathrm{ns}$. The spectra are modeled using the ARZ parameterization of a hadronic shower observed from viewing-angle offsets of $\Delta \theta_{VC} = |\theta_{V} - \theta_{C}| = 3.5^{\circ}, \, 5.0^{\circ} \, \& \, 7.5^{\circ}$ \cite{Alvarez_Mu_iz_2000}. Parameters such as the bandwidth and time spread of the pulses are summarized in Table \ref{tab:signal_properties}.}
    \label{fig:source_waveform}
\end{figure}

A cylindrically symmetric simulation geometry is used, reducing the problem to two dimensions: radial distance $x$ and depth $z$. This choice, taken to minimize computational costs, can be justified by invoking the relatively flat terrain at Summit, the absence of large crevasses that can act as planar reflectors, and the rapidly diminishing power of pipe-like and point-like reflectors, with the power $P_{rx}$ diminishing with distance $L$ described with $ P_{rx} \propto L^{-3}$ \& $P_{rx} \propto L^{-4}$ respectively. The ice-air boundary is defined as perfectly flat at $z = 0$, with air occupying the region $z < 0$ and ice at $z \geq 0$. The geometry consists of two distinct regions: an ice cylinder extending to a radius of $X_{ice} = 520\,\mathrm{m}$ and depth of $Z_{ice} = 200\,\mathrm{m}$, with a refractive index profile given by the corresponding $n_{y,m}(z)$; an overlying air cylinder of homogeneous refractive index $n = 1$ with the same radial extent and a height of $H_{air} = 10\,\mathrm{m}$. 
\\~\\
To exclude any artificial reflections caused at the edges of the simulation domain, an outer `buffer' radius of 30 m is defined, in addition to a 40 m PML. As a further precaution from artificial reflections, two buffer layers of width 40 m were used for the upper and lower edge of the domain respectively, in addition to a 40 m PML layer applied to the top and bottom side of the respective buffer layers. The computational grid had a cell size of $\Delta x_{cell} = 0.033\,\mathrm{m}$, resulting in a FDTD time step of $\Delta t = 0.056\,\mathrm{ns}$. As the in-simulation sampling rate is higher than that of radio neutrino experiments, only every $5^{th}$ time step was sampled and recorded, resulting in an effective sampling rate of $3.6 \, \mathrm{GHz}$ and sampling time interval of $0.28 \, \mathrm{ns}$. Given the simulation's resolution and size specifications, frequencies above $0.5\,\mathrm{GHz}$ experienced noticeable numerical dispersion, so the usable signal spectrum was limited to well below this threshold when using FDTD. The in-simulation time-window has a length of $t_{max} = 3200 \,\mathrm{ns}$, ensuring that all signal arrivals and their relevant decay tails were captured at the receiver locations.
\\~\\
The emission of a band-limited RF impulse was simulated from a short dipole source located at a depth $z_{tx}$ within the ice as a test pulse. The frequency dependence of its spectrum follows the Alvarez-Romero-Zas (ARZ) parametrization for the Askaryan spectrum of a $E_{sh} = 10^{18}\,\mathrm{eV}$ hadronic shower, viewed at an angle offset by $\Delta\theta_{VC} = |\theta_{V} - \theta_{C}|$ from the Cherenkov angle \cite{ARZ_model}, referred to hereafter as the viewing-angle offset. This parameterization is used here solely as a form factor to define a relatively `Askaryan-like' pulse envelope; we do not explicitly model the cascade nor the near field emission. To explore the wavelength dependence of ice-induced signal modulation we utilized spectra with viewing angle offsets $\Delta \theta_{VC} = 3.5^{\circ}$, $5.0^{\circ}$  and $7.5^{\circ}$, respectively. As the viewing angle deviates further from the Cherenkov angle, the coherence breaks down for successively longer wavelengths and the signal power diminishes, leading to the successively narrower and dimmer spectra seen in Fig \ref{fig:source_waveform}. Given the aforementioned issue of numerical dispersion in FDTD, only the latter signal ($\Delta \theta_{VC} = 7.5^{\circ}$) is used in the \texttt{MEEP} simulations whereas all 3 are utilized to initialize the \texttt{paraProp} simulations. The emitted signal for $\Delta \theta_{VC} = 7.5^{\circ}$ has a bandwidth of $156\,\mathrm{MHz}$, defined using 3 dB lower frequency cutoff at $f_{low} = 19\,\mathrm{MHz}$ and a high frequency cutoff at $f_{high} = 175\,\mathrm{MHz}$, and has its maximum spectral power at $f_{max} = 82\,\mathrm{MHz}$. The resulting pulse has a full width half-maximum (FWHM) time spread of $t_{\mathrm{FWHM}} = 6.41\,\mathrm{ns}$, with the pulse and spectrum shown in Fig.\,\ref{fig:source_waveform}. The low frequency cutoff $f_{low}$, frequency maximum amplitude $f_{max}$, bandwidth $B$, and FWHM time spread $t_{FWHM}$ of the signals for the different observer angles are listed in Table \ref{tab:signal_properties}. As the source was modeled as a short dipole antenna, a single simulation run was able to illuminate most of the receiver array, allowing the investigation of ice-induced propagation effects across a broad range of paths. 
\begin{table}[t]
\centering
\renewcommand{\arraystretch}{1.2}
\begin{tabular*}{\columnwidth}{@{\extracolsep{\fill}}|c|c|c|c|c|}
    \hline
    $\Delta \theta_{VC}\, [^{\circ}]$  & $f_{low} \, [\mathrm{GHz}]$  & $B \, [\mathrm{GHz}]$ & $f_{max}\, [\mathrm{GHz}]$ & $t_{\mathrm{FWHM}} \,[\mathrm{ns}]$ \\
    \hline
    3.5 & 0.076 & 0.53 & 0.167 & 3.07 \\
    \hline
    5.0 & 0.056 & 0.37 & 0.120 & 4.31 \\
    \hline
    7.5 & 0.038 & 0.25 & 0.82 & 6.41 \\
    \hline
\end{tabular*}
\caption{The spectral properties of the transmitted signal (using different inputs of $\Delta \theta_{VC}$): the low cutoff frequency $f_{low}$, the bandwidth $B$, the frequency of maximum amplitude $f_{max}$, and the FWHM spread of the pulse $t_{\mathrm{FWHM}}$. The corresponding pulses and spectra are displayed in Fig. \ref{fig:source_waveform}.}
\label{tab:signal_properties}
\end{table}
\\~\\
The transmitted pulse was sampled at an array of receivers positioned at radial distances $x_{rx}$ ranging from $160\,\mathrm{m}$ to $520\,\mathrm{m}$ in increments of $\Delta x_{rx} = 10\,\mathrm{m}$, and at depths $z_{rx}$ from $4\,\mathrm{m}$ to $160\,\mathrm{m}$ in increments of $\Delta z_{rx} = 4 \,\mathrm{m}$. These receiver positions were chosen to broadly overlap the depths sampled by past and existing radio and radar neutrino detectors, including ARIANNA, RNO-G, RET-CR, as well as the proposed depths of the IceCube-Gen2 Radio Array. The computational constraints of running \texttt{MEEP} at this resolution and scale prevented us from simulating these effects at depths $z_{rx} > 160 \, \mathrm{m}$, utilized by the RICE and ARA detectors, so we use ray-tracing arguments to extrapolate our results to greater depths.
\\~\\
For simplicity, each refractive index profile $n_{y,m}(x,z)$ was assumed to be range-independent, such that $n_{y,m}(x,z) = n_{y,m}(z)$. Although this may not reflect the true horizontal variability present at Summit, it offers a useful approximation for isolating vertical profile effects. The refractive index profiles were discretized with a vertical resolution of $0.1\,\mathrm{m}$, finer than the shortest simulated wavelength $\lambda_{min} \approx 0.33\,\mathrm{m}$, ensuring accurate representation of refractive index gradients within the simulation. Finally, the refractive index was modeled with no imaginary (conductive) component, and thus we assume no dielectric attenuation across the domain. The reason for this is that seasonal variation in the attenuation constant is driven by the temperature oscillation within the firn, which primarily occurs in the upper 5 m to 10 m, and thus the attenuation variation is only sampled by a relatively small portion of the ray path. An examination of this effect is included in Appendix \ref{atten_appendix} where it was found that the largest variation in signal attenuation was on the order of 0.1 dB for secondary signals, and is generally lower by orders of magnitude.
\subsection{Signal Observables}
Given an ice model described by the two-part exponential function expressed in Eq. \ref{firn_exponential}, the received signal is composed of two pulses: a primary pulse traveling from the source to the receiver which arrives earlier and generally has a larger amplitude, and a secondary pulse that arrives later, generally has a lower amplitude owing to its greater travel path.
\\~\\
The latter has been reflected by the air-snow interface at the surface, leading to an inversion of polarity and further amplitude decay via transmission loss, or has been refracted without reaching the surface. Indeed, for most of the receiver domain, the signal power is overwhelmingly composed of two pulses which correspond to these two paths through the firn and ice. In keeping with convention, we describe the primary, least-time signal as `direct' even though it may have undergone refraction along its trajectory. This simple bi-pulse description does not hold true within the so-called shadow zone where emission is forbidden under classical ray tracing, discussed in greater detail in the following section. Outside the shadow zone, the primary and secondary pulses are referred to as `D signals' and `R signals' respectively. We define the following signal parameters to characterize our simulated signals:
\begin{itemize}
    \item Propagation times of the D and R signals: $t_{D} \, [\mathrm{ns}]$ \& $t_{R}\,[\mathrm{ns}]$: are defined as the time of maximum amplitude of the D and R signals,
    \item Time delay $\Delta t_{DR} = t_{R} - t_{D} \, [\mathrm{ns}]$:\\ The relative propagation time delay between $t_{D}$ and $t_{R}$,
    \item Fluence: $\phi^{E} = \epsilon c \int^{t_{1}}_{t_{0}} E_{RX}(t)^{2} dt$ $[\mathrm{eV/m^{2}}]$:\\ The integrated RF energy flux between times $t_{0}$ and $t_{1}$, with $\epsilon$ being the absolute local permittivity, $E_{RX}(t)$ the electric field at the receiver, and $c$ the speed of light.
\end{itemize}
The fluence $\phi^{E}$ can be estimated for the entire waveform or can be estimated independently for the D signals and R signals, allowing us to disentangle effects from different components of the ice-sheet. Hence we will define `total fluence' $\phi_{tot}^{E}$ as distinct from the fluence of the individual D signal $\phi^{E}_{D}$ or R signal $\phi^{E}_{R}$. The D fluence $\phi^{E}_{D}$ is integrated over the time window $t_{D} - t_{res} \, \mathrm{ns}$ to $t_{D} + t_{res} \, \mathrm{ns}$. Here $t_{res}$ is set to a value of $t_{res} = 2  t_{\mathrm{FWHM}} = 12.82 \, \mathrm{ns}$, selected to correspond to twice the FWHM time spread of the source pulse generated for a viewing-angle offset of $\Delta \theta_{VC} = 7.5^{\circ}$. The same time window is used to calculate $\phi_{R}^{E}$ by integrating about $t_{R}$.
\\~\\
To identify the D  and R pulses for a given RX position, we find the time of the two brightest peaks in the Hilbert envelope of the waveform. If ray tracing (RT) predicts two solutions for the RX, and if the peaks are separated by at least $t_{res}/2$, we designate the time of the earliest peak as $t_{D}$, and the following peak as $t_{R}$. If RT predicts only one solution (such as near the shadow zone boundary), or no solutions, then the time of brightest emission of the waveform is designated $t_{D}$. We caution the reader that while we define the brightest signal as the D signal for all cases where an R signal can not be identified with ray tracing, this relationship is not necessarily true in the shadow zone. Furthermore, we define another region of the geometry as the `overlap' region; where ray tracing predicts two solutions such that $\Delta t_{DR} \leq t_{res}$. In this region, the time-spread of the pulses generally results in overlap and interference between the D and R pulses, meaning that the variation in time of flight and fluence can not be viewed in isolation for the two propagation paths. Thus, one must be cautious to interpret the results for time and fluence variation within the shadow and illuminated zones.
\\~\\
To quantify the magnitude of variation in the signals from different times for signal parameter $x_{i}$ for refractive index scenario $n_{y,m}$, we calculate the \textit{variation coefficient} $\delta x/x$:
\begin{equation}
    \frac{\delta x}{x} = \frac{\sigma(x)}{\mu(x)},
\end{equation}
Where $\mu(x)$ is the mean parameter of $x$ across all CFM scenarios and $\sigma(x)$ the standard deviation. Given that fluence can vary by orders of magnitude in our simulation domain due to propagation losses, it is prudent to examine the variation coefficient which we also refer to as the \textit{relative variation}, as outlined in Section \ref{section:fluence_variance}. In other cases, we examine the seasonal distribution of the relative fluence residuals at specific receivers. In Section \ref{section:propagation_time_variance}, we examine the \textit{absolute variation} of the absolute propagation times $\delta t = \sigma(t)$ for the D and R signals and their relative offset $\Delta t_{DR}$.
\subsection{Propagation Domains}\label{prop-domain}
The degree and nature of seasonal fluctuation in received RF signals differs considerably depending on the relative position of the receiver to the transmitter, and hence on the path taken by the signal through the ice according to ray tracing. To illustrate this, we divide the receivers by depth into \textit{shallow} receivers and \textit{deep} receivers, the former located at $z_{rx} < 15 \, \mathrm{m}$ and the latter at $z_{rx} > 15 \, \mathrm{m}$. We further divide the deep geometry by depth and range into three distinct propagation zones as defined by ray tracing; the \textit{reflection}, \textit{refraction} and \textit{shadow} zones.
\\~\\
Signals observed by deep receivers in the reflection and refraction zones are characterized by the two signal components; the primary (D) signal, which may propagate in a straight line or may have had its path curved through refraction, followed by the secondary signal (R) that is either reflected off the ice-air interface, or undergoes refraction above the depth of the receiver. The reflection zone extends from the surface to the deepest reaches of the ice, and extends from the transmitter range $x_{tx} = 0$ to the critical reflection boundary describe by a range $x_{L}(z)$. Beyond this boundary is the refraction zone, where the secondary signal propagates from the source in an up-going arc but is refracted to a down-going trajectory before it can reach the surface, instead reaching a minimum depth or depth of closest-approach to the ice-air interface. It is instructive to subdivide this zone into the shallow and deep refraction zones, where the secondary signal's closest approach to the surface is within the shallow firn and below it, respectively. 
\\~\\
The separation between the shallow and deep refraction zones is hence designated by the \textit{shallow refraction boundary} range $x_{R,15}(z)$. Beyond the \textit{shadow boundary} $x_{S}(z)$ is the shadow zone, where no valid ray path is permitted under classical ray tracing. In fact, shadow zone RF signals have been observed in multiple field experiments and, in the presence of density fluctuations, are predicted in this zone by FDTD and PE simulation methods, including in this study \cite{Deaconu_2018, paraprop_prohira}. We will also refer to the aforementioned `overlap region' where the $\Delta t_{DR}$ predicted from ray-tracing is less than or equal to twice the FWHM spread time of the pulse.
\section{Results}\label{results-section}
The principal result of this study is that seasonal density fluctuations in the firn, examined over the decade 2010 to 2020, produce the strongest variation in received RF fluence along shallow refracted paths, while direct and deep-reflected signals remain comparatively stable. In contrast, the impact on signal arrival times is generally on a scale of nanosecond or sub-nanosecond scale outside of the shadow zone. As will be demonstrated in Section \ref{discussion}, the variation in both quantities result in uncertainty in neutrino reconstruction.
\subsection{Fluence by domain}
\begin{figure}[t]
    \centering
    \begin{subfigure}[t]{\columnwidth}
        \centering
        \includegraphics[width=\linewidth]{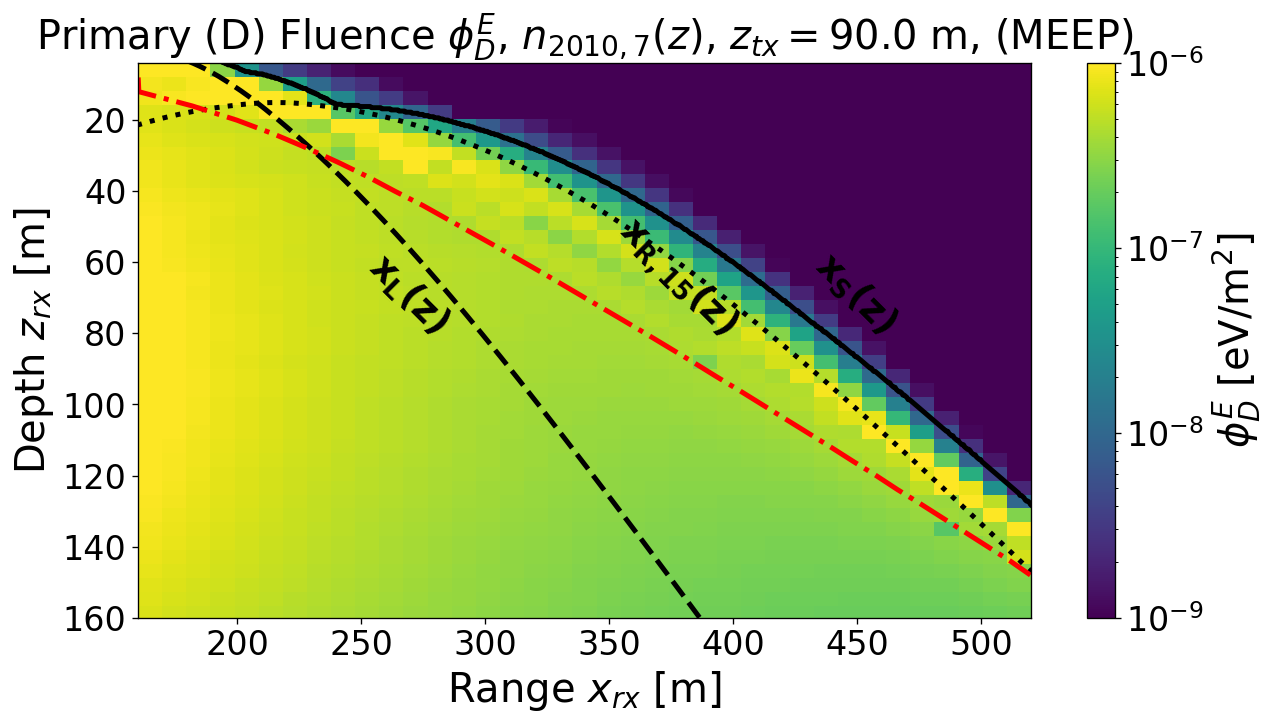}
        \caption{D-Fluence $\phi^{E}_{D}$}
        \label{fig:fluence_D_years}
    \end{subfigure}
    \begin{subfigure}[t]{\columnwidth}
        \centering
        \includegraphics[width=\linewidth]{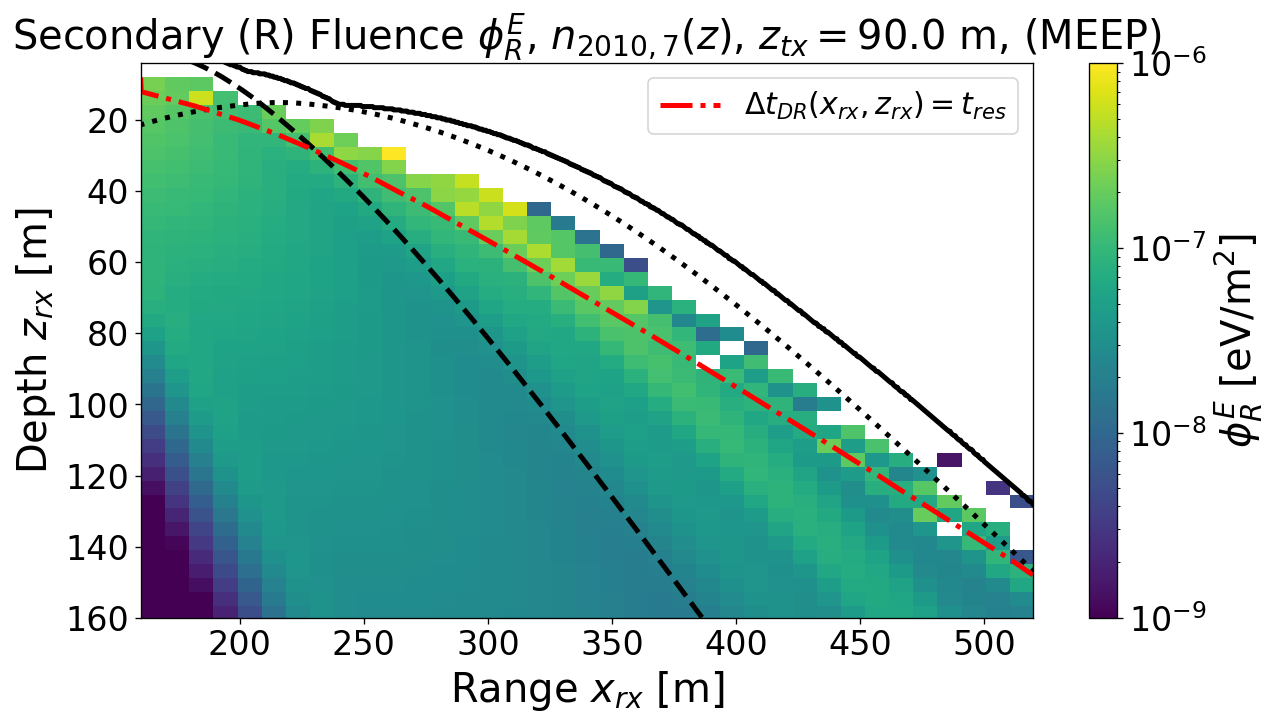}
        \caption{R-Fluence $\phi^{E}_{R}$}
        \label{fig:fluence_R_years}
    \end{subfigure}
    \caption{A heatmap showing the received signal fluence $\phi^{E}$ (Fig. \ref{fig:fluence_D_years}: D signal, Fig. \ref{fig:fluence_R_years} R signal) as a function of receiver position with spline interpolation, from a source at $z_{tx} = 90 \, \mathrm{m}$ propagating through the refractive index profile $n_{2015,7}(z)$. The different propagation regions are indicated with the boundary lines $x_{L}$, $x_{R,15}$ and $x_{S}$. An additional line indicates the overlap region, where the D and R signals have a relative delay of $\Delta t_{DR} < t_{res} = 12.82 \, \mathrm{ns}$.}
    \label{fig:fluence_D_and_R_years}
\end{figure}

To display the characteristics of the RF signal in the different propagation zones, in Fig.\,\ref{fig:fluence_D_and_R_years} we show heatmaps of the signal fluence $\phi^{E}$ for the 2015 July firn model $n_{2015,7}(z)$ the primary fluence $\phi^{E}_{D}$ in Fig.\,\ref{fig:fluence_D_years} and the secondary fluence $\phi^{E}_{R}$ in Fig.\,\ref{fig:fluence_R_years}. The latter heatmap excludes the shadow zone by construction due to the breakdown of a simple `bi-pulse' model in the shadow zone, the former simply defines the brightest peak in the waveform as being associated with the primary signal. A refrozen ice layer located at $z \approx 2 \, \mathrm{m}$ is a major feature of $n_{2015,7}(z)$, as shown in Fig.\,\ref{fig:rho_prof_cfm_both}. The critical reflection boundary $x_{L}(z)$, the shallow refraction boundary $x_{R,15}(z)$ and the shadow zone boundary $x_{S}(z)$ are also indicated using lines. The shadow zone is evident as a dark region in the top right of Fig. \ref{fig:fluence_D_years}, where the fluence is reduced by at least two orders of magnitude compared to the non-shadow zone, but non-zero. Relatively bright fluence `hotspots' are present between the shadow zone boundary and are clearly evident in both maps. As these are contained within the `overlap' region, its possible that we are seeing artificial enhancement of the D and R fluence due to positive interference between the distinct pulses. This boundary band may also be associated with the inflection point in the refractive index at $z_{550} = 14.9 \, \mathrm{m}$, causing multiple parallel ray-paths to constructively interfere. Between the reflection and shallow refraction boundaries, there are bands of relatively high fluence visible in Fig.\,\ref{fig:fluence_R_years}, with the peak fluence at $z = 160 \, \mathrm{m}$ located at a range $400 \, \mathrm{m} <x< 420 \, \mathrm{m}$ and the other brighter band's peak located at ranges $450 \, \mathrm{m} < x < 500 \, \mathrm{m}$. These bands suggest constructive and destructive interference within the shallow refracted zone resulting from the presence of the refrozen layer at $z \approx 2 \, \mathrm{m}$.
\subsection{Spatial Variation}
To quantify seasonal fluctuations in RF fluence on a year to year scale, we compare simulated outputs from July of each year between 2010 and 2020, and examine the average variation in the fluence and propagation time for the individual pulse components. Appendix \ref{signal-variation} contains descriptions of the modification of the waveforms for receivers located in the different propagation domains.
\subsubsection{Fluence}\label{section:fluence_variance}
\begin{figure}[t]
    \centering
    \begin{subfigure}[t]{\columnwidth}
        \centering
        \includegraphics[width=\linewidth]{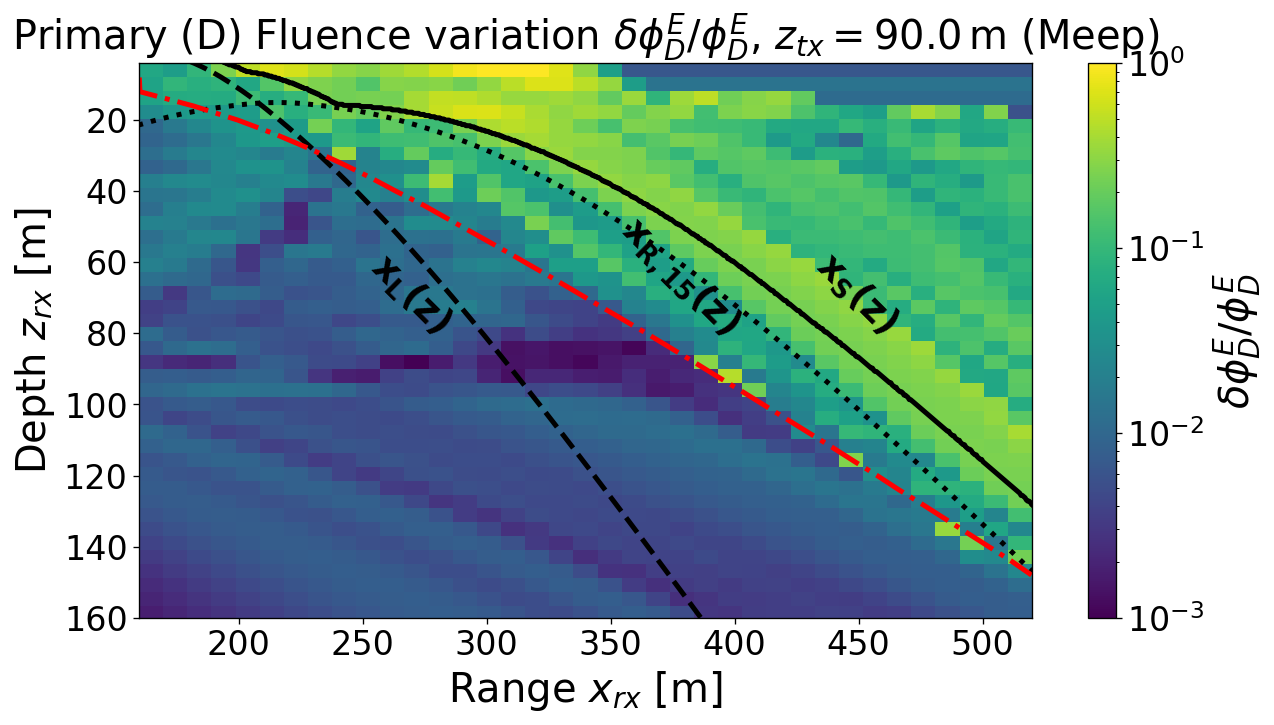}
        \caption{D-fluence variation coefficient: $\delta \phi^{E}_{D}/\phi^{E}_{D}$}        
        \label{fig:fluence_var_D_years}
    \end{subfigure}
    \begin{subfigure}[t]{\columnwidth}
        \centering
        \includegraphics[width=\linewidth]{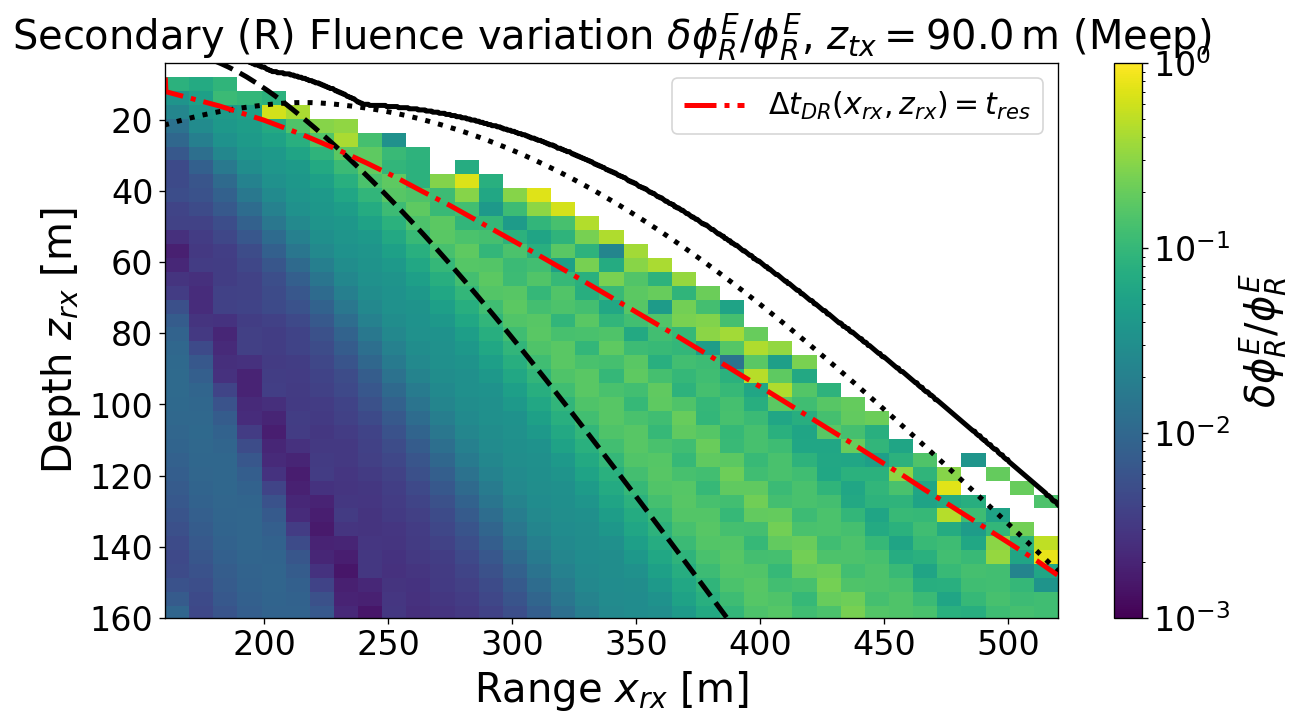}
        \caption{R-fluence variation coefficient: $\delta \phi^{E}_{R}/\phi^{E}_{R}$}
        \label{fig:fluence_var_R_years}
    \end{subfigure}
    \caption{Heatmaps showing the variation coefficient of the fluence $\delta \phi^{E}/\phi^{E}$ for the ice models for month 7 and years 2010 through to 2020 (a: D-fluence, b: R-fluence).}
    \label{fig:fluence_var_D_and_R_years}
\end{figure}

\begin{figure}[t]
    \centering
    \includegraphics[width=0.95\linewidth]{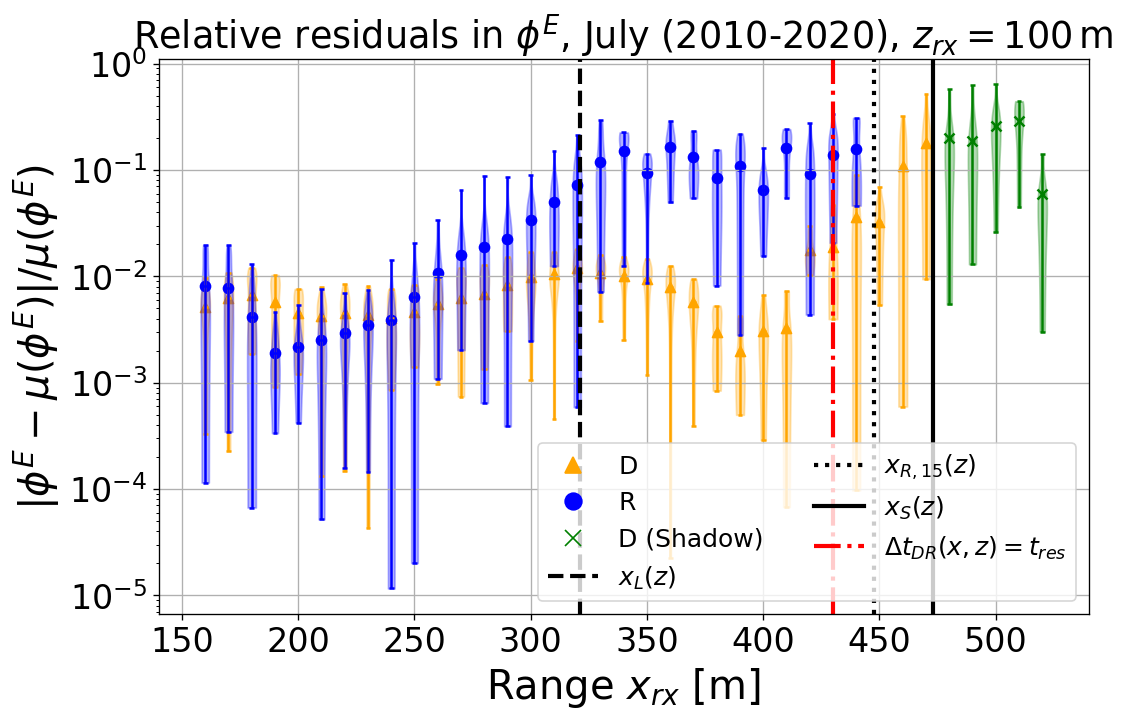}
    \caption{The violin plots show the distribution of relative residuals in the signal fluence $|\phi^{E} - \mu(\phi^{E})|/\mu(\phi^{E})$, for the July ice models over the years 2010–2020, shown at receiver depth $z_{rx} = 100$ m.  The lines indicate the boundaries between the propagation zones. Within the refraction zone, the variation in the R signal fluence exceeds that of the D signal by an order of magnitude. Within the shadow zone $x > x_{L}$, the definition of the primary `D' signal is decoupled from ray-tracing.}
    \label{fig:rx_variance_at_100m}
\end{figure}

\begin{figure}[t]
    \centering
    \includegraphics[width=\linewidth]{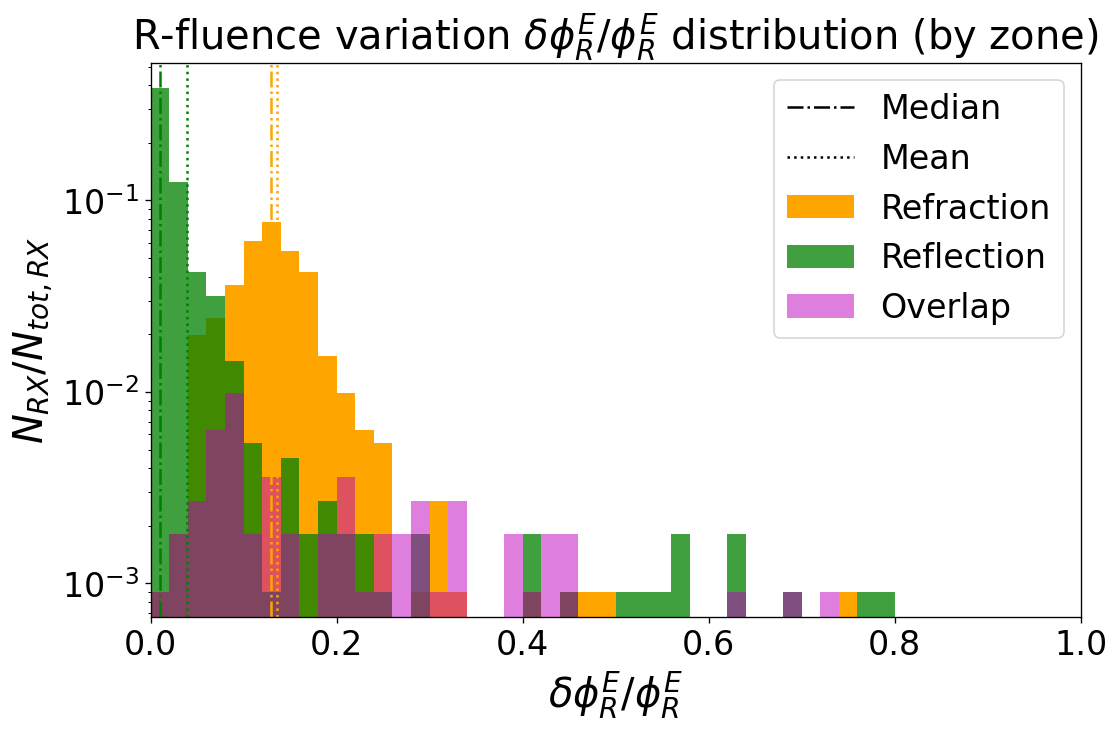}
    \caption{The distribution of R-fluence variation coefficient $\delta \phi_{R}^{E}/\phi_{R}^{E}$ by propagation zone, with the mean and median values displayed (and summarized in Table \ref{tab:rf_variation}).}
    \label{fig:rx_fleunce_dist_1d}
\end{figure}

The resulting spatial distributions of the fluence variation coefficient $\delta \phi^{E}/\phi^{E}$ are shown in Fig.\,\ref{fig:fluence_var_D_and_R_years}, with separate heatmaps of the primary signal (Fig.\,\ref{fig:fluence_var_D_years}), and the secondary signal (Fig.\,\ref{fig:fluence_var_R_years}). In both cases the variation is lowest for deep receivers in the reflection zone, remaining below $10^{-2}$. The three propagation zones stand out with distinct patterns with low variation in the reflection zone for both the primary and secondary fluence, and enhanced variation of the secondary fluence relative to the primary in the refraction zone.
\\~\\
Moving outward from the reflection boundary $x_{L}$, the secondary signal variation rises sharply, exceeding $10^{-1}$ across much of the refraction zone. The heatmaps reveal banded structures within the shallow firn that align with the fluence interference patterns of Fig.\,\ref{fig:fluence_R_years}. These arise from diffraction on refrozen layers, whose downward migration with surface accumulation causes the interference bands to shift. As a result, signals exiting the shallow refraction zone acquire range-dependent amplitudes and reach deep receivers with quasi-random strengths. Three clear bands of high variation appear between the reflection and shallow-refraction boundaries, with a fourth beginning near $z_{rx} = 160 \,\mathrm{m}$. Additional hotspots occur close to the shallow refraction boundary, including a bright region with relative variation above 0.1 between ranges of 200–400 m, and generally elevated values for shallow receivers in the shadow zone. However once again, as this lies in the `overlap region', these results are difficult to disentangle from the primary signal.
\\~\\
In contrast, the primary signal remains stable, with a variation coefficient below $10^{-2}$ throughout the reflection and refraction zones. Because these pulses travel mainly through deeper firn and glacial ice, where density fluctuations are small ($\Delta \rho \lesssim 1 \,\mathrm{kg/m^{3}}$), their amplitudes are largely unaffected. This is illustrated in Fig.\,\ref{fig:rx_variance_at_100m}, which shows the distribution of relative fluence residuals to the mean at 100 m receiver depth across receiver ranges of $160 \, \mathrm{m} \leq x_{rx} \leq 520 \, \mathrm{m}$. Within the refraction zone, the secondary signal exhibits fluctuations that are at least an order of magnitude larger than those of the primary signal, with the spread of the violin plots indicating the range of variability between years. This elevated variation reflects the greater sensitivity of the refracted/reflected paths to small changes in the firn density profile, whereas the direct path remains comparatively stable until approaching the shadow boundary. Beyond the shadow boundary the primary's signal variation rises rapidly, approaching a value similar to the shallow refracted signals, an effect likely related to interference of the direct and secondary signal, or multiple signal paths. We remind the reader that beyond the shadow boundary $x> x_{S}$, the primary (D) signal is simply defined as the brightest peak in the waveform and may not necessarily correspond the shortest earliest arriving signal component in the waveform, a fact indicated with a change of color in the violin plot.
\\~\\
Fig.\,\ref{fig:rx_fleunce_dist_1d} presents the distribution of the variation coefficient of the secondary fluence $\delta \phi^{E}_{R}/\phi^{R}_{E}$ for each receiver, classified by reflection and refraction zones, and normalized to the total number of receivers (we exclude receivers that are located within the overlap zone). The reflection zone's distribution reaches its peak at $\delta \phi^{E}_{R}/\phi^{E}_{R} \approx 9.97 \cdot 10^{-3}$, corresponding to the median value of the distribution. In comparison, the refraction zone distribution peaks at the median value of $\delta \phi^{E}_{R}/\phi^{E}_{R} \approx 0.125$. The mean and median values of the respective distributions of fluence relative variation coefficient, for the direct, secondary and total waveforms, are summarized in Table \ref{tab:rf_variation}. We further investigated the variation in fluence for wider bandwidth signals, defined in section \ref{geometry-and-source-modelling}, and it was found that the variation in fluence is proportional to the bandwidth, implying that these variations are frequency dependent. These results are discussed in the Appendix \ref{section:bandwidth_variance}.
\begin{table*}[t]
\centering
\renewcommand{\arraystretch}{1.2}
\begin{tabular*}{\linewidth}{@{\extracolsep{\fill}}|c|c|c|c|c|c|c|}
    \hline
    \textbf{Zone} & \textbf{Geometry} & $\delta\phi^{E}_{D}/\phi^{E}_{D} \, [\%]$  & $\delta \phi^{E}_{R}/\phi^{E}_{R} \, [\%]$ & $\delta\phi^{E}_{\mathrm{tot}}/\phi^{E}_{\mathrm{tot}} \, [\%]$ & $\delta t_{D} \, [\mathrm{ns}]$ & $\delta \Delta t_{DR}\,[\mathrm{ns}]$ \\
    \hline
    Reflection & $ x_{rx} < x_{L}(z_{rx})$ & 0.89 (0.63) & 3.86 (0.97) & 0.84 (0.61) & 0.19 (0.18) &  0.72 (0.58) \\
    \hline
    Refraction & $x_{L}(z_{rx}) < x_{rx} < x_{s}(z_{rx})$  & 2.89 (0.81) & 13.58 (12.91) & 2.89 (2.04) & 0.42 (0.30) & 1.05 (0.86) \\
    \hline
    Shadow & $x_{rx} > x_{S}(z_{rx})$& 19.43 (15.5) & - & 16.63 (12.60) & 9.31 (2.18) & - \\
    \hline
    Overlap & $\Delta t_{DR}(x_{rx},z_{rx}) < t_{res} $ & 7.58 (5.23) & 20.58 (14.32) & 7.568 (5.2282) & 0.53 (0.39) & 1.45 (0.86) \\
    \hline
\end{tabular*}
\caption{The mean and median fluctuation in RF parameters in different propagation domains. The mean average of the relative variation coefficient (relative for fluence and absolute for time) is shown outside the brackets while the median is shown within the brackets.}
\label{tab:rf_variation}
\end{table*}
\subsubsection{Propagation Time}\label{section:propagation_time_variance}
\begin{figure}[t]
    \centering
    \begin{subfigure}[t]{\columnwidth}
        \centering
        \includegraphics[width=\linewidth]{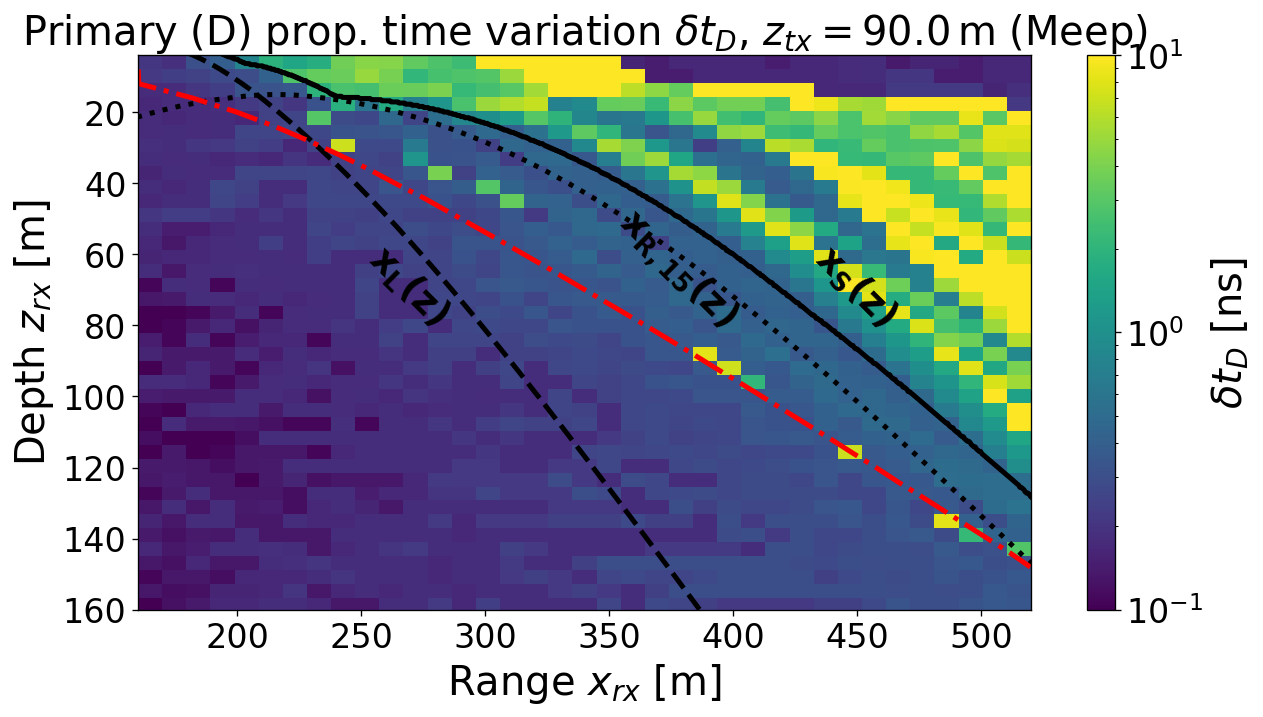}
        \caption{$t_{D}$ distribution over the receiver plane.}
        \label{fig:tD_years_hist2d}
    \end{subfigure}
    \begin{subfigure}[t]{\columnwidth}
        \centering
        \includegraphics[width=\linewidth]{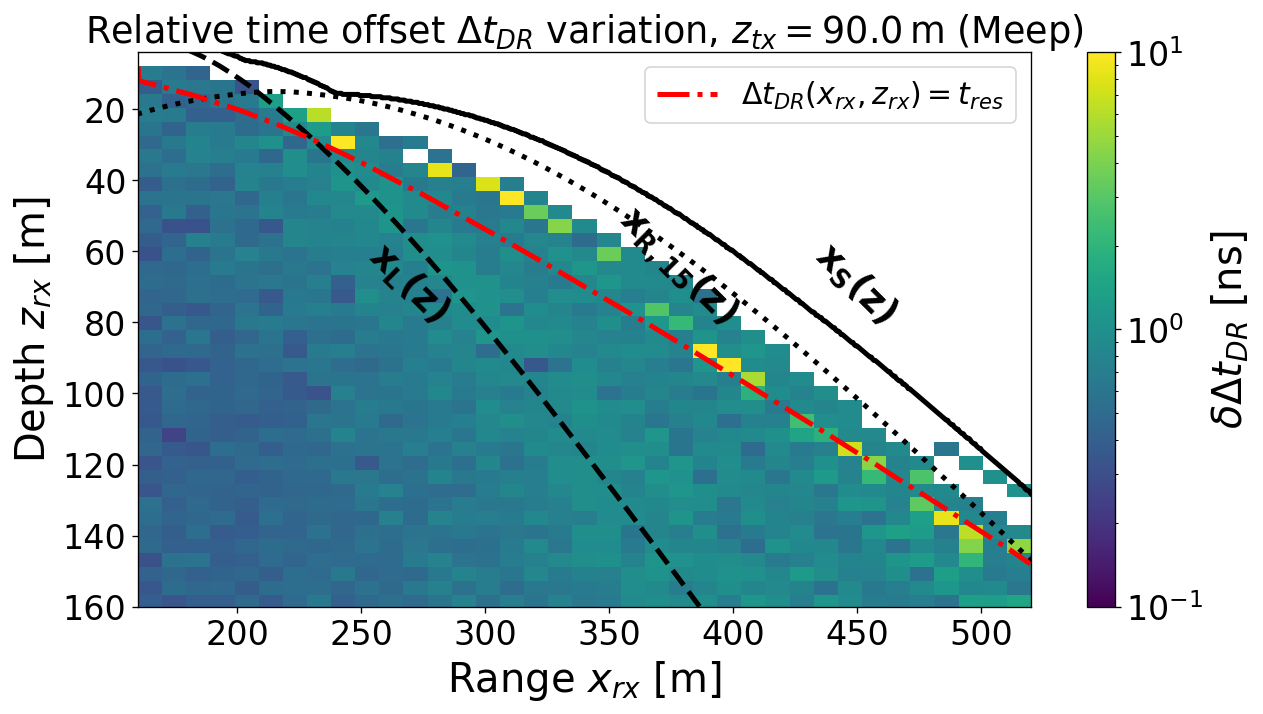}
        \caption{$\Delta t_{DR}$ distribution over the receiver plane.}
        \label{fig:dtDR_years_hist2d}
    \end{subfigure}
    \caption{Heatmap of the absolute variation of the direct propagation time $t_{D}$ relative time delay $\Delta t_{DR}$ at a transmitter at $z_{tx} = 90 , \mathrm{m}$. Overlaid ray trajectories $x_{L}(z)$, $x_{R,15}(z)$, and $x_{S}(z)$ mark the boundaries between the reflected, refracted and shadowed regions, while the an additional line in red indicates the threshold beyond which $\Delta t_{DR} < t_{res} = 12.82 \, \mathrm{ns}$.}
    \label{fig:t_years_hist2d}
\end{figure}

\begin{figure}[t]
    \centering
    \begin{subfigure}[t]{\columnwidth}
        \centering
        \includegraphics[width=\linewidth]{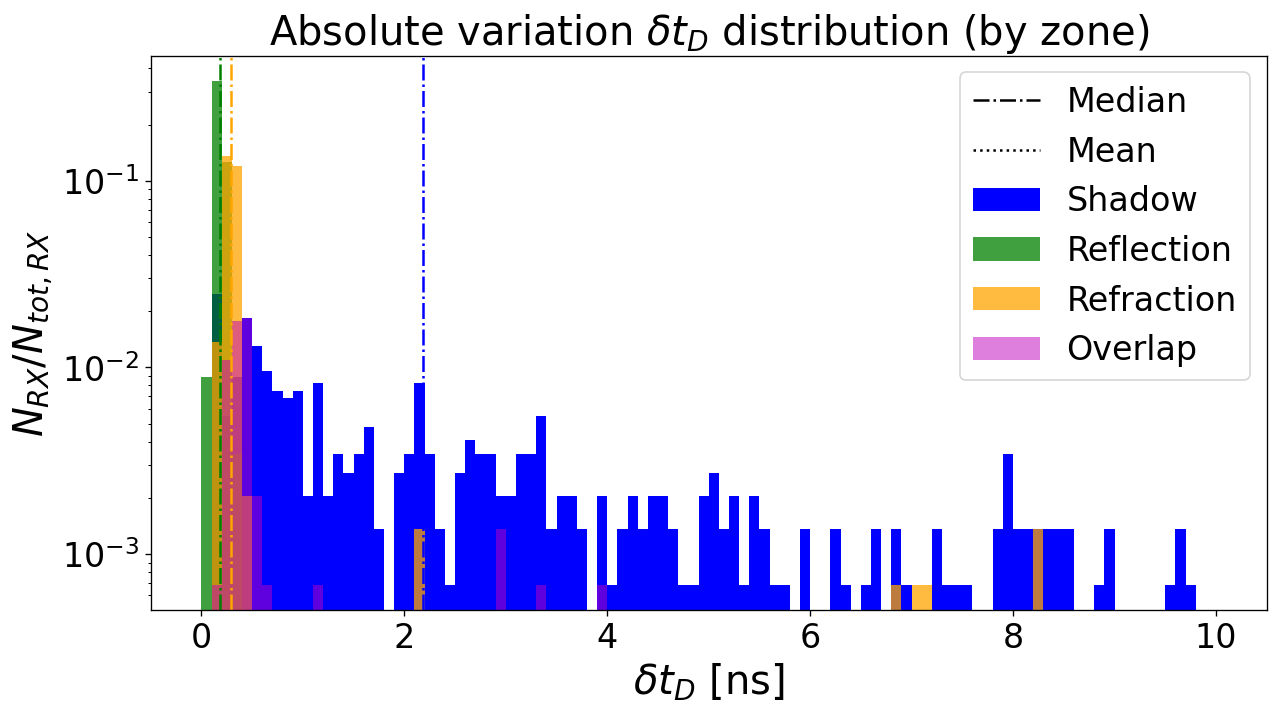}
        \caption{distribution by propagation zone $\delta t_{D}$}
        \label{fig:tD_year_hist1d}
    \end{subfigure}
    \begin{subfigure}[t]{\columnwidth}
        \centering
        \includegraphics[width=\linewidth]{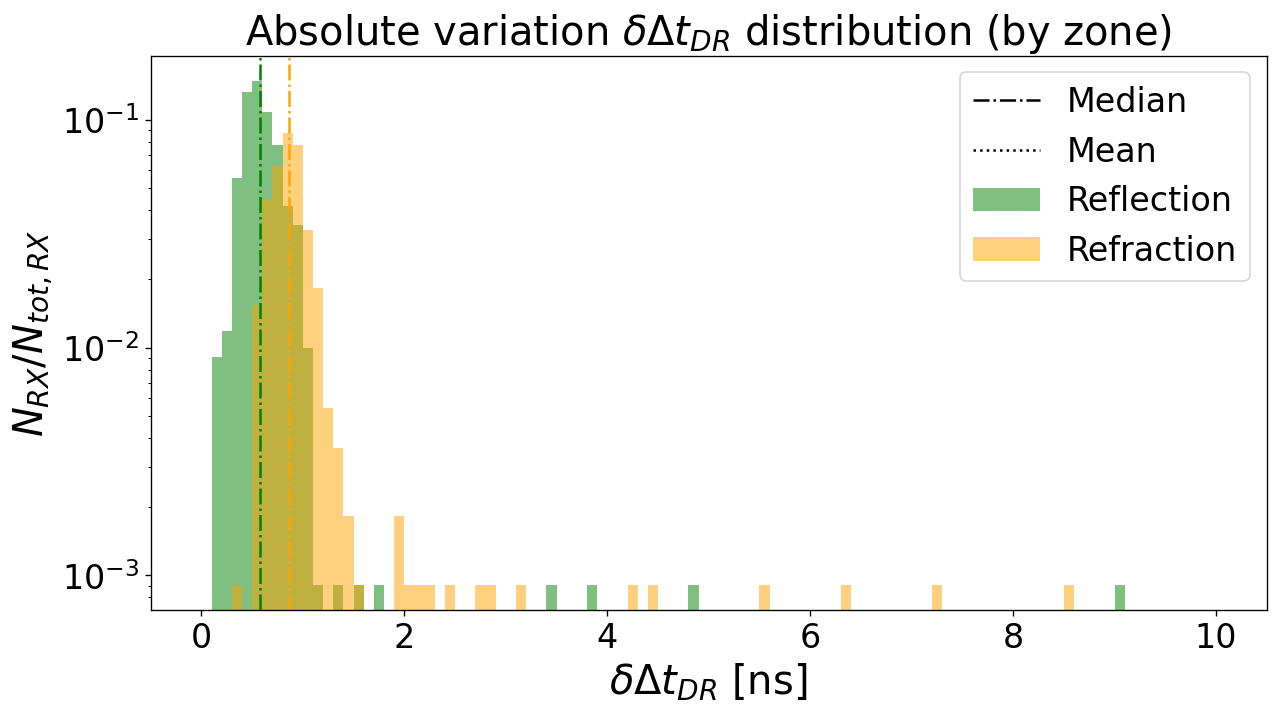}
        \caption{$\Delta t_{DR}$ distribution by propagation zone}
        \label{fig:dtDR_hist1d}
    \end{subfigure}
    \caption{Normalized distribution by of the absolute variation in $t_{D}$ (top) and $\Delta t_{DR}$ (bottom) by propagation zone; reflection, refraction and shadow.}
    \label{fig:t_years_hist1d}
\end{figure}

We investigated variation in the propagation time of the direct (D) and secondary pulse (R) components, and by extension the relative time between them $\Delta t_{DR}$, as defined by the time of maximum amplitude for each waveform component. Here we quantify the absolute variation (or standard deviation) in the absolute propagation time $t_{D}$ and the relative delay time $\Delta t_{DR}$ over 2010–2020. As noted before within the shadow zone it's not possible to confidently identify a solitary reflected or refracted pulse, with the shadow zone waveforms appearing as a superposition of many different ray paths. Thus, we do not draw conclusions about $\Delta t_{DR}$ within the shadow zone. 
\\~\\
The spatial distribution of $\delta t_{D}$ and $\delta \Delta t_{DR}$ are shown in the heatmaps in Fig.\,\ref{fig:t_years_hist2d} (Fig.\,\ref{fig:tD_years_hist2d} and Fig.\,\ref{fig:dtDR_years_hist2d} respectively) and the variation by zone is summarized in Table \ref{tab:rf_variation}. The normalized distribution across receivers of $\delta t_{D}$ and $\delta \Delta t_{DR}$, grouped by propagation zone, are displayed in Fig.\,\ref{fig:t_years_hist1d} (Fig.\,\ref{fig:tD_year_hist1d} \& \ref{fig:dtDR_hist1d} respectively). In comparison to the signal amplitude and fluence, the density fluctuations have an apparently minor effect on the propagation time outside the shadow zone. The distribution of $\delta t_{D}$ within the illuminated zone is almost entirely on the sub-nanosecond level, with mean values of $0.19 \, \mathrm{ns}$ and $0.42 \, \mathrm{ns}$ for the reflection and refraction zones, respectively. By contrast, the shadow zone  shows a much wider distribution extending from $0 \lesssim \delta t_{D} \lesssim 10 \, \mathrm{ns}$, with a mean value of $9.31 \, \mathrm{ns}$.  However, it should be noted that our estimates of $t_{D}$ correspond to the maximum amplitude time rather than the time of the shortest path from the source to the receiver within the shadow zone, complicating the interpretation of this value. Given the comparatively low amplitude of shadow zone signals, they are comparatively difficult to detect given instrumental and environmental noise. In Appendix \ref{shadow-zone}, we also examine the variation of $t_{D}$ using an alternative identification method, based on the time that the cumulative amplitude rises above a threshold. We found that this alternative definition of arrival time results in a lower degree of seasonal variation as compared to the maximum-amplitude definition.
\\~\\
The distribution of the relative time variations $\delta \Delta t_{DR}$ for the reflection and refraction zones includes the increased fluctuations in the path and time of the secondary signals relative to the primary, with mean values of $0.72 \, \mathrm{ns}$ and $1.05 \, \mathrm{ns}$ respectively. These fluctuations are well below the ${\sim}20 \, \mathrm{ns}$ width of the pulses produced assuming a $\Delta \theta_{VC} = 7.5^{\circ}$ Askaryan spectrum. In contrast to $t_{D}$, these fluctuations are directly observable for a single receiver. While we also display the variation of $\delta \Delta t_{DR}$ within the shadow zone for completeness, the physical interpretation of the estimated $\Delta t_{DR}$ in the shadow zone is unclear for reasons already stated.
\\~\\
In a ray optics picture, the magnitude of $\Delta t_{DR}$ fluctuations within the illuminated zones, 0.1 ns to 1.8 ns, appears to correspond to propagation path changes on the order of $\mathcal{O}(10 \, \mathrm{cm})$, accumulated on propagation paths $\mathcal{O}(100 \, \mathrm{m})$ from the source to the receivers. The seasonal perturbations in the density profile can yield subtle changes in propagation path. The direct pulse propagates along a relatively stable refracted path, while the secondary pulse interacts more strongly with refrozen layers, grazing near-critical angles where small seasonal variations in $n(z)$ can induce larger relative shifts. This difference leads to the observed 0.1–1.8 ns scatter in most of the illuminated domain, with localized enhancements along the shadow boundary. Nonetheless, while these values are small compared to the overall pulse width they are potentially observable in radio neutrino detectors.
\subsection{Zenith dependence}\label{section:zenith_variance}
\begin{figure}[t]
    \centering
    \includegraphics[width=\linewidth]{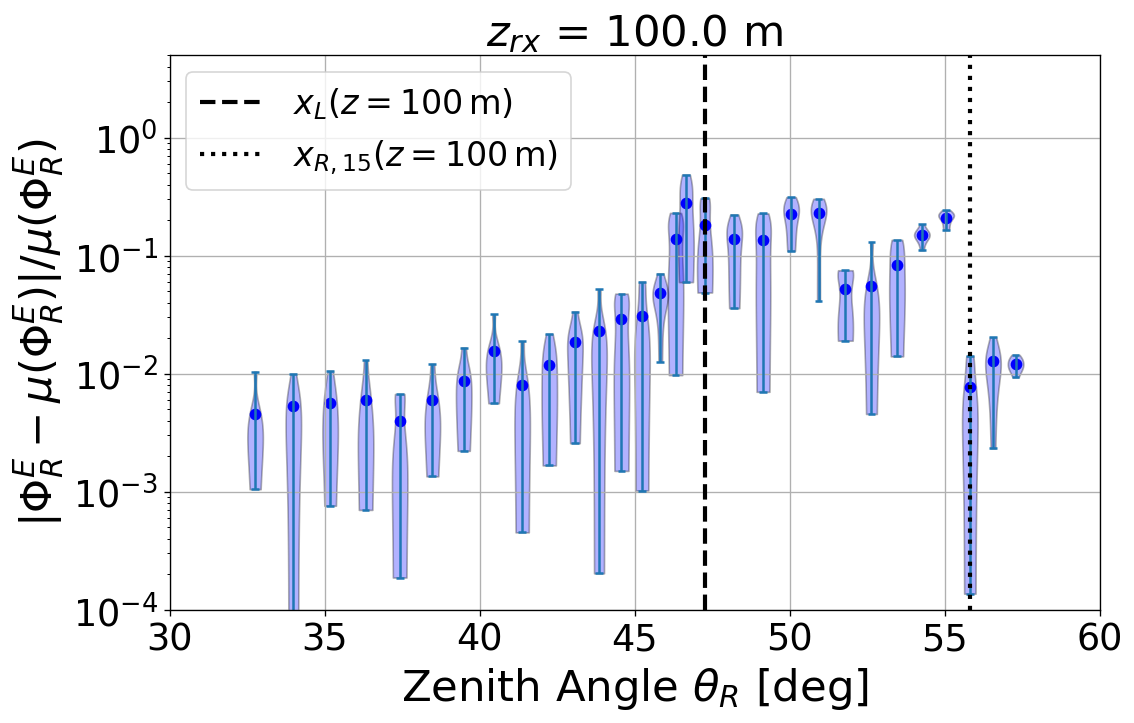}
    \caption{Variation of fluence for $z_{rx} = 100 \, \mathrm{m}$ as a function of zenith angle $\theta_{R}$.}
    \label{fig:fluence_theta}
\end{figure}

From the point of view of an in-ice radio neutrino observatory, the dependence of the amplitude and fluence variation on the path taken through the firn is useful, as it means that for a given receiver depth, there exists a volume of the ice for which the aforementioned signal variation is expected when the neutrino vertex is located within this zone, as illustrated in Fig. \ref{fig:shaded-regions}. Assuming a cylindrical fiducial volume with a depth of $2700 \, \mathrm{m}$ and radius $3000 \, \mathrm{m}$ and receiver at $z_{rx} = 100 \, \mathrm{m}$, approximately $18.5 \%$ of the volume would be located in the receiver's shallow refraction zone, such that ray tracing predicts a path from an observable neutrino-induced radio signal to the receiver that undergoes refraction within the shallow firn without undergoing reflection. This corresponds to $\approx22\%$ of the illuminated (non-shadowed) volume.
\\~\\
In principle, one can use the RF arrival angle at the receiver to discriminate which neutrino events are subject to these effects. If the neutrino vertex is deeper than the receiver depth $z_{rx} = 100 \, \mathrm{m}$, rays arriving at a zenith angle of $\theta_{zen} < 46^{\circ}$ will have undergone reflection off the surface, rays arriving between $46^{\circ} < \theta_{R} < 56^{\circ}$ will have undergone refraction in the shallow firn, while rays $56^{\circ} < \theta_{R} < 90^{\circ}$ underwent refraction between $z_{550}$ and the receiver depth, with all other rays below these angles arriving directly from the neutrino vertex. The dependence of the fluence variation with zenith angle can be seen in Fig.\,\ref{fig:fluence_theta}, with events with a mean $\delta \phi^{E}_{R}/\phi^{E}_{R} > 10^{-2}$ arriving within a range of zenith angles $40^{\circ} < \theta_{R} < 55^{\circ}$, and a narrower region $48^{\circ} < \theta_{R} < 55^{\circ}$ constraining events with  $\delta \phi^{R}_{R}/\phi^{E}_{R} > 10^{-1}$.
\section{Implications for Neutrino Detectors}\label{discussion}
The RF simulations discussed in Section \ref{results-section} suggest that for a significant fraction of neutrino events, $> 18 \%$, seasonal firn fluctuations will result in a relative uncertainty of $>0.1$ for the power of the secondary signal (reflecting/refracting) for a receiver antenna located at depths utilized by past and current in-ice Askaryan detectors, when compared over a decade.
\\~\\
This effect appears to be frequency-dependent, suggested by the increase in variation for higher bandwidth source pulses seen in Fig.\,\ref{fig:3_fluence_variations_x} in Appendix \ref{section:bandwidth_variance}. The properties of interest in neutrino physics and astronomy are the primary neutrino's energy $E_{\nu}$, flavor and arrival direction $(\varphi_{\nu}, \theta_{\nu})$ (with $\varphi$ being the azimuthal direction and $\theta$ the polar direction). Together these may indicate the location of the neutrino sources, the nature of these sources, and the nature of ultra-high energy cosmic rays, for example, from mass composition measurements at the highest energies. Additionally, it has been shown that the neutrino-nucleon cross-section can be constrained to within a factor of 2 of the Standard Model cross-section above 10 PeV with only 10 neutrino events for a generic detector with 1 degree angular resolution and one decade of energy resolution  \cite{UHE_cross_section}. An understanding of seasonal effects is thus important to eventual measurements of neutrino properties.
\\~\\
The accuracy of neutrino energy, vertex and arrival direction reconstruction is contingent on the ability to reconstruct radio emission properties at the source given the available information at the detector. This includes the location of the source of the radio emission (which is in close proximity to the interaction vertex), the radiation pattern, and the total emitted energy. These can in principle be discerned from the RF signal amplitude, arrival direction, polarization, and spectral content. In this section we estimate the approximate uncertainty in neutrino energy and arrival direction resulting from the seasonal variation in RF fluence within the illuminated zone. We assume perfect reconstruction of the electric field $E(RX,t)$ at the receiver $RX$ from the received voltage $V(RX,t)$, neglecting detector-specific antenna effects. Since the PE simulation explicitly assumes a vertically polarized signal under cylindrical symmetry and the FDTD code used in this study makes the same assumption, our current simulations are unable to comment on seasonal influences on polarization reconstruction; that topic is deferred for future work.
\\~\\
To demonstrate the seasonal variation in the reconstructed neutrino parameters, we model a string of 6 vertically polarized antennas in a phased array between depths of $z_{rx,min} = 96 \, \mathrm{m}$ to $z_{rx,max} = 120 \, \mathrm{m}$, spaced at $4 \, \mathrm{m}$ intervals, located at some range $x_{rx,ph}$ from the source and within the illuminated zone. The vertex position ($x_{tx}, z_{tx}$) is calculated by finding an optimal fit to the arrival times of the \texttt{MEEP}-simulated waveforms at the receivers using ray tracing, described in greater detail below in Section \ref{vertex-reco}. A signal arrival direction $\theta_{RX}$ and signal propagation distances $L$ are then estimated for the direct and secondary signals arriving at an equivalent antenna located in the center of the phased array, and the uncertainty is found from the variation in the parameters across the ice models. The spectrum-derived quantities; the viewing angle offset $\Delta \theta_{VC}$, fluence $\phi^{E}$ and ultimately shower energy $E_{sh}$ for the phased array, are calculated from the coherent average of the D and R waveform components at each antenna in the array, with the coherent averages estimated independently for each component, and the variation in the resulting averaged waveform as described in Section \ref{energy-reco}. Additionally, we address the impact of seasonal variation on neutrino reconstruction within the shadow zone in the Appendix \ref{shadow-zone}.
\subsection{Vertex Reconstruction}\label{vertex-reco}
Given a model of the ice, when the phased array is within the illuminated zone, the source position may be inferred via ray tracing. The accuracy of the reconstruction is dependent on the accuracy of the ice model as well as the experimental timing resolution. Consequently, the seasonal variation in the arrival times discussed in Section \ref{section:propagation_time_variance} impose a limit on the resolution of the source reconstruction. As we approximated the neutrino source with a dipole source, all rays arriving at the receiver are convergent on the position of the source at ($x_{tx} = 0, z_{tx} = 90 \, \mathrm{m}$). 
\\~\\
Utilizing the \texttt{radiopropa} numerical ray-tracer implemented in the \texttt{NuRadioMC} module \texttt{SignalProp}, we calculate an array of arrival times $t^{RT}_{D}$ and $t^{RT}_{R}$ for an array of transmitting antennas located within depths, $10 \, \mathrm{m} < z_{tx} < 200 \, \mathrm{m}$, ranges $ -100 \, \mathrm{m}< x_{tx} < 100 \,\mathrm{m} $, with spacings of $\Delta z_{tx} = 2 \, \mathrm{m}$ in depth and $\Delta x_{tx} = 10 \, \mathrm{m}$. The ray tracing solutions were estimated using a double-exponential profile (quantified by Eq.\,\ref{firn_exponential}) fit to the CFM profile $n_{CFM,2010,1}(z)$, with the fit parameters described in the Appendix \ref{vertex-appendix}. The rays are sampled at receiver antennas with depths corresponding to the aforementioned phased array and located at ranges $150 \, \mathrm{m} < x_{rx} < 500 \, \mathrm{m}$ with spacing $\Delta x_{rx} = 20 \, \mathrm{m}$. For each point in this array, we use \texttt{MEEP} to estimate a penalty score $\chi^{2}$ from the offset of the arrival times estimated of the $D$ and $R$ signals from ray-tracing, relative to the time of the first arriving direct signal at the highest receiver $t_{0} = t_{D}(z_{rx,min})$. The penalty score for $\Delta t_{D} = t_{D}-t_{0}$ is then:
\begin{multline}\label{penalty-score}
    \chi^{2}_{D}(x_{tx}, z_{tx},x_{rx}) =\\ \sum^{N_{ph}}_{i} |\Delta t^{RT}_{D}(x_{tx},z_{tx},x_{rx},z_{rx,i}) - \Delta t^{meep}_{D}(x_{rx},z_{rx,i})|^{2},
\end{multline}
Where $N_{ph}$ is the number of antennas in the phased array and $i$ corresponds to the number of the phased array channel, indexed from $z_{rx,min}$ to $z_{rx,max}$. The best fit source ($x_{tx,reco}, z_{tx,reco}$) is therefore found from the minimum penalty score. The penalty score $\chi^{2}_{R}$ is estimated similarly for $t_{R}$, which is still offset from the earliest arrival time $t_{0} =t_{D}(z_{rx,min})$. Finally, a combined penalty score $\chi^{2}_{DR}$ is estimated by multiplying the two: $\chi^{2}_{DR} = \chi^{2}_{D}\cdot\chi^{2}_{R}$. We estimate the penalty scores across the ensemble of sources and receivers, for each ice model $n_{CFM}(z)$ utilized in the Results Section \ref{results-section}, and thereby estimate the variation in the reconstructed source position. We include an example of penalty scores $\chi^{2}$ mapped over the candidate TX space, as well as a plot of reconstructed TX positions in Appendix \ref{vertex-appendix}.
\\~\\
By taking the best-fit source position, we can further estimate the variation in propagation lengths $L_{D}$ \& $L_{R}$ from the source to average depth of the receiver string $\hat{z}_{rx} = (z_{rx,min} + z_{rx,max})/2$. In the same way, we also estimate launch $\theta_{TX,D}$ \& $\theta_{TX,R}$ and receiver angles $\theta_{RX,D}$ \& $\theta_{RX,R}$ at the average receiver depth. Since variations in the launch angles and receiver angles exhibit absolute correlation, it is sufficient for our purposes to consider the variation in the receiver angle alone.
\begin{figure}[t]
    \centering
    \includegraphics[width=\linewidth]{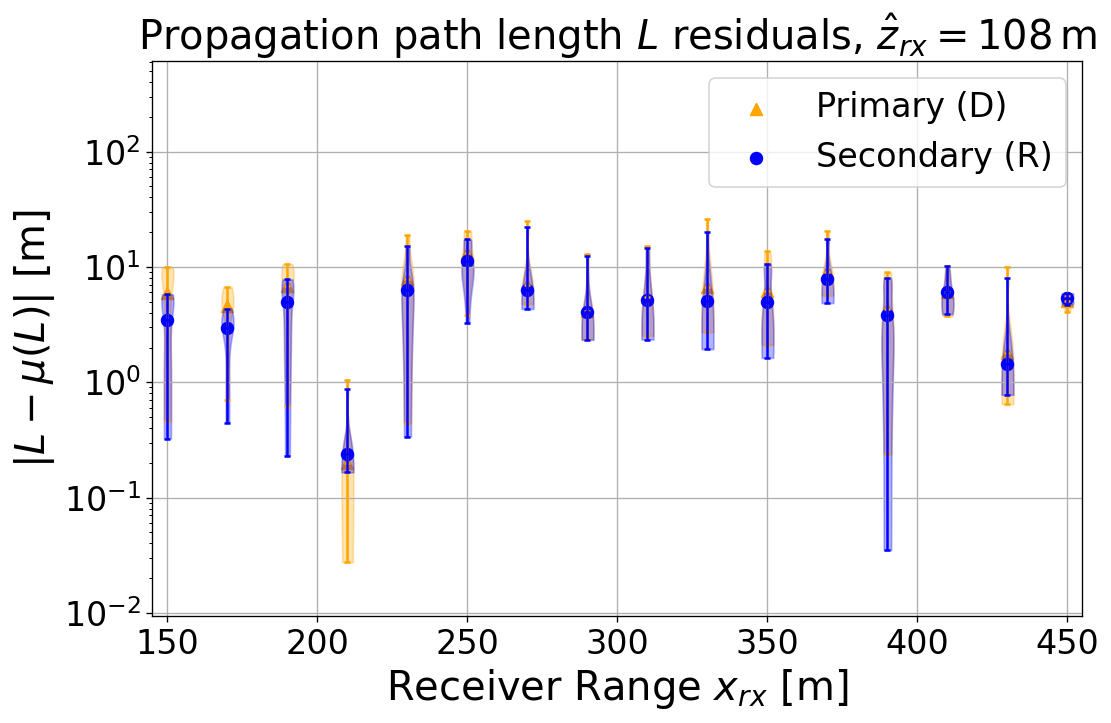}
    \caption{Propagation length $L$ reconstruction.}
    \label{fig:distance_reco_L}
\end{figure}
The variation in the reconstructed propagation path length $L$, both for the direct path and the secondary path are shown as a function of range $x_{rx}$ in Fig. \ref{fig:distance_reco_L}. In comparison to variation in the fluence, the variation in the propagation path is relatively flat, with a change $1 \, \mathrm{m} \lesssim \delta L \lesssim 10 \, \mathrm{m}$ across the receiver range until the shadow zone, with minimal difference between D and R signal components. This property follows from the vertex reconstruction being estimated from the propagation time, and the relatively minor difference in the variation of arrival times within the reflection and refraction zones. The change in the propagation time appears to be a function of the vertical path taken by the ray between the source and the receiver. A similar picture emerges for the derived receiver angle, as outlined in Section \ref{theta-reco}. The residual errors in the path length additionally contribute to the energy uncertainty, albeit as a sub-leading uncertainty as outlined below.
\subsection{Energy Reconstruction}\label{energy-reco}
\begin{figure}[!ht]
    \begin{subfigure}[t]{\columnwidth}
        \centering
        \includegraphics[width=\linewidth]{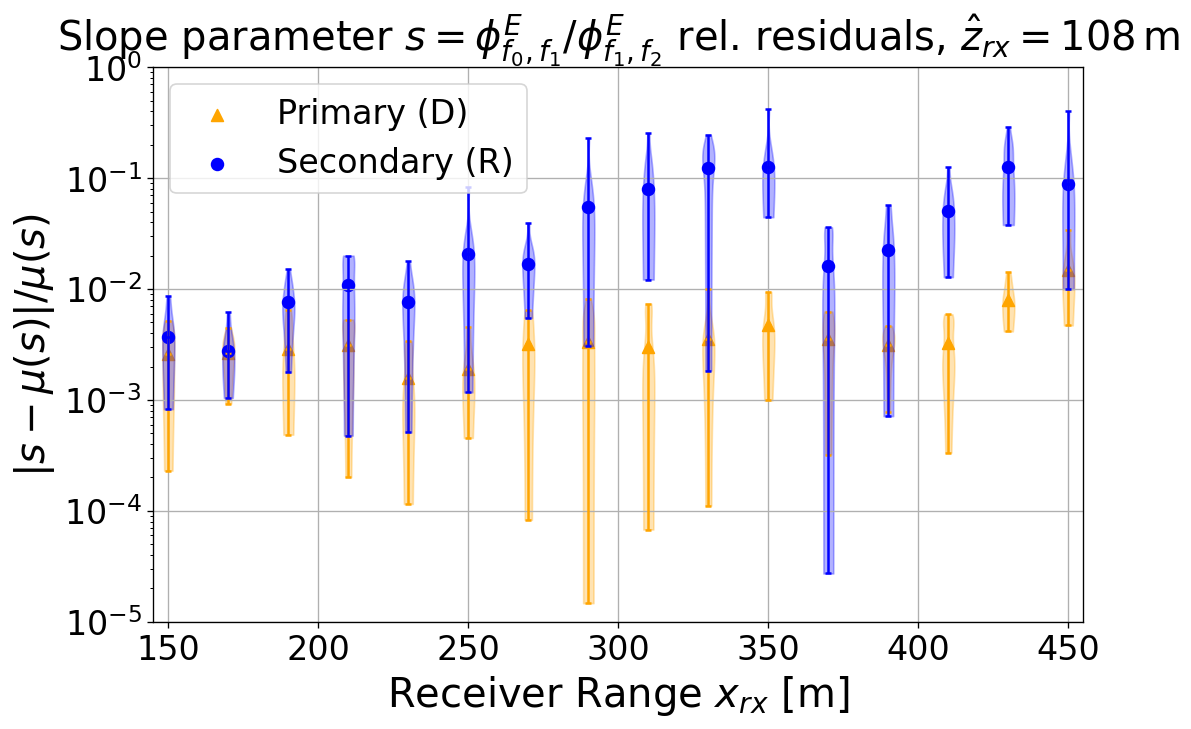}
        \caption{Slope parameter $s=\phi^{E}_{f_{0},f_{1}}/\phi^{E}_{f_{1},f_{2}}$}
    \end{subfigure}
    \begin{subfigure}[t]{\columnwidth}
        \centering
        \includegraphics[width=\linewidth]{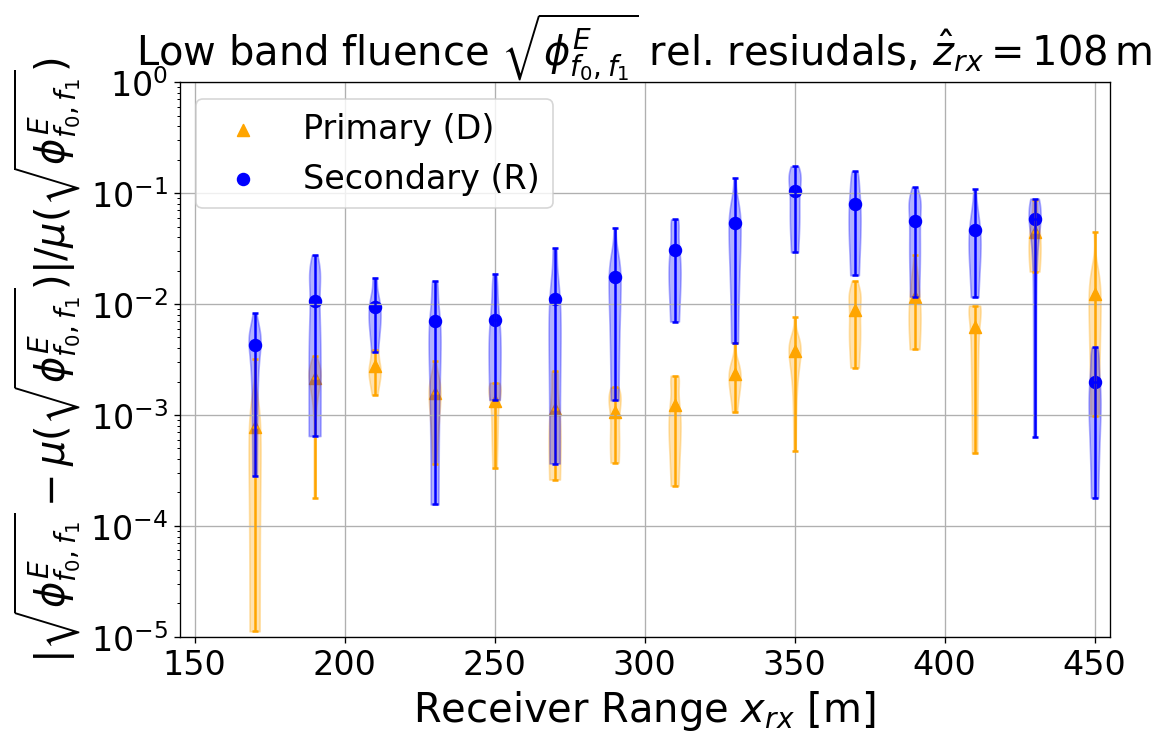}
        \caption{Square root of low band variation $\sqrt{\phi^{E}_{f_{0},f_{1}}}$}
    \end{subfigure}
    \begin{subfigure}[t]{\columnwidth}
        \centering
        \includegraphics[width=\linewidth]{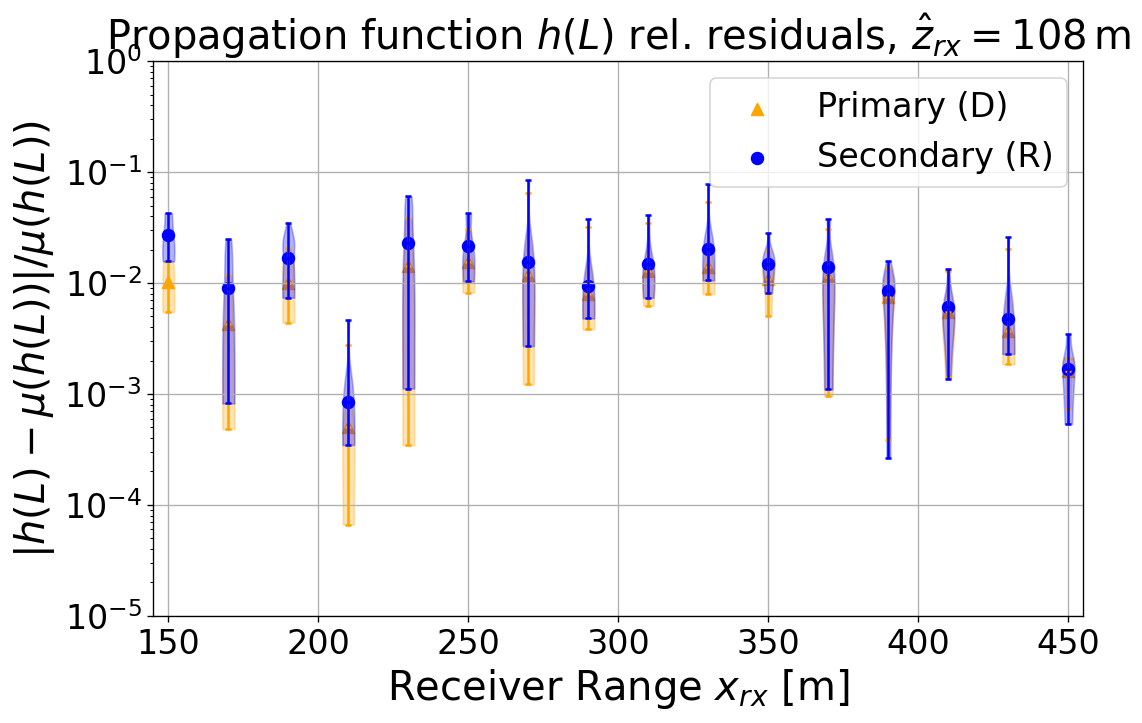}
        \caption{Path loss function $h(L)$}
    \end{subfigure}
    \caption{Relative variation in the spectrum-derived parameters: the slope parameter $s$, the square root of the low-band fluence $\sqrt{\phi^{E}_{f_{0},f_{1}}}$, and the path loss function $h(L)$}
    \label{fig:spectral-parameters}
\end{figure}

\begin{figure}[!ht]
    \centering
    \includegraphics[width=\linewidth]{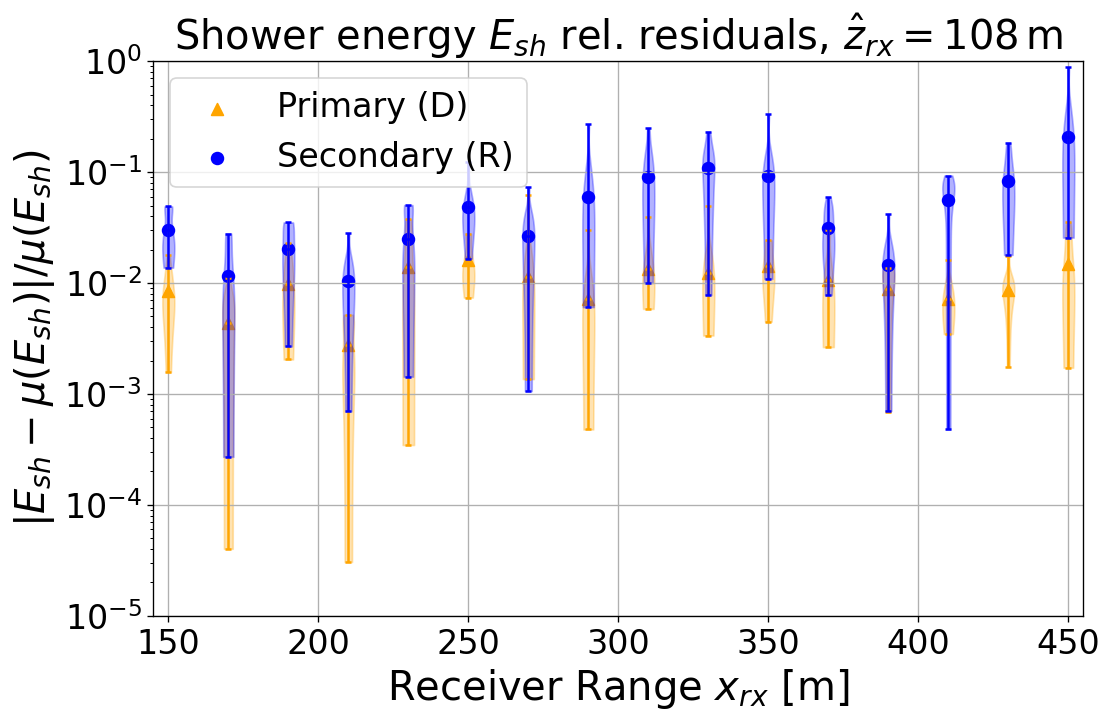}
    \caption{Relative variation in the reconstructed shower energy $E_{sh}$}
    \label{fig:shower-energy}
\end{figure}

Reconstructing energy from a received Askaryan signal is challenging due to the dependence of the observed spectrum on the viewing angle $\theta_{V}$ with respect to the Cherenkov angle $\theta_{C}$. As $\Delta\theta_{VC}$ increases, the overall signal strength diminishes, and the spectral bandwidth also diminishes as shorter wavelengths lose coherence faster than longer wavelengths. For illustration, we can suppose that we observe a single pulse (direct, reflected or refracted) originating from the neutrino vertex, and we measure the fluence $\phi^{E}$. The square root of the fluence is directly proportional to the deposited shower energy $\sqrt{\phi^{E}} \propto E_{sh} = E_{\nu} \cdot \kappa$, where $E_{\nu}$ is the primary neutrino's energy and $\kappa$ is the fraction of the primary's energy that goes into the shower. For purely hadronic showers, $\kappa$ is equal to the neutrino inelasticity $y$, however if the interaction is charged current and results in a lepton depositing energy into the observed shower, then $\kappa > y$. Additionally, the fluence diminishes with the propagation distance $L$ of the wavefront due to geometric spreading and attenuation losses, described with a path loss function $h(L) = \exp(-L/L_{\alpha})/L$. The fluence of a single-component Askaryan pulse is expressed then as \cite{Aguilar_2022_RNOG_E_RECO}:
\begin{equation}
    \sqrt{\phi^{E}} \propto E_{sh} \cdot h(L) \cdot f(\Delta\theta_{VC}),
\end{equation}
where $f(\Delta\theta_{VC})$ parametrizes the root of the fluence with respect to the viewing angle offset. If we consider a two-component (D \& R) signal, each wavefront will have propagated over distances $L_{D}$ and $L_{R}$ and have different spectra corresponding to different effective viewing angles $\theta_{V_{D}}$ and $\theta_{V_{R}}$. We therefore express $\phi^{E}$ as the sum of $\phi^{E}_{D}$ and $\phi_{R}^{E}$:
\begin{multline}
        \phi^{E} = \phi^{E}_{D} + \phi^{E}_{R} \\ \phi^{E}_{D} \propto [E_{sh}  \cdot h(L_{D}) \cdot f(\Delta\theta_{VC_{D}})]^{2} \\  \phi^{E}_{R} \propto [E_{sh} \cdot h(L_{R}) \cdot f(\Delta\theta_{VC_{R}})]^{2}.
\end{multline}
Consequently, the uncertainty in shower energy estimation is a function of the uncertainties in the measured fluence as well as the reconstructed viewing angle of each signal component and the reconstructed propagation distances. It is clear from inspection that for a single component signal the relative uncertainty in fluence $\delta\phi^{E}/\phi^{E}$ is linearly proportional to that of the relative energy uncertainty:
\begin{equation}
    \frac{\delta{E_{sh}}}{E_{sh}} \propto \frac{1}{2}\frac{\delta \phi^{E}}{\phi^{E}}.
\end{equation}
Given the frequency dependence of coherent radio emission, the shape of the frequency spectrum can be used as a proxy for the viewing angle. An established technique is to quantify the frequency spectrum using a slope parameter $s$ \cite{Aguilar_2022_RNOG_E_RECO}, which is the ratio of the integrated fluence for two distinct bands, with frequency limits $f_{0}$ \& $f_{1}$, and $f_{1}$ \& $f_{2}$, respectively:
\begin{equation}
    s = \frac{\phi^{E}_{f_{0},f_{1}}}{\phi^{E}_{f_{1},f_{2}}},
\end{equation}
such that generally, the slope parameter will grow larger as the viewing angle offset increases, as the high frequency band diminishes in power relative to the lower band. As an example, we adopt the band limits of $f_{0} = 130 \, \mathrm{MHz}$, $f_{1} = 200 \, \mathrm{MHz}$ \& $f_{2} = 300 \, \mathrm{MHz}$, and the parameterization of the shower energy \cite{Aguilar_2022_RNOG_E_RECO}:
\begin{equation}
    E_{sh} \propto \sqrt{\phi^{E}_{130,300}} \cdot g(s),
\end{equation}
Where $g(s)$ is:
\begin{equation}
    g(s) = \frac{1}{p_{2}(\ln{s})^{2} + p_{1}\ln{s} + p_{0}},
\end{equation}
and $p_{2}$, $p_{1}$ and $p_{0}$ are parameters derived from simulation. In the case of the source spectrum corresponding to $\Delta\theta_{VC}=7.5^{\circ}$ the values $p_{0} = 6.46$, $p_{1} = -16$ \& $p_{2}=12.45$ are used \cite{Aguilar_2022_RNOG_E_RECO}. The relative uncertainty in shower energy is therefore a linear combination of the relative uncertainties in the root of the lower band fluence $\sqrt{\phi^{E}_{f_{0},f_{1}}}$, the slope parameter function $g(s)$, and in the path loss function $h(L)$. While $h(L)$ is an independent parameter, there is obvious correlation and covariance between $s$ and $\phi^{E}_{f_{1},f_{2}}$. To derive the uncertainty for a given phased array observation, we divide the waveforms at each receiver into the primary (D) and secondary (R) components. The coherent average of each waveform component is then taken independently, and the slope parameter and fluence estimated from the resulting spectrum for the equivalent central antenna at $\hat{z}_{rx}$. Combining the spectrum-derived quantities with the variation in the path length, we then estimate the variation in the reconstructed energy $\delta E_{sh}/E_{sh}$ across the ice models. The relative variation in each of the spectrum-derived quantities and the spread loss are shown in Fig. \ref{fig:spectral-parameters}, and the derived result for the shower energy in Fig. \ref{fig:shower-energy}. It is seen that, for the secondary signal, the variation in the slope parameter is the dominant contribution to the energy uncertainty, with the shape and magnitude $\mathcal{O}(10^{-1})$ of the variation matching closely between the slope parameter and shower energy and the largest variations occurring in the refraction zone as expected. The path length loss is a sub-leading source of variability for the secondary signal, however it forms a relatively flat floor for the shower-energy variation for the direct signal $\delta E_{sh}/E_{sh} \sim \mathcal{O}(10^{-2})$.
\subsection{Arrival Direction}\label{theta-reco}
\begin{figure}[!ht]
    \begin{subfigure}[t]{\columnwidth}
        \centering
        \includegraphics[width=\linewidth]{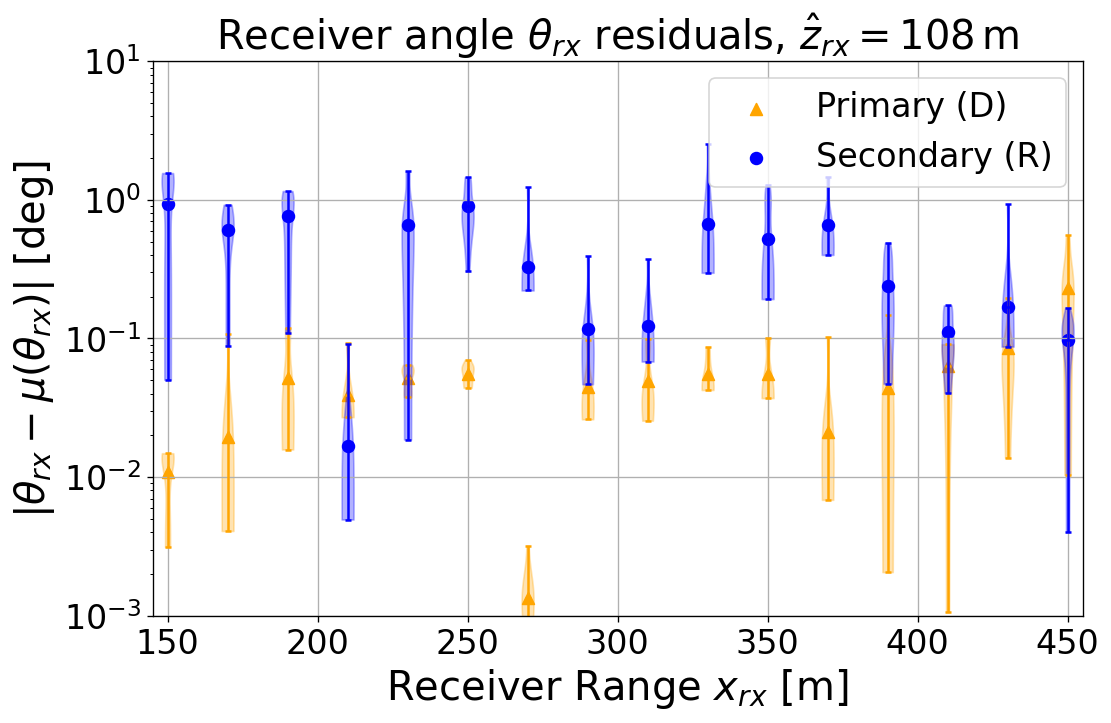}
        \caption{Receiver angle $\theta_{rx}$}
        \label{fig:rx_arrival_dir}
    \end{subfigure}
    \begin{subfigure}[t]{\columnwidth}
        \centering
        \includegraphics[width=\linewidth]{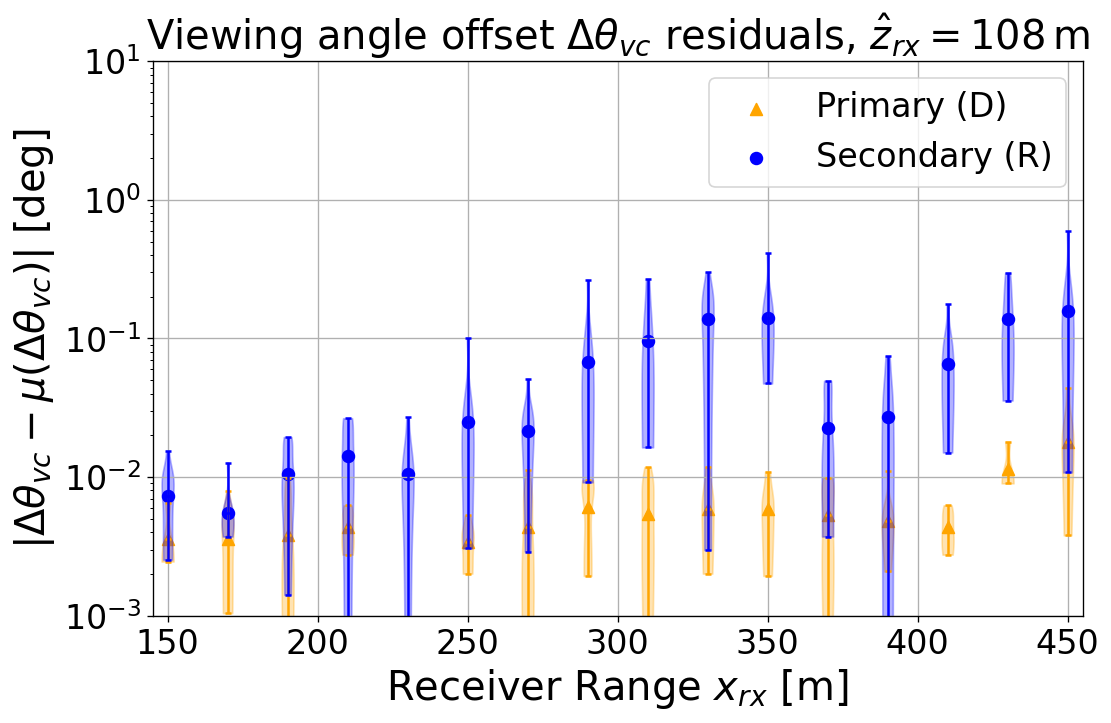}
        \caption{Viewing angle to Cherenkov angle offset $\Delta \theta_{VC}$}
        \label{fig:dtheta_vc_dir}
    \end{subfigure}
    \begin{subfigure}[t]{\columnwidth}
        \centering
        \includegraphics[width=\linewidth]{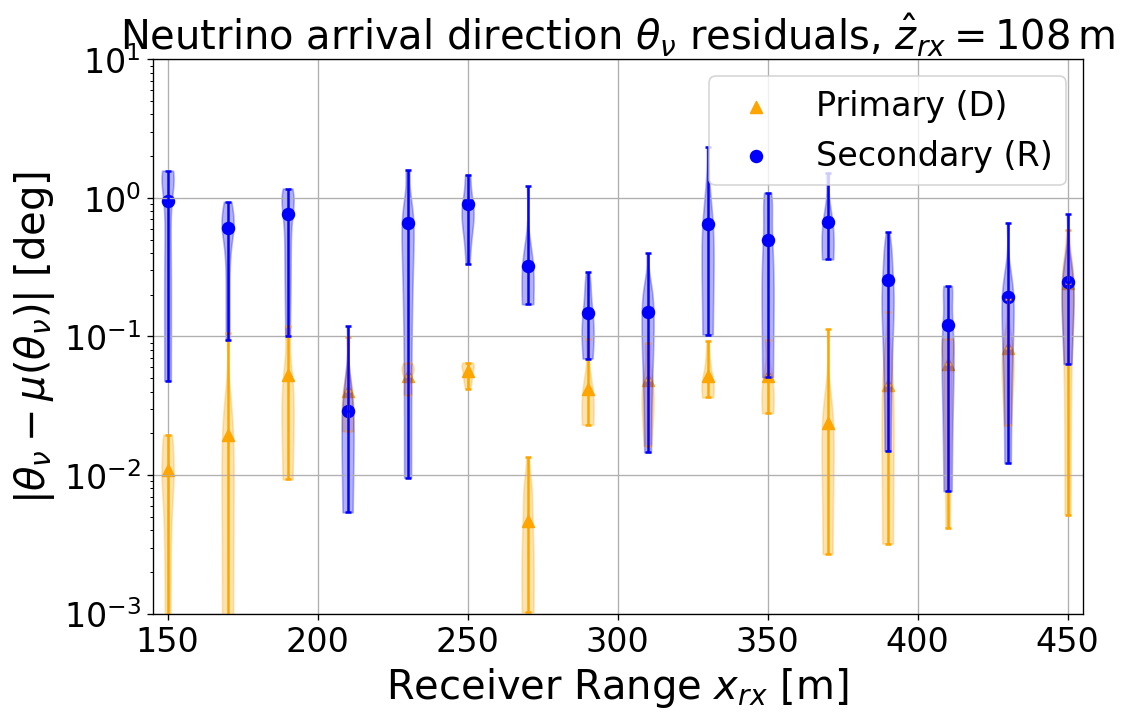}
        \caption{Neutrino arrival direction (polar) $\theta_{\nu}$}
        \label{fig:nu_arrival_dir}
    \end{subfigure}
    \label{fig:arrival-dir-params}
    \caption{Variation in the receiver arrival angle $\theta_{rx}$, the viewing angle to Cherenkov angle offset  $\Delta \theta_{VC}$ and the neutrino polar arrival direction $\theta_{\nu}$.}
\end{figure}

Again taking the case of a single component Askaryan pulse measured at a given receiver, the reconstruction of the primary neutrino's arrival direction $(\varphi_{\nu},\theta_{\nu})$ requires measurement of the arrival direction of the wavefront $(\varphi_{RX},\theta_{RX})$, the aforementioned viewing angle offset from the Cherenkov angle $\Delta \theta_{VC}$, and the polarization angle $\alpha_{p}$ \cite{Plaisier_2023_arrival_dir}, with the latter being the argument of the ratio of the vertical $E_{\theta}$ and horizontal $E_{\phi}$ components of the field:
\begin{equation}
    \alpha_{p} = \arctan(\frac{E_{\phi}}{E_{\theta}}).
\end{equation}
The uncertainty in the neutrino arrival direction will be a linear combination of the uncertainties of the receiver arrival direction, the polarization angle and the viewing angle offset from the Cherenkov angle, as these values are independent of one another. We comment on polarization in Section \ref{polarization-section}.
\\~\\
From the vertex reconstruction exercise in Section \ref{vertex-reco}, we can estimate the variation in the receiver arrival direction $\theta_{RX,D}$ and $\theta_{RX,R}$ from the variation in the reconstructed source position $(x_{tx}, z_{tx})$. The results found by minimizing $\chi^{2}_{DR}$ are displayed as a function of range $x_{rx}$ in Fig. \ref{fig:rx_arrival_dir}. To obtain the residual error $\Delta \theta_{VC}$, we take the variation in the slope parameter $s$ from Section \ref{energy-reco} and, by exploiting the linear proportionality between $\log(s)$ and $\Delta \theta_{VC}$, obtain the variation as a function of range as shown in Fig \ref{fig:dtheta_vc_dir}. The total uncertainty in the polar angle component of the  neutrino arrival direction is shown in Fig. \ref{fig:nu_arrival_dir}. The variation in the receiver angle, derived from vertex reconstruction, is relatively flat across the spatial range of 150 m to 400 m, but is an order of magnitude greater for the secondary signal $0.1^{\circ} \lesssim \delta \theta_{RX,R} \lesssim 1^{\circ}$ as compared to the primary signal $\delta \theta_{RX,D} \lesssim 0.1^{\circ}$. It also forms the leading source of error for the derived neutrino angle variation. The viewing angle offset uncertainty is greater for the secondary signal and is largest within the refraction zone, as expected from our observations of the spectrum, and is generally an order of magnitude lower than the uncertainty for the receiver arrival angle. Overall, the magnitude and shape of the neutrino angle variation as a function of range is effectively identical to that of the receiver arrival angle, and we can expect variations on the order of $\delta \theta_{\nu,R} \sim0.5^{\circ}$ when considering the secondary signal alone and $\delta \theta_{\nu,D} \sim0.05^{\circ}$ when considering the direct signal alone. 
\subsection{Polarization}\label{polarization-section}
Previous works have noted that the relatively large systematic errors in the polarization angle reconstruction $\delta \alpha_{p} \sim \mathcal{O}(10^{
\circ})$ are typically the dominant source of uncertainty in neutrino arrival-direction reconstruction for deep in-ice receiver arrays \cite{Plaisier_2023_arrival_dir}. Whereas shallow arrays have achieved polarization angle resolutions of $\delta \alpha_{p} \sim 1.3^{\circ}$ \cite{Anker_2022}, the lower gain, cross-polarization isolation, and signal-to-noise ratio of deep-borehole Hpol antennas makes it difficult to accurately reconstruct the horizontally polarized field $E_{\phi}$ from the measured voltage trace \cite{Plaisier_2023_arrival_dir}. 
\\~\\
By construction, the PE method used by \texttt{paraProp} can not estimate $E_{\phi}$ as it is derived for a cylindrically symmetric medium. This symmetry was also utilized in the FDTD simulations using \texttt{MEEP}. The limitations of the simulations and consequent neglect of birefringence effects meant that we could not make meaningful statements about the influence of seasonal ice variation  on the horizontally polarized field $E_{\phi}$, nor on resulting errors in the polarization and azimuth angle reconstruction. Given that the aforementioned $\alpha_{p}$ uncertainty exceeds our estimates of neutrino zenith angle variation $\theta_{\nu}$ by at least an order of magnitude, a dedicated study of seasonal ice variation on $E_{\phi}$ may be warranted, but would require a different modeling approach than that used in this study and must explicitly incorporate birefringence into the ice model.
\FloatBarrier
\section{Conclusion}
We have found that in-ice radio neutrino detectors are sensitive to seasonal changes in the density of the upper firn layer, particularly the formation of refrozen ice layers, which lead to fluctuations in refracted and reflected signals on the order of
$\delta \phi_{R}^{E}/\phi_{R}^{E} \gtrsim \mathcal{O}(10^{-1})$
for signals arriving with zenith angles in the range $45^{\circ} < \theta_{RX} < 55^{\circ}$ and measured over the decade of 2010 to 2020. Although this study has focused on detectors searching for Askaryan emission, similar variations in signal fluence and arrival time are expected for radar signals traversing the shallow firn. The implications for parameter reconstruction in future radar echo detectors, however, will differ and depend strongly on the depths spanned by the detector, and are therefore beyond the scope of this work. 
\\~\\
For an Askaryan detector with a fiducial volume of $2700\,\mathrm{m}$ depth and $3000\,\mathrm{m}$ radius, approximately $18\%$ of UHE neutrino event vertices are expected to produce shallow refracted events that must pass through the shallow firn, a fraction that rises to $22 \%$ if one excludes the shadow zone from the volume. In addition, secondary signals observed in both the reflection and refraction
zones exhibit arrival-time variations on the scale of $\mathcal{O}(0.1$--$1.8\,\mathrm{ns})$. The primary consequence for event reconstruction is a systematic uncertainty in the reconstructed shower energy $E_{sh}$ for events with both observable primary and secondary signals, at the level of $0.01 \lesssim \delta E_{sh}/E_{sh} \lesssim 0.1$, with the magnitude of the uncertainty depending on the relative strengths of the direct and secondary signals.
Stronger fluctuations at higher frequencies ($f > 250\,\mathrm{MHz}$) may also introduce uncertainty in the estimation of the viewing-angle offset from the Cherenkov angle, $\Delta \theta_{VC}$. Background variations in signal arrival times further limit vertex reconstruction resolution and, consequently, the estimated arrival direction and propagation length. We find arrival-direction uncertainties of $\delta \theta_{RX,R} \sim \mathcal{O}(0.1^{\circ}$--$0.5^{\circ})$ for refracted signals and $\delta \theta_{RX,D} \sim \mathcal{O}(0.05^{\circ})$ for direct signals, which, in the absence of an up-to-date ice model, represent a small but irreducible systematic uncertainty. It is worth commenting that the estimates of energy and arrival direction variation are done under optimistic assumptions of perfect field reconstruction, the detection and identification of both D and R signal components, and are done in the absence of noise.
\\~\\
Variations in signal fluence and time of flight within the
refraction zone appear stochastic, while correlating with the presence of
refrozen layers along the ray path. This suggests that {\it in-situ} ice measurements at a given detector site, combined with forward glaciological modeling of snow accumulation and temperature, may allow for the construction of an up-to-date ice model that enables identification of the shallow refractive propagation zone relative to receiver arrays. It is not clear whether the detailed amplitude and timing fluctuations within this zone can be fully mitigated through modeling, particularly given uncertainties in receiver positioning within the ice and horizontal variation in the firn.
\\~\\
Finally, variations in signal fluence may reduce antenna multiplicity and thereby degrade reconstruction quality. Detector-specific analyses are therefore required to fully quantify
reconstruction systematics. The influence of firn density anomalies on polarization-angle reconstruction was not examined in this study, but certainly merits further investigation. As polar temperatures continue to warm in both the Arctic and Antarctic, density and temperature anomalies in the firn are expected to become increasingly important considerations for neutrino event reconstruction.
\\~\\
\begin{acknowledgments}
We sincerely thank our colleagues in the RET collaboration for their comments, particularly I. Esteban, I. Loudon, J. Ralston and K. de Vries. We recognize support from the National Science Foundation under grant numbers 2306424, 2019597 and 2012989, as well as from the John D. and Catherine T. MacArthur Foundation.
\end{acknowledgments}
\bibliography{references}
\medskip
\nocite{*}

\appendix
\section{MERRA-2 Input Data for CFM}\label{CFM_explained}
\begin{figure}[!ht]
    \centering
    \begin{subfigure}[t]{\columnwidth}
        \includegraphics[width=\linewidth]{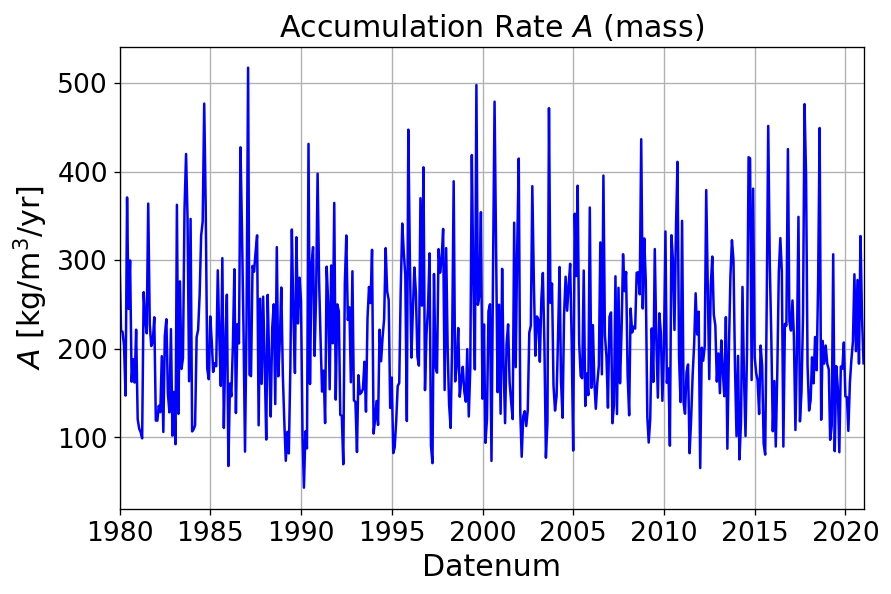}
        \caption{Mass accumulation rate $A$ per unit area in units of $\mathrm{kg/m^{2}/yr}$}
        \label{fig:adot}
    \end{subfigure}
    \begin{subfigure}[t]{\columnwidth}
        \includegraphics[width=\linewidth]{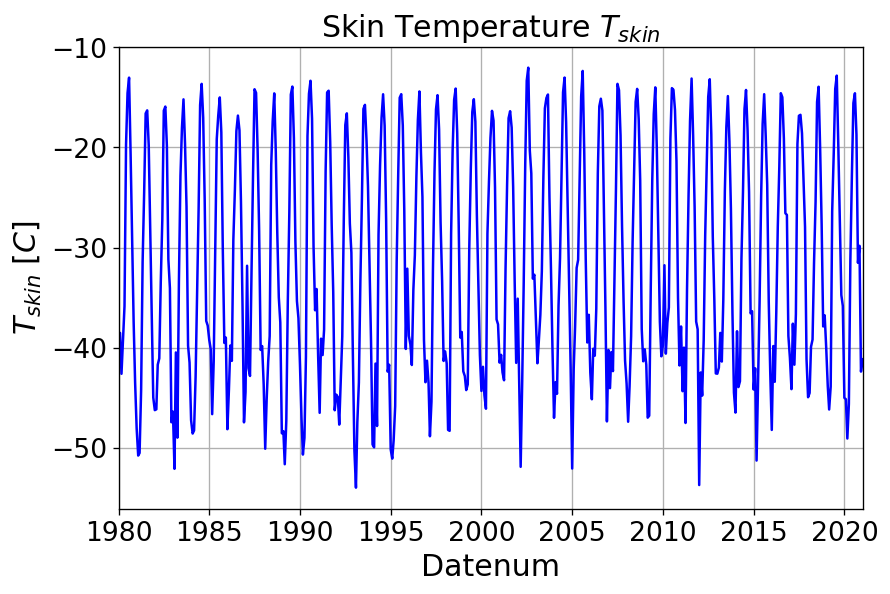}
        \caption{The skin temperature $T_{skin}$}
        \label{fig:tskin}
    \end{subfigure}
    \begin{subfigure}[t]{\columnwidth}
        \includegraphics[width=\linewidth]{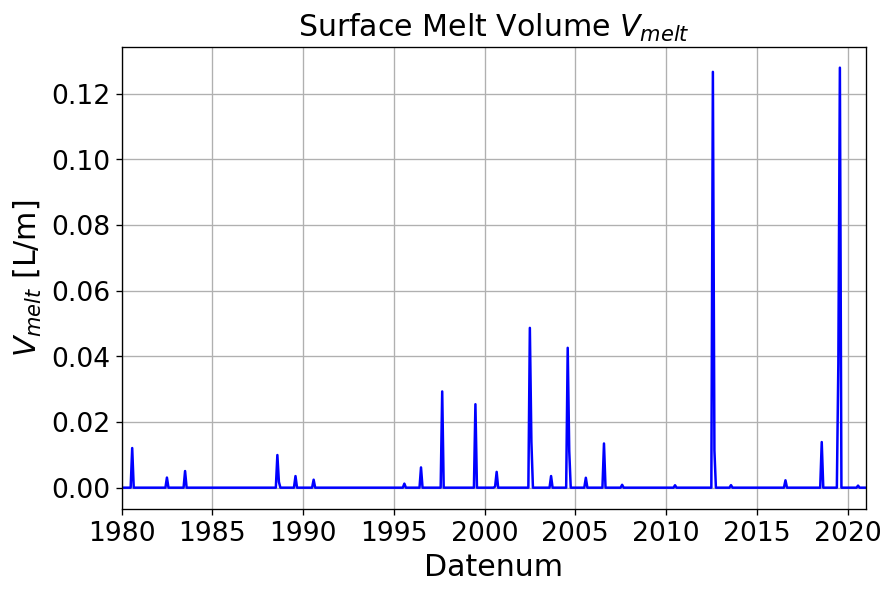}
        \caption{The melt volume per unit area $V_{melt}$}
        \label{fig:smelt}
    \end{subfigure}
    \caption{A time series of the monthly climate data (1980-2021) from MERRA-2 used to drive the evolution of the firn in CFM. From top to bottom: the mass accumulation rate $A$, skin temperature $T_{skin}$ and melting volume per unit area $V_{melt}$.}
    \label{fig:input-cfm-data}
\end{figure}

The input data used to drive the CFM simulation is derived from the Modern-Era Retrospective analysis for Research and Application (version 2) dataset or MERRA-2 \cite{MERRA_2}.  MERRA-2 is a global atmospheric reanalysis produced by NASA’s Global Modeling and Assimilation Office, spanning from 1980 to the present. It combines a wide range of satellite and in-situ observations with a numerical weather prediction model to generate a continuous, physically consistent record of atmospheric and surface climate. In the dataset, the Earth is divided into a grid with each cell having resolution $0.5^{\circ}$ in latitude by $0.66^{\circ}$ longitude. MERRA-2 provides surface air temperature, precipitation, accumulation rate, and other fields at monthly resolution, which we use as boundary conditions to force the evolution of the firn density profile. In Fig.\,\ref{fig:input-cfm-data} we show the monthly input data for the mass accumulation rate $A$ (Fig.\,\ref{fig:adot}), the skin temperature $T_{skin}$ (Fig.\,\ref{fig:tskin}), as well as the melting events which are quantified by the volume of surface melt water per unit area $V_{melt}$ (Fig.\,\ref{fig:smelt}). 
\section{RF signal variation with variable surface density}\label{appendix:rho_s_var}
\begin{figure}[!ht]
    \centering
    \includegraphics[width=\linewidth]{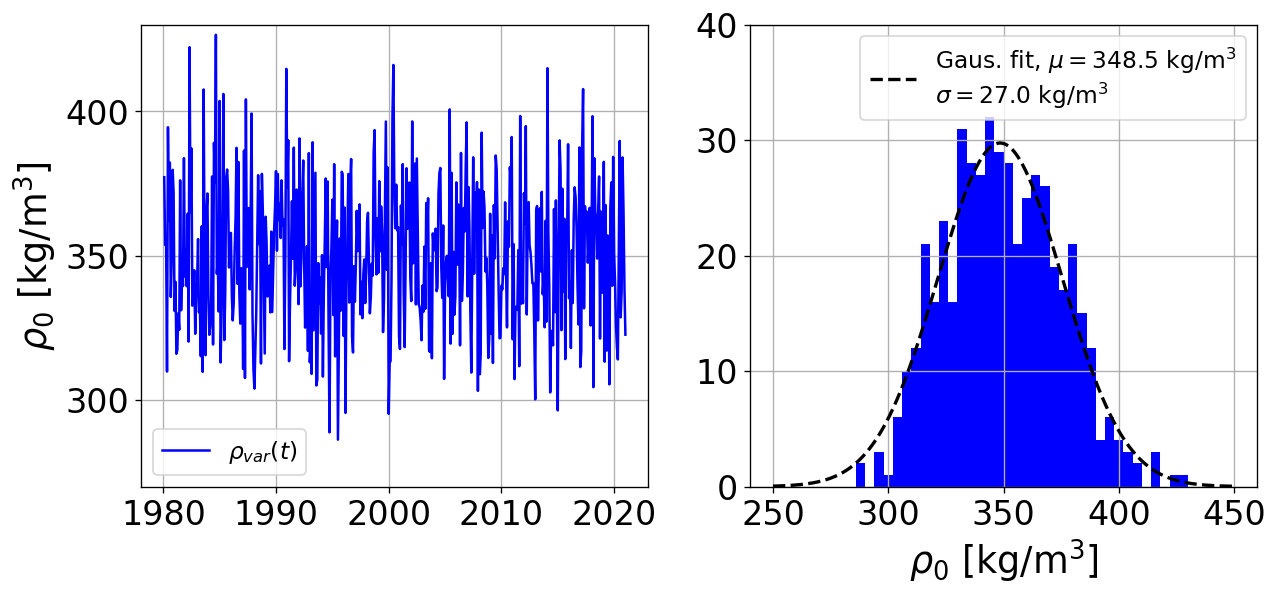}
    \caption{Left: Variable surface density of the CFM models over the epoch 1980-2020 $\rho_{0}(t)$. Right a histogram of the $\rho_{0}$ values with a Gaussian fit.}
    \label{fig:rhos_var}
\end{figure}

\begin{figure}[!ht]
    \centering
    \begin{subfigure}[t]{\columnwidth}
        \centering
        \includegraphics[width=0.96\linewidth]{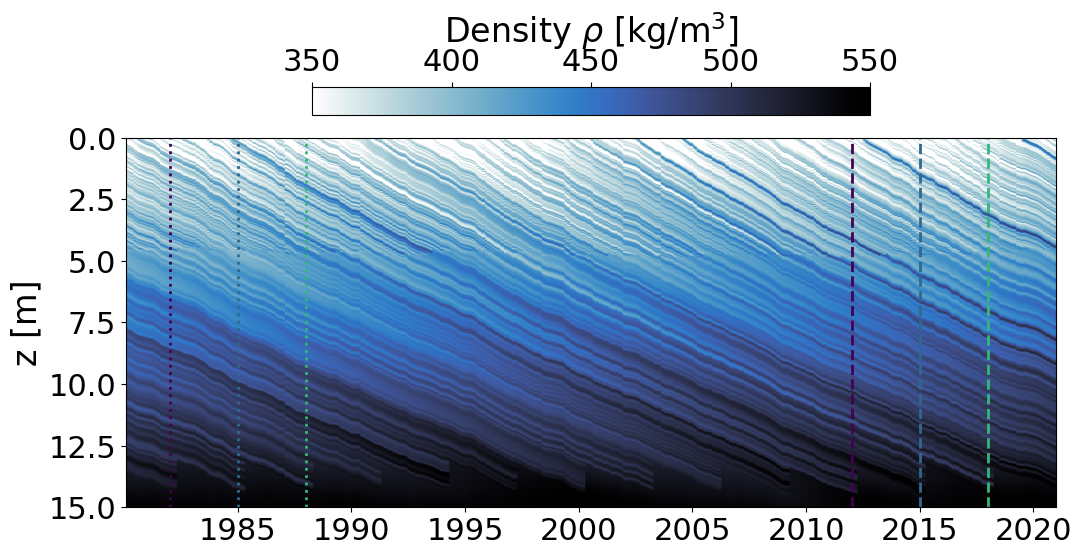}
        \caption{Density evolution map $\rho_{CFM}(z,t)$}
        \label{fig:cfm_map_rho_var}
    \end{subfigure}
    \begin{subfigure}[t]{\columnwidth}
        \centering
        \includegraphics[width=0.96\linewidth]{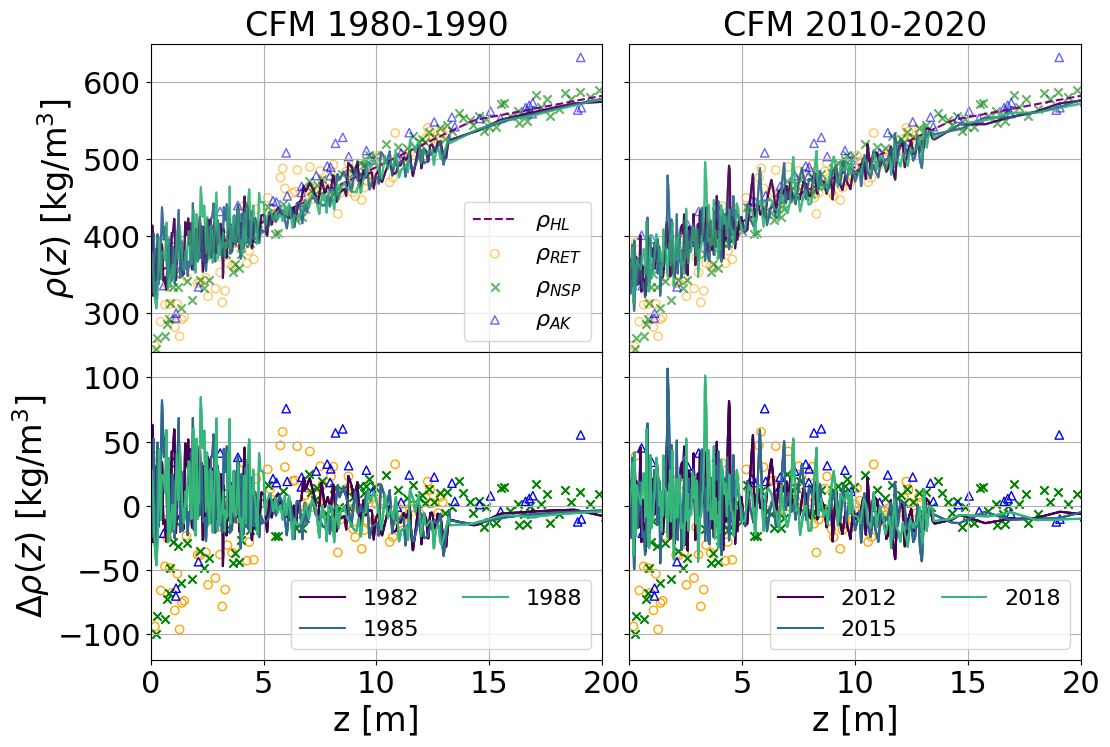}
        \caption{The shallow firn density \& residuals (left: 1980-1990 \& right: 2010-2020)}
        \label{fig:rho_prof_cfm_both_rho_var}
    \end{subfigure}
    \begin{subfigure}[t]{\columnwidth}
        \centering
        \includegraphics[width=0.96\linewidth]{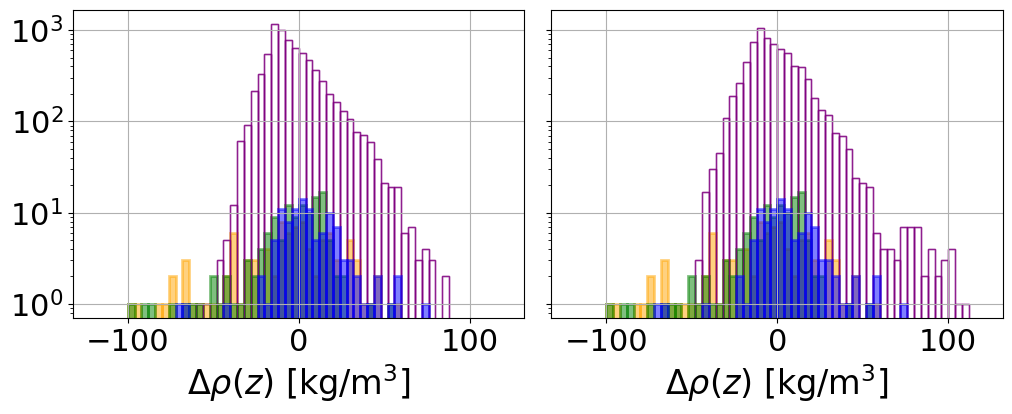}
        \caption{Histogram of density residuals.(\ref{fig:rho_prof_cfm_both})}
        \label{fig:rho_prof_cfm_hist_rho_var}
    \end{subfigure}
    \caption{Above: A color-map of the time-dependent density profile of the shallow firn from 1980 through to 2020, with an assumption of variable surface density.}
    \label{fig:cfm_model_rho_var}
\end{figure}

\begin{figure}[!ht]
    \centering
    \includegraphics[width=0.95\linewidth]{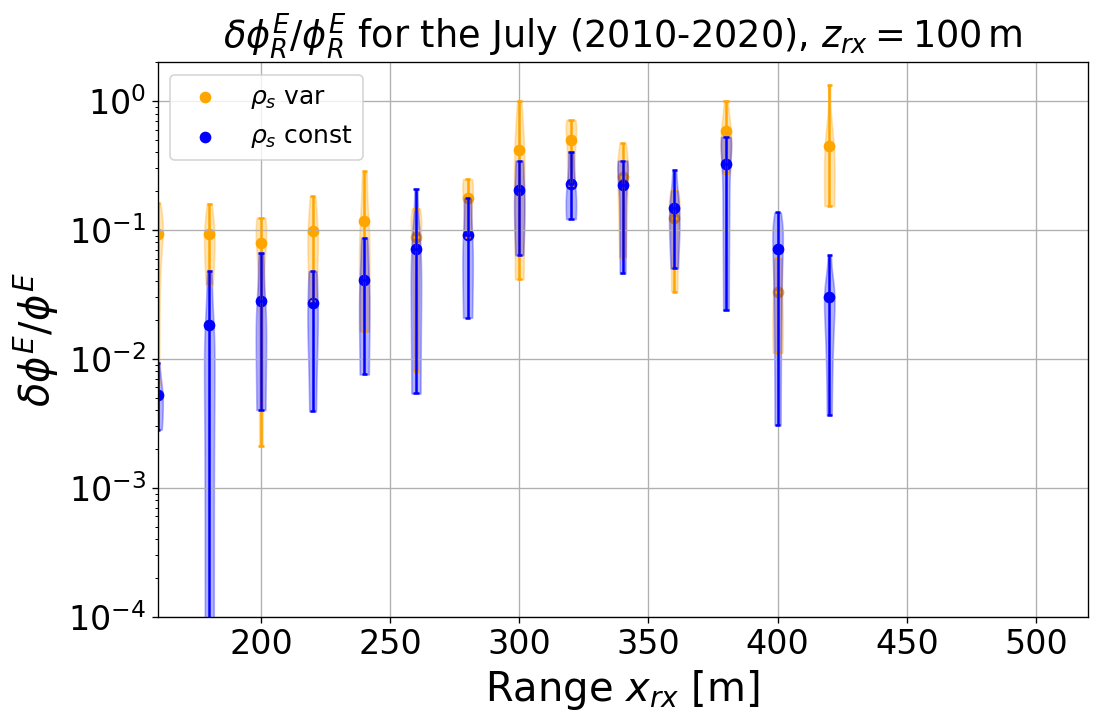}
    \caption{The variation coefficient of the secondary fluence $\delta \phi_{R}/\phi_{R}$ for the constant surface density case ($\rho_{s}$ const) and compared to the variable density case ($\rho_{s}$ var).}
    \label{fig:phiR_var_rhos_var}
\end{figure}

As discussed in section \ref{CFM_results}, we ran an ancillary simulation of the Summit firn by allowing the surface density to vary randomly instead of holding it fixed. At each epoch, the surface density value is drawn randomly from a gaussian distribution centered on $350 \, \mathrm{kg/m^{3}}$ and with a standard deviation of $\sigma = 27 \, \mathrm{kg/m^{3}}$, as shown in Fig. \ref{fig:rhos_var}. This value of the spread was chosen to reflect the variation in the upper 10 cm of surface snow density $\rho_{0}$ that was measured at Summit from 2017 to 2019 by \cite{Howat_2022}, although it should be noted that an average value of $\rho_{0} = 305 \, \mathrm{kg/m^{3}}$ was found. The output of the density models are shown in Fig. \ref{fig:cfm_model_rho_var}.
\\~\\
Using the $\rho_{0}$-varying CFM density to define our ice models, we repeated the simulation study performed in Section \ref{radio-sim}, utilizing the same source signal and using \texttt{paraProp}. We examined the variation in the secondary signal fluence $\phi^{E}_{R}$ for a receiver at a depth of $z_{rx} = 100 \, \mathrm{m}$ as a function of receiver range for both the constant and variable $\rho_{0}$ cases, as displayed in Fig. \ref{fig:phiR_var_rhos_var}. We find that in both cases, the variation in fluence peaks above a value of $\delta \phi_{R}^{E}/\phi_{R}^{E} \geq 10^{-1}$ when the receiver is within the shallow refraction propagation zone, demonstrating that this result is robust with respect to the choice of surface density as a variable or as a constant. A notable difference in the outcome between the CFM models is the noticeably higher variation, by an order of magnitude, within the reflection zone for the variable $\rho_{s}$ case as compared to the constant $\rho_{s}$ case. This suggests the potential for larger reconstruction uncertainties in energy and arrival direction for neutrino reconstruction when using the secondary signal within the reflection zone, although these signals are generally weaker compared to the primary.
\section{Variation of signal waveforms}\label{signal-variation}
We examine the change in the characteristics of the pulses in the different propagation regions by showing example waveforms at ranges of $x_{rx} =$ 160, and 350$\mathrm{m}$ (designated A and B respectively) and at depths of $z_{rx} = 8 \, \mathrm{m}$ (shallow) and at $100 \, \mathrm{m}$ (deep). 
\subsubsection{Shallow Receivers}
\begin{figure*}[!ht]
    \centering
    \begin{subfigure}{\columnwidth}
        \centering
        \includegraphics[width=0.9\linewidth]{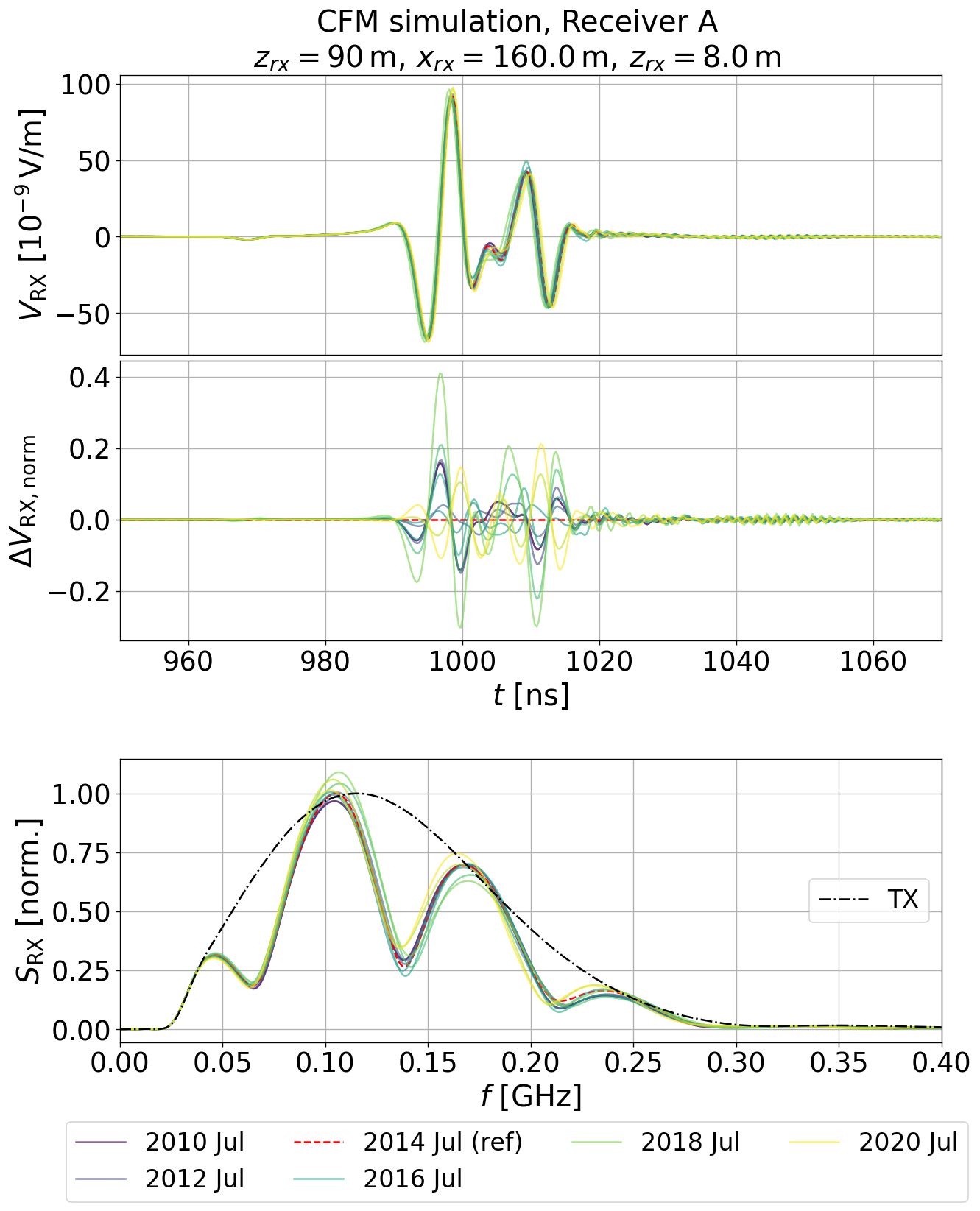}
        \caption{Receiver A ($x_{rx} = 160 \, \mathrm{m}, \, z_{rx} = 8 \, \mathrm{m}$)}
        \label{fig:rx_pulse_sh_160}
    \end{subfigure}
       \begin{subfigure}{\columnwidth}
        \centering
        \includegraphics[width=0.9\linewidth]{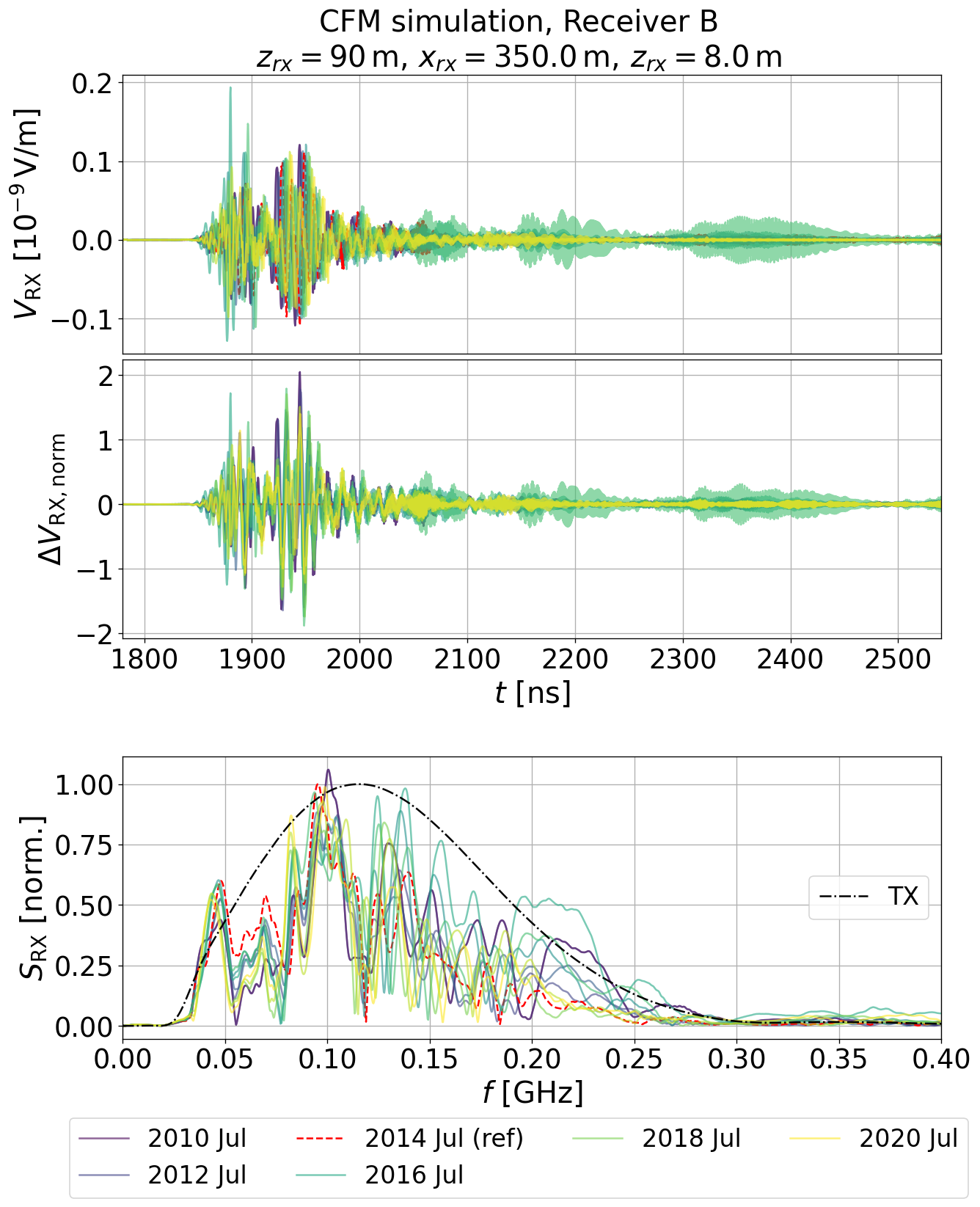}
        \caption{Receiver B ($x_{rx} = 350 \, \mathrm{m}, \, z_{rx} = 8 \, \mathrm{m}$)}
        \label{fig:rx_pulse_sh_350}
    \end{subfigure}
    \caption{Waveforms measured by a receiver at $z_{rx} = 8 \, \mathrm{m}$ and $x_{rx} = 160$ \& $350 \, \mathrm{m}$. For the trace at receiver A on the left most plots ($x_{rx} = 160$), it is possible to distinguish between two wavelets, and the spectral peaks and troughs are a result of the combination of overlapping wavelets. For receiver B, the waveform is some combination of multiple wavelets that can not be isolated and identified.}
    \label{fig:rx_pulse_sh_DR}
\end{figure*}

Waveforms within the shallow firn experience lensing effects due to the presence of over-dense layers. These layers can also give rise to second-order reflections, so that waveforms in this domain may have more than two components. The simulated pulses received at $z_{rx} = 8 \, \mathrm{m}$ are shown in the upper pannels of Fig.\,\ref{fig:rx_pulse_sh_DR}, with receiver A ($x_{rx} = 160 \, \mathrm{m}$) shown in Fig. \ref{fig:rx_pulse_sh_160} and receiver B ($x_{rx} = 350 \, \mathrm{m}$) in Fig. \ref{fig:rx_pulse_sh_350} alongside the normalized residuals of the waveforms relative to a reference, for which we use the simulation of July 2016 and their spectra, which are normalized to the peak of the reference spectrum. The simulated outputs from the ice models of July for the years from 2010 to 2020 are overlaid. Whereas receiver A is located within the illuminated regime, receiver B falls within the shadow zone, and their signals are therefore significantly attenuated relative to A and cannot be described by a simple double-pulse model. Receiver A exhibits a clear primary pulse with normal polarity at $t_{D} = 998 \, \mathrm{ns}$ that partially overlaps with a secondary, inverted pulse peaking at $1010 \, \mathrm{ns}$. Likewise, spectra at receiver A show similar shapes, with 4 distinct peaks owing to the combination of the spectra of the D and R signals. The amplitude of these peaks vary on the order of a few percent. Across the different ice models, the resulting fluence fluctuations reach $4.47\%$, while arrival-time fluctuations remain below the sub-ns level. Receiver B, in contrast, shows a complex dispersive structure spanning 1800–2500 ns that varies in overall fluence and shape. The spectra at receiver B do not well match the shape of the origin, being composed of a large number of peak structures that are indicative of the combination of multiple wavelets. The attenuated and broadened waveform at receiver B (1900–2500 ns) exhibits long-lived oscillatory tails whose morphology differs between ice models, highlighting the sensitivity of shadow-zone propagation to small seasonal variations in the firn profile. 
\subsubsection{Deep Receivers}
\begin{figure*}[!ht]
    \centering
    \begin{subfigure}[t]{\columnwidth}
        \centering
        \centering
        \includegraphics[width=0.9\linewidth]{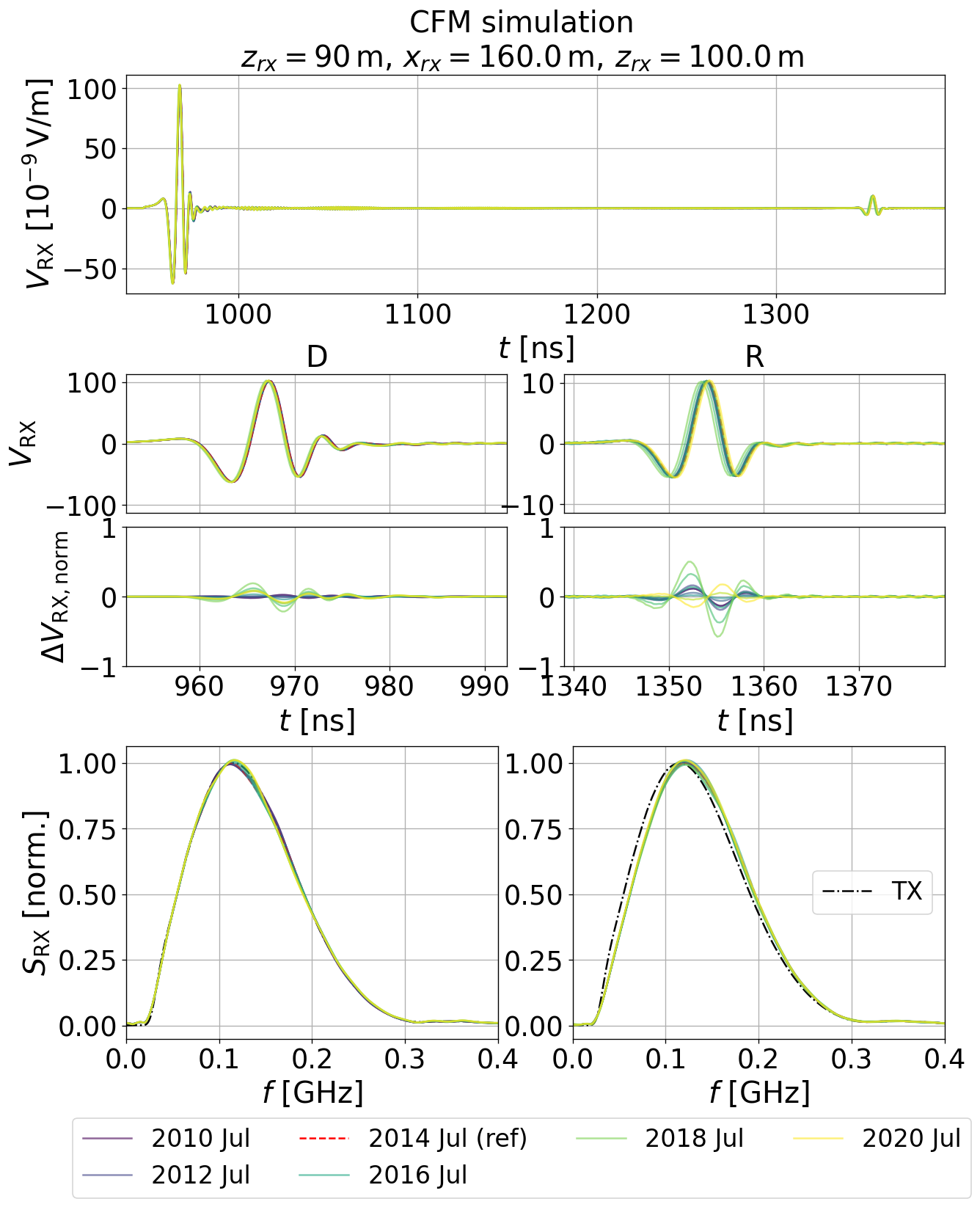}
        \caption{Receiver A ($x_{rx} = 160 \, \mathrm{m}, \, z_{rx} = 100 \, \mathrm{m}$)}
        \label{fig:rx_pulse_deep_160}
    \end{subfigure}
    \begin{subfigure}[t]{\columnwidth}
        \centering
        \centering
        \includegraphics[width=0.9\linewidth]{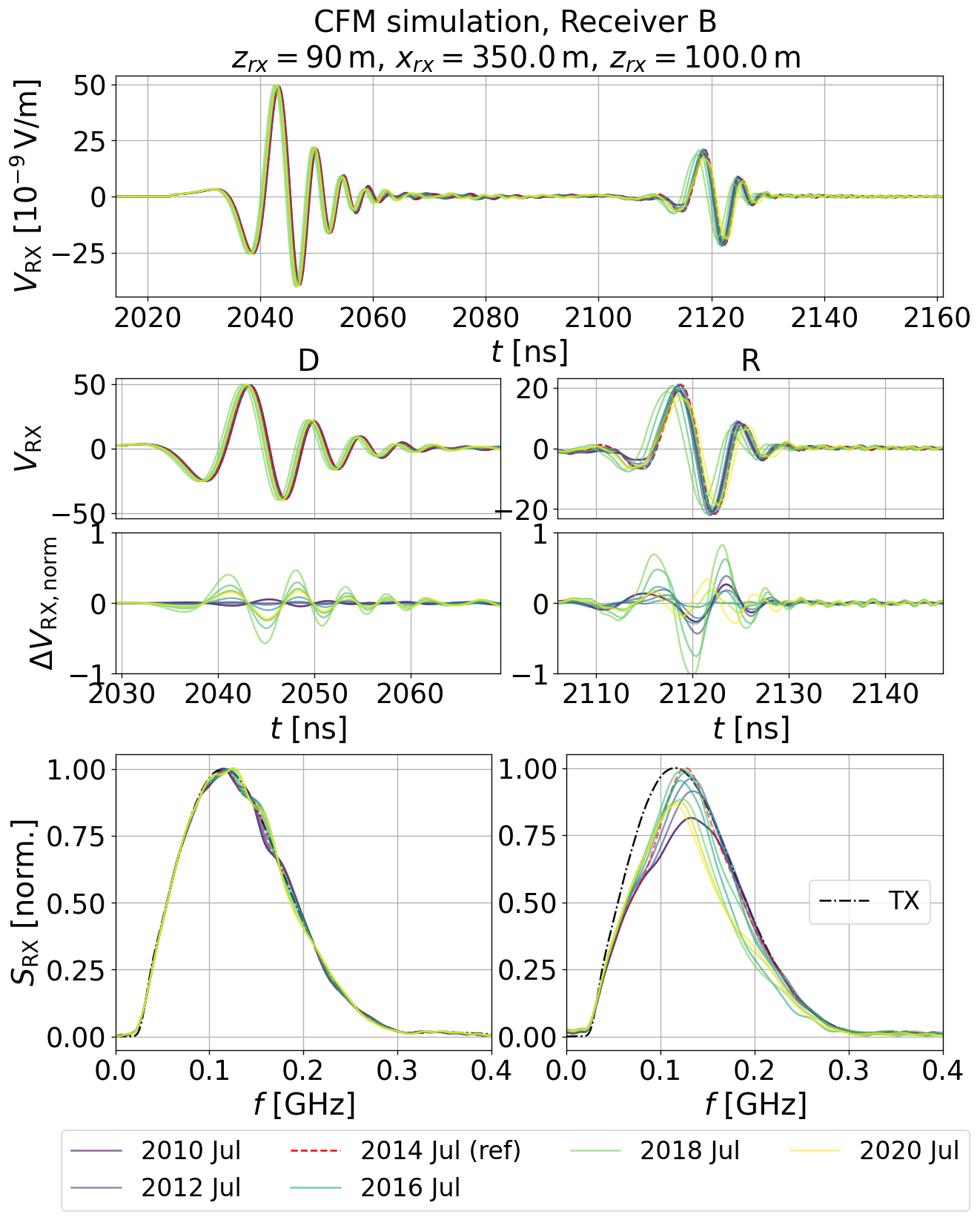}
        \caption{Receiver B ($x_{rx} = 350 \, \mathrm{m}, \, z_{rx} = 100 \, \mathrm{m}$)}
        \label{fig:rx_pulse_deep_350}
    \end{subfigure}
    \caption{Signals observed for a `deep' receiver at $z_{rx} = 100 \, \mathrm{m}$ and $x_{rx} = 160$ \& $ 350 \, \mathrm{m}$.}
    \label{fig:rx_pulses_deep}
\end{figure*}

Fig. \ref{fig:rx_pulses_deep} shows the simulated pulses, their residuals, and spectra recorded at receivers located at $z_{rx} = 100 , \mathrm{m}$ for horizontal baselines of 160 m and 350 m, once again labeled A (shown in \ref{fig:rx_pulse_deep_160}) and B (shown in \ref{fig:rx_pulse_deep_350}) respectively. Here the waveform, residuals and spectra have divided to show the D and R components of the signal, and their residuals and spectra are normalized to the maximum of a `reference signal' (defined using July of 2014). Furthermore, the receiver spectra are compared with the normalized spectrum of the field at the transmitter. 
\\~\\
For this depth ($z_{rx} = 100 \, \mathrm{m}$), receiver A is within the reflection zone and receiver B is in the shallow refraction zone. For both receivers we see a distinct double pulse signature, with the direct (D) pulse having a brighter amplitude and arriving earlier, and a secondary pulse (R) delayed relative to the direct. On inspection, the amplitude and morphology of the direct pulse and their spectra shows negligible change (indeed the amplitude change is below the level of $10^{-2}$) for receiver A and B, but there there are small phase offsets between the waveforms at different times, leading to the residual changes between the pulses. The arrival times of the primary signal are consistent to the sub-ns level between ice models. In both cases, the direct pulse has propagated through the glacial ice and deep firn, and thus encounter minimal change in the ice properties along their travel path.
\\~\\
For the secondary signals, there are marked differences between receiver A and B, as in the former case the secondary signal has reflected off the surface and in the latter, it has undergone refraction in the shallow firn without experiencing reflection, reaching a minimum depth of 3 m. For receiver A, the amplitude and morphology of the R-signal and spectrum is essentially between the different CFM models, similar to the case for the D-signals. However, the phase offsets between the R-signals is larger than for the D-signal, leading to larger residuals. For receiver B, the changes to the amplitude and morphology of the R-pulse and are visible to the eye, along with shifts in arrival time to the order of $\mathcal{O}(\mathrm{ns})$. The spectra also exhibit marked differences, with changes to the frequency of maximum amplitude between the RX spectra and that of the TX, as well as bumpiness. The relative residuals of the spectral amplitude are $\mathcal{O}(10^{-1})$ and higher. The changes in the spectral shape are due to the diffraction of the secondary pulse as it traverses the shallow firn.
\\~\\
For example in receiver B, we find from ray-tracing that the D signal has undergone refraction within the deep firn, reaching a minimum depth of 85 m. The D signal pulse experiences minimal changes to its amplitude and shape between the different ice models, with the waveforms looking identical by eye. Minor amplitude shifts between the spectra are visible on the left hand plot, with a residual spectral amplitude of $10^{-2}$ relative to the average of the spectra. The R signal has reached a minimum depth of 3 m, and has necessarily passed through the variable shallow firn, and the changes to the amplitude and shape of the pulse are visible to the eye, along with shifts in arrival time to the order of $\mathcal{O}(\mathrm{ns})$. The spectra also exhibit marked differences, with changes to the frequency of maximum amplitude between the RX spectra and that of the TX, as well as bumpiness. The relative residuals of the spectral amplitude are $\mathcal{O}(10^{-1})$ and higher. The changes in the spectral shape are due to the diffraction of the secondary pulse as it traverses the shallow firn.
\section{Bandwidth dependence}\label{section:bandwidth_variance}
\begin{figure}[!ht]
    \centering
    \includegraphics[width=0.9\linewidth]{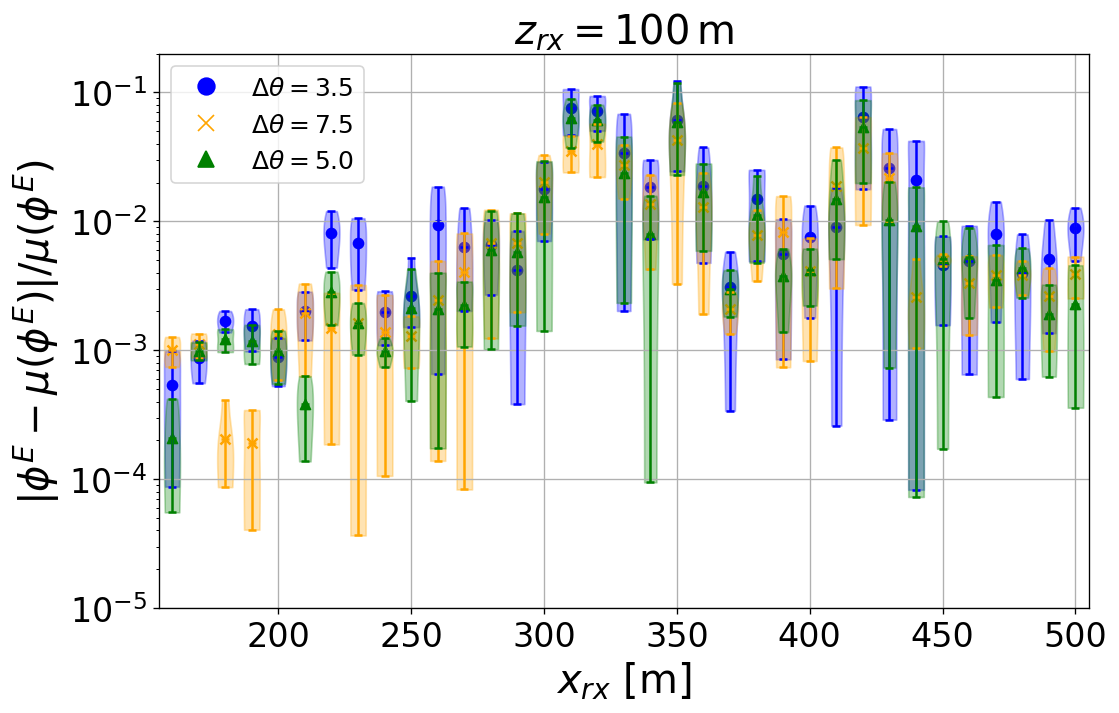}
    \caption{The total fluence variation $\delta \phi^{E}/\phi^{E}$ at $z_{rx} = 100 \, \mathrm{m}$ as a function of receiver range $x_{rx}$ given different source signal bandwidths, as sampled from different observer angle offsets to the Cherenkov angle. A higher degree of variation is seen for wider bandwidth signals. The source signal properties are defined in table \ref{tab:signal_properties} and the waveforms and spectra are shown in Fig. \ref{fig:source_waveform}.}
    \label{fig:3_fluence_variations_x}
\end{figure}

Since the magnitude and spectral width of an Askaryan pulse is dependent on the observer angle relative to the Cherenkov angle, and given that the fluence fluctuations we have discussed are a product of inter-layer diffraction, we investigated the dependence of the fluence variation of shallow refracted signals on the bandwidth of the original signal. To accomplish this, we repeated the RF simulation procedure with two source signals of successively wider bandwidth, corresponding to a Cherenkov angle offset of $\Delta \theta = 3.5^{\circ}$ \& $\Delta \theta = 5.0^{\circ}$ in addition to the $\Delta\theta = 7.5^{\circ}$ signal used thus far. The resulting pulses were both brighter and had greater spectral width and narrower time-width, with their properties summarized in Table \ref{tab:signal_properties} and their spectra shown in Fig.\,\ref{fig:source_waveform}. Because of the limitations of the FDTD method for higher frequencies at this scale, we utilize the PE method to simulate the pulses across the same domain used for the FDTD simulations.
\\~\\
The variation coefficient of the total pulse fluence $\delta \phi^{E}_{tot}/\phi^{E}_{tot}$ for the different input signal bandwidths is shown for a receiver at a depth of 100 m in Fig.\,\ref{fig:3_fluence_variations_x}, showing that the magnitude of the signal variation indeed increases for larger bandwidth signals. This is further evidence that the fluctuations in fluence are a result of the self-interference of signals traversing the shallow firn. Higher bandwidth pulses are sensitive to smaller length scales of the refractive index, and hence are subject to higher degrees of variability.
\section{Attenuation}\label{atten_appendix}
\begin{figure}[!ht]
    \begin{subfigure}[t]{\columnwidth}
        \centering
        \includegraphics[width=0.95\linewidth]{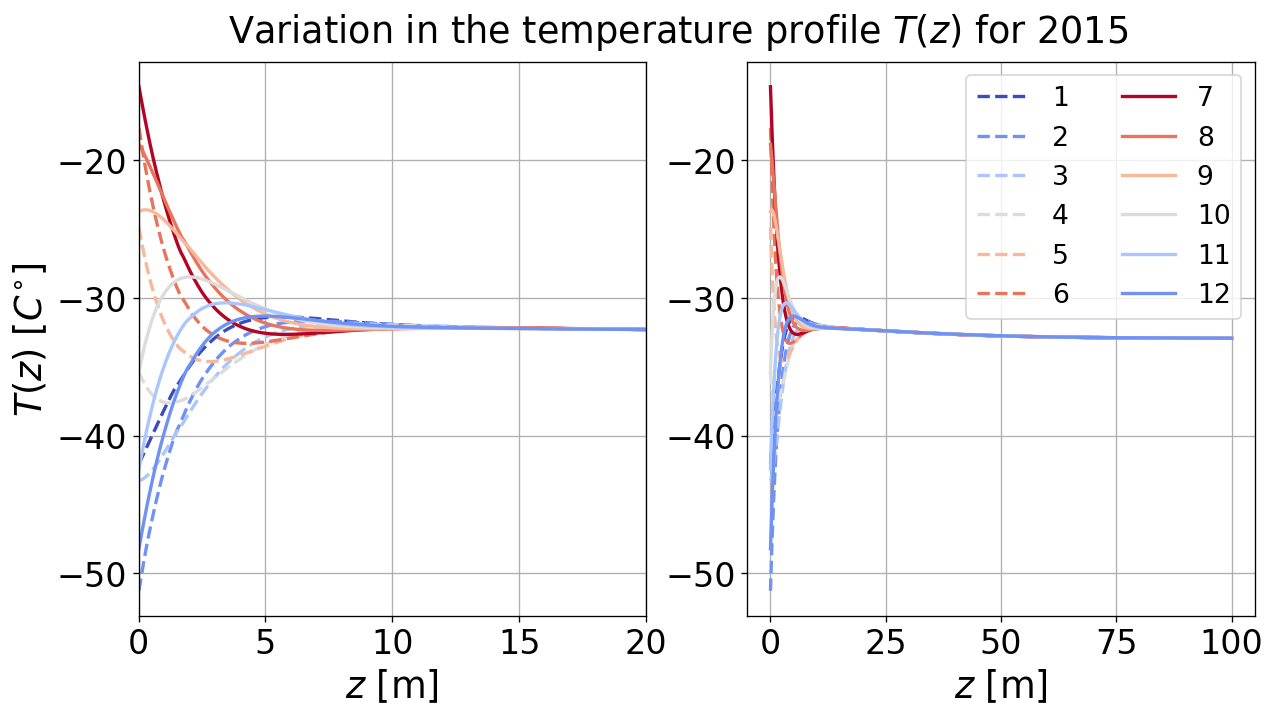}
        \caption{Temperature profile $T(z)$ in Celsius.}
        \label{fig:temp_prof}
    \end{subfigure}
    \begin{subfigure}[t]{\columnwidth}
        \centering
        \includegraphics[width=0.95\linewidth]{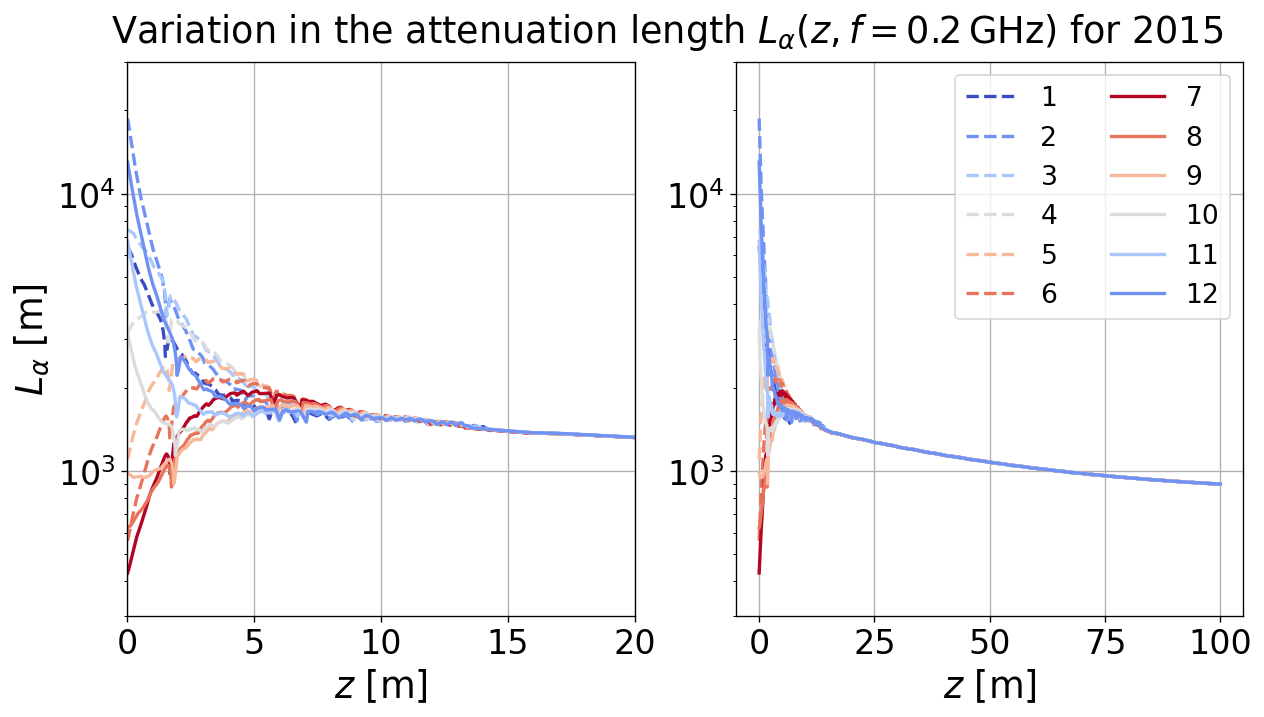}
        \caption{Attenuation length $\alpha(z)$.}
        \label{fig:atten_length}
    \end{subfigure}
    \caption{Above: the temperature profiles $T(z)$ of the firn as predicted by the CFM model for the year 2015 and months 1 (January) through to 12 (December), zoomed in to the upper 20 m n the left, and showing down to 100 m on the right. Below: the corresponding attenuation length profiles $L_{\alpha}(z)$.}
    \label{fig:temp_and_attenuation}
\end{figure}

\begin{figure}[!ht]
    \centering
    \includegraphics[width=0.95\linewidth]{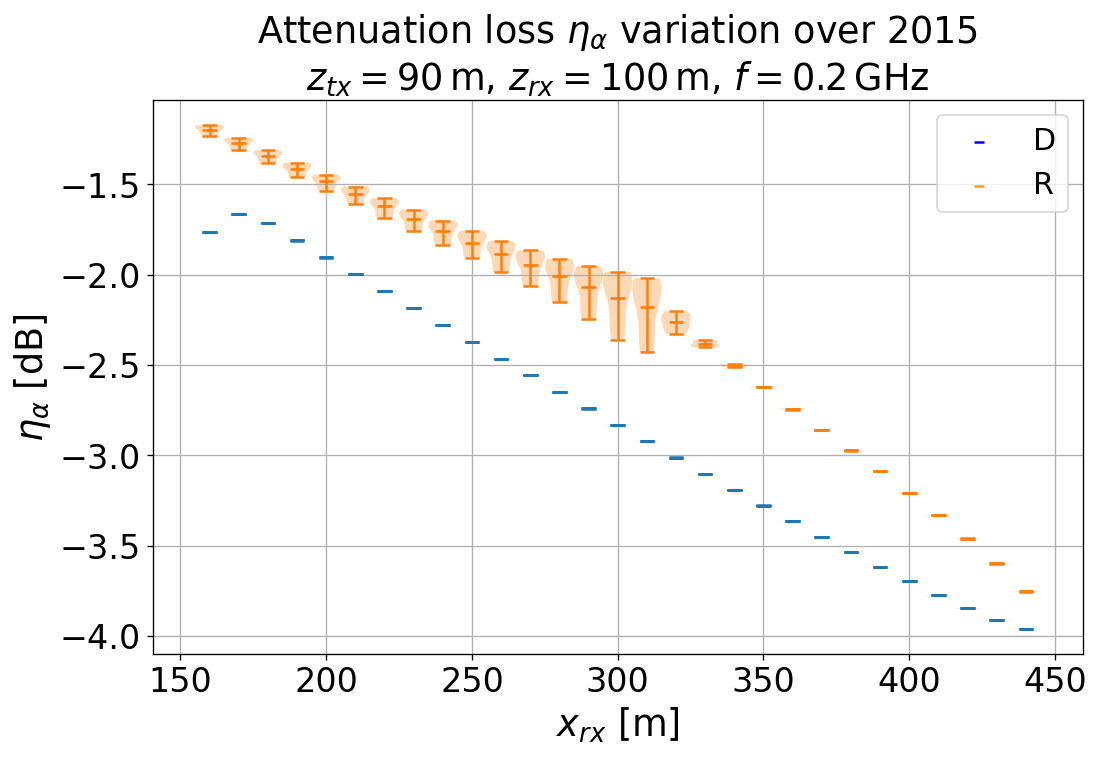}
    \caption{A violin plot showing the attenuation loss $\eta_{\alpha}$ in dB-scale for signals propagating from a source at $z_{tx} = 90 \, \mathrm{m}$ to receivers at $z_{rx} = 100 \, \mathrm{m}$ and a set of ranges $160 \, \mathrm{m} \leq x_{rx} < 450 \,\mathrm{m}$. The vertical spread of the violins indicate the seasonal variation in the attenuation factor.}
    \label{fig:attenuation_loss}
\end{figure}

Signal loss due to dielectric attenuation is an important consideration for neutrino reconstruction, and therefore we examine if the seasonal changes in the firn result in variation in the attenuation loss $\eta_{\alpha}$, defined in terms of signal power:
\begin{equation}
    \eta_{\alpha} = e^{-2 \alpha L} = e^{-2L/L_{\alpha}}
\end{equation}
Where $P_{RX}$ is the RF power at the receiver, $P_{TX}$ is the power emitted at the transmitter, $L$ is the propagation path of the ray through the ice, $\alpha$ is the attenuation factor and $L_{\alpha} = 1/\alpha$ is the attenuation length. It is generally understood that in-ice attenuation is strongly dependent on the ice temperature and the presence of chemical impurities, and thus we can expect that variation in attenuation should primarily occur in the upper part of the shallow firn where seasonal temperature variation is measurable. For a dielectric medium, attenuation is related to the imaginary component of the permittivity $\epsilon^{''}$, where the complex permittivity is: $\epsilon = \epsilon^{'} + j \epsilon^{''}$. $\epsilon^{''}$ defines the conductivity $\sigma$ of the medium $\epsilon^{''} = \sigma/\omega$. The attenuation factor $\alpha$ is calculated from the imaginary component $\kappa$, of the complex refractive index $m = n + j \kappa$ (which is simply the square root of the complex relative permittivity $\epsilon_{r} = \epsilon/\epsilon_{0} = m^{2}$):
\begin{equation}
    \alpha = \frac{2 \omega \kappa}{c}
\end{equation}
We make the assumption here that the conductivity at the frequencies of interest is fully described by the infinite frequency limit conductivity $\sigma_{ice,\infty}$. This value is related to chemical impurities (for $N$ distinct species) and temperature via the following equation \cite{Aguilar_broad_band_atten}:
\begin{multline}
    \sigma_{ice, \infty}(z) = \sigma_{pure} \exp[\frac{E_{pure}}{k} (\frac{1}{T_{ref}} - \frac{1}{T(z)})] +\\ \sum^{N}_{i} \mu_{i} C_{i}(z) \exp[\frac{E_{i}}{k}(\frac{1}{T_{ref}} - \frac{1}{T(z)})]
\end{multline}
Where $\sigma_{pure} = 9.2 \mathrm{\mu S /m}$ is the infinite frequency limit conductivity of pure ice \cite{aguilar_RF_attenuation}, $T_{ref} = -21^{\circ}$ C is the reference temperature, $T(z)$ is the ice temperature as a function of depth, $k$ is the Boltzmann constant, $\mu_{i}$ is the molar conductivity for species $i$, and $C_{i}(z)$ is the depth dependent molar concentration of species $i$. For this exercise however, we assume no impurities. In addition, the firn layer is not a solid crystal of ice with impurities, but a porous combination of ice and air. The conductivity of the bulk firn $\sigma_{firn}$ is found from the volume filling factor $v$ \cite{conductivity_firn}:
\begin{equation}
    \sigma_{firn}(z) = \sigma_{ice, \infty} v(z) (0.68 + 0.32v(z))^2
\end{equation}
Where $v(z)$ is simply the ratio of the firn density to that of solid ice $v =\rho(z)/\rho_{ice}$. 
\\~\\
To estimate the variation in signal attenuation, we select the year 2015 as our test case, utilizing the CFM output temperature $T_{2015,m}(z)$ and density profiles $\rho_{2015,m}(z)$ for each month $m$ to estimate the conductivity and hence attenuation factor of the firn for a monochromatic signal frequency of $f = 0.2 \, \mathrm{GHz}$. The estimated attenuation length profiles, alongside the temperature profiles are shown in Fig. \ref{fig:temp_and_attenuation}. We find that the temperature is stable at depths of around 10 m and deeper. At the surface the temperature oscillates between $-10^{\circ}$ in the summer months (peaking for month 7), and $-50^{\circ}$ in the winter months (the coldest in month 2), and the oscillation dampens with depth, a result that agrees well with field measurements at Summit \cite{Giese_Hawley_2015}. The same oscillation trend can be seen in the attenuation length profiles, varying at the surface from a maximum attenuation length of $L_{\alpha} \sim 2 \cdot 10^{4} \, \mathrm{m}$ in month 2, to a minimum attention length of $\sim 400 \, \mathrm{m}$ in month 7, with the variation damping rapidly with depth, stabilizing at around $2 \cdot 10^{3} \, \mathrm{m}$ at $z=5 \, \mathrm{m}$. For greater depths, the attenuation length declines, reaching a value of $L_{\alpha} \sim 10^{3}$ at depths of 75 m. Thus, we expect the attenuation variation to be proportional to the length of the signal path segment contained within depths of 5 m or shallower.
Using the source depth $z_{tx} = 90 \,\mathrm{m}$ utilized in the main work, we use ray-tracing to estimate the propagation path $s$ through the ice from the source to a set of receiver at a depth of $z_{rx} = 100 \, \mathrm{m}$ and receiver ranges spaced in 10 m intervals from a starting range of 160 m to a maximum of 520 m. The attenuation loss $\eta_{\alpha}$ is found by integrating the attenuation along ray path segments $ds$:
\begin{equation}
    \eta_{\alpha}(s) = \exp(-2 \int \alpha(s) ds )
\end{equation}
The resulting attenuation loss as a function of receiver range $\eta_{\alpha}(x_{rx})$ is shown in the violin-plot Fig. \ref{fig:attenuation_loss} in dB scale. The mean attenuation loss across the models is indicated by the markers, while the spread of the attenuation is indicated by the width of the brackets. It can be seen for the secondary (R) signals, that the spread in attenuation losses increases from $x_{rx} = 160 \, \mathrm{m}$ and peaks at occurs at $x_{rx} = 300 \, \mathrm{m}$ reaching levels of $\delta \eta_{\alpha,dB} \sim 0.1$, a result that can be explained by the increased path length of the ray contained with depths above 5 m. Beyond 300 m, as the receiver leaves the reflection zone and enters the refraction zone, the maximum height of the ray path is below the surface and progressively decreases, thus resulting in a decrease of the path length located at depths above 5 m. With the largest variation in the attenuation being on the level $\delta \eta_{\alpha,dB} \sim 0.1 \, \mathrm{dB}$, we draw the conclusion that variation in attenuation is a second order and negligible effect on RF signals, owing to relative stability in temperature and hence attenuation length for depths below 5 - 10 m.
\section{Neutrino reconstruction in the Shadow Zone}\label{shadow-zone}
Our discussion of event reconstruction in section \ref{discussion} has assumed unambiguous detection of the D and R pulses along a phased array of Vpol antennas within the illuminated zone. In the shadow zone, this condition is not satisfied and it is exceedingly difficult, if not impossible, to estimate the true vertex position via ray-tracing. As a consequence, the propagation distance $L$ from the receiver to the source is ambiguous, with an uncertainty in the distance on the scale of the attenuation length $\delta L \sim \mathcal{O}(L_{\alpha}) \sim\mathcal{O}(\mathrm{km})$, leading to $\delta g(L)$ dominating the shower energy uncertainty, which would span multiple energy-decades.
\\~\\
To address the situation for arrival direction, we make the crude approximation that the rays have propagated in a straight line from the source to a pair of receivers at $z_{rx} > 80 \, \mathrm{m}$ within the shadow zone. Under this assumption, we estimate the angle $\alpha_{12}$ of this straight line path with respect to the horizontal between antennas $RX_{1}$ and $RX_{2}$:
\begin{equation}
    \tan(\alpha_{12}) = \frac{n(z_{\mathrm{mid}}) \Delta z_{12} }{c \Delta t_{12}},
\end{equation}
where $z_{\mathrm{mid}} = (z_{rx,1} + z_{rx,2})/2$,, $\Delta z_{12}$ is the depth interval between the antennas and $\Delta t_{12} = t_{D,2} - t_{D,1}$ is the relative arrival time of the waveforms at $RX_{1}$ and $RX_{2}$. We can define $t_{D}$ in two ways. The first is the `leading edge time' $t_{D,0.05}$, which we quantify here as the time at which the cumulative amplitude of the waveform reaches 5\% of the sum total. The second is the time of maximum amplitude of the waveform $t_{D,max}$. The shadow waveforms displayed in Fig. \ref{fig:rx_pulse_sh_DR} are a combination of complex multi-path propagation and mutual interference between different waveforms. This would suggest that $t_{D,0.05}$ would be more likely to correspond to the `earliest arriving' signal. On the other hand, $t_{D,max}$ is likely to be easier to identify in a real-world detection where environmental and system noise must be considered. In this study we do not include noise in the waveforms and thus the leading edge is easier to identify than in real world conditions.
\begin{figure}[!ht]
    \begin{subfigure}[t]{\columnwidth}
        \centering
        \includegraphics[width=0.9\linewidth]{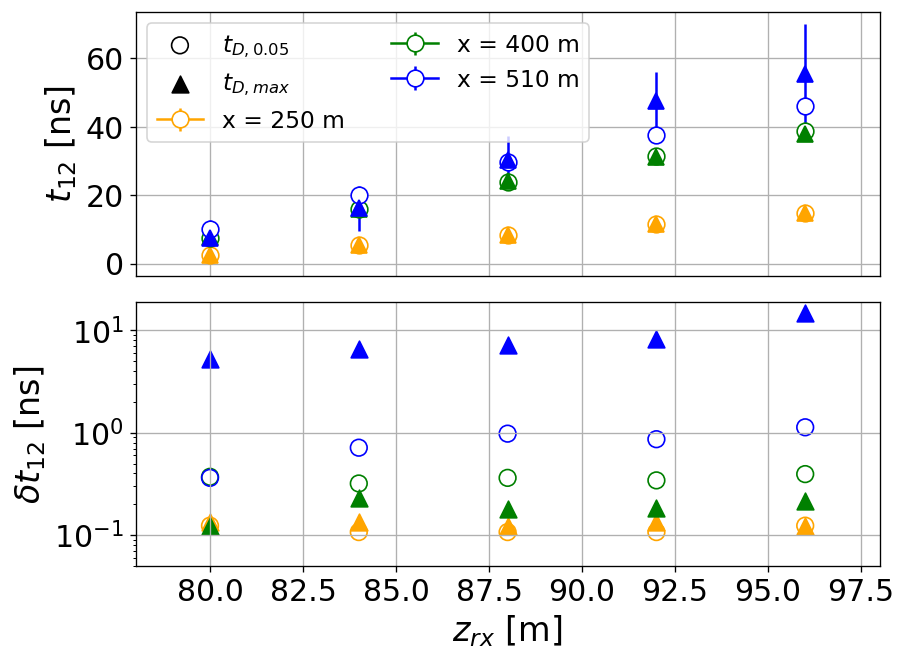}
        \caption{Relative arrival times $t_{12}$ between $\mathrm{RX_{1}}$ and $\mathrm{RX_{2}}$.}
        \label{fig:shadow_zone_t}
    \end{subfigure}
    \begin{subfigure}[t]{\columnwidth}
        \centering
        \includegraphics[width=0.9\linewidth]{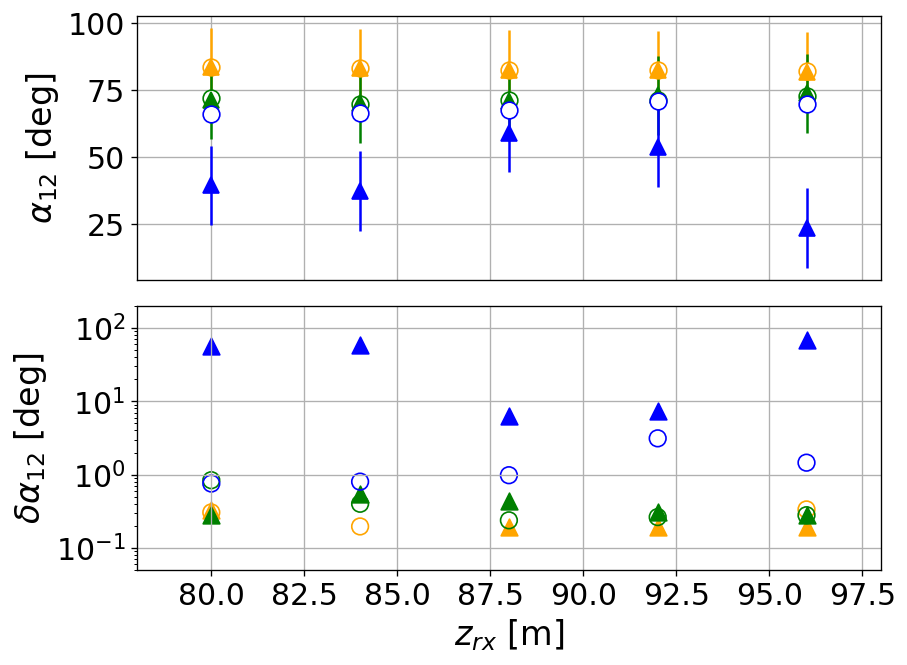}
        \caption{Arrival angle $\alpha_{12}$ between $\mathrm{RX_{1}}$ and $\mathrm{RX_{2}}$.}
        \label{fig:shadow_zone_alpha}
    \end{subfigure}
    \caption{Relative arrival time $t_{12}$ (\ref{fig:shadow_zone_t}) and inferred arrival angle $\alpha_{12}$ (\ref{fig:shadow_zone_alpha}) between neighboring receiver pairs as a function of receiver depth $z_{rx}$. Results are shown for receiver arrays at $x_{rx}=250\,\mathrm{m}$ (reflection zone), $x_{rx}=400\,\mathrm{m}$ (refraction zone), and
    $x_{rx}=510\,\mathrm{m}$ (shadow zone). In the top panel of each subplot the points indicate mean values across ice models, with error bars showing the standard deviation, using both the leading-edge time $t_{D,0.05}$ and the time of maximum amplitude $t_{D,\mathrm{max}}$. The bottom panels directly plot the variation in the respective quantities among the ice models.}
    \label{fig:shadow_zone}
\end{figure}
As shown in Section \ref{section:propagation_time_variance}, the propagation time variation within the shadow zone, using $t_{D,max}$, was on a scale $\delta t \sim \mathcal{O}(10 \, \mathrm{ns})$, which would imply RX arrival direction uncertainty  $\delta \theta_{RX} \gtrsim \mathcal{O}(10^{\circ})$, over an order of magnitude greater than within the illuminated zone. To better quantify this, we take an array of 6 receivers (RX) located at depths $76 \, \mathrm{m} \leq z_{rx} \leq 96 \, \mathrm{m}$, and for RX from $z_{rx} = 80 \, \mathrm{m}$ downwards, we calculate $t_{12}$ between the RX and its neighbor using both $t_{D,0.05}$ and $t_{D,max}$, which, equivalently, allows us to calculate $\alpha_{12}$. We then find the variation among the estimated values of $t_{12}$ and $\alpha_{12}$. 
\\~\\
In Fig. \ref{fig:shadow_zone}, we plot $t_{12}$ (\ref{fig:shadow_zone_t}) and $\alpha_{12}$ (\ref{fig:shadow_zone_alpha}) and their variations as a function of depth, where the range to the receiver array is $x_{rx} = 250 \, \mathrm{m}$ (reflection zone), $x_{rx} = 400 \, \mathrm{m}$ (refraction zone) and $x_{rx} = 510 \, \mathrm{m}$ (shadow zone). For the reflection and refraction zone examples, the variation in the time offset between RX pairs is $\delta t_{12} < 1 \, \mathrm{ns}$, and the variation in the arrival angle at the pair is close to $\delta \alpha_{12} \sim 0.1^{\circ}$, consistent with our results in the previous Section. With a clearly identifiable direct pulse, there is little difference between the results when using $t_{D} = t_{D,max}$ and $t_{D} = t_{D,0.05}$. By contrast, the choice of $t_{D,max}$ and $t_{D,0.05}$ has a significant effect on both the variation in $t_{12}$ and $\alpha_{12}$. For $t_{D} = t_{D,0.05}$, we find variation in the arrival times on the scale of a nanosecond $\delta t_{12} \sim 1 \, \mathrm{ns} $ and hence in arrival angle of $\delta \alpha_{12} \sim 1^{\circ}$. For $t_{D} = t_{D,max}$ both variations are increased by an order of magnitude or greater, with the worst-case example of $\delta \alpha_{12} \approx 67^{\circ}$ for $z_{rx} = 96 \, \mathrm{m}$. In addition, the mean value of $\alpha_{12}$ differs by 10 to 20 degrees between the two $t_{D}$ methods. In this example, we find that the variation in receiver arrival direction, using receiver pairs, is on the scale of $0.8^{\circ} \lesssim \delta\alpha_{12} \lesssim 3^{\circ}$ for the leading-edge time estimate, compared to $6^{\circ} \lesssim \delta\alpha_{12} \lesssim 70^{\circ}$ using the maximum time method.
\section{Illustration of transmitter source reconstruction}\label{vertex-appendix}
\begin{figure}[!ht]
    \centering
    \includegraphics[width=\linewidth]{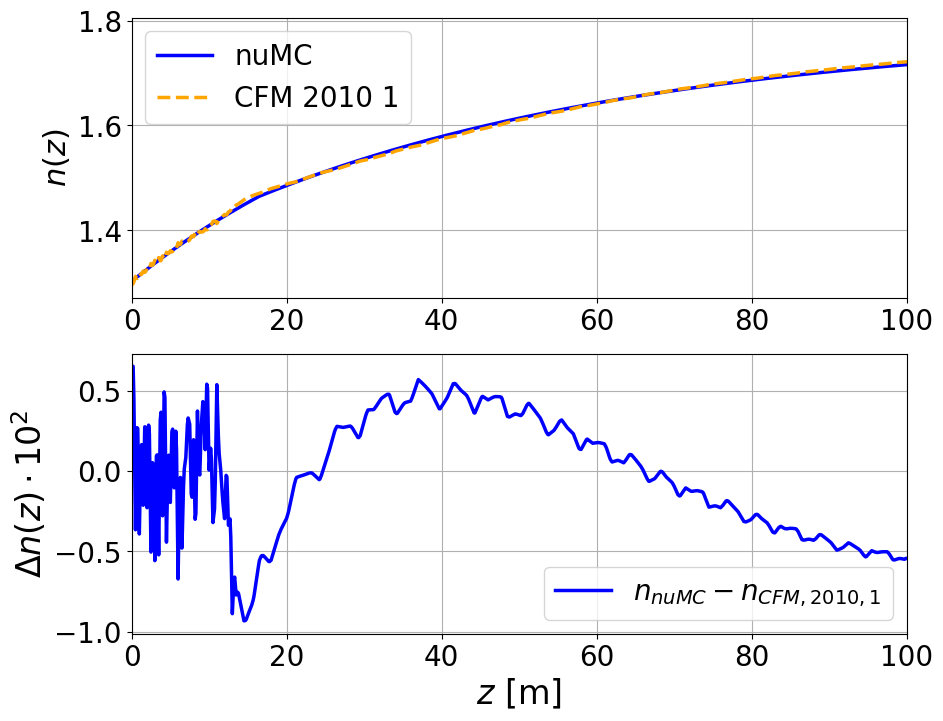}
    \caption{The refractive index profile $n_{nuMC}(z)$ used to generate ray tracing solutions for the TX reconstruction. Top $n_{nuMC}(z)$ shown alongside a CFM profile $n_{CFM,2010,1}(z)$ corresponding to the model estimated for January 2010. The offset between the two profiles is shown in the bottom plot.}
    \label{fig:n_nuMC}
\end{figure}

\begin{figure*}[!ht]
\begin{subfigure}{\columnwidth}
    \raggedright
    \centering
    \includegraphics[width=0.9\linewidth]{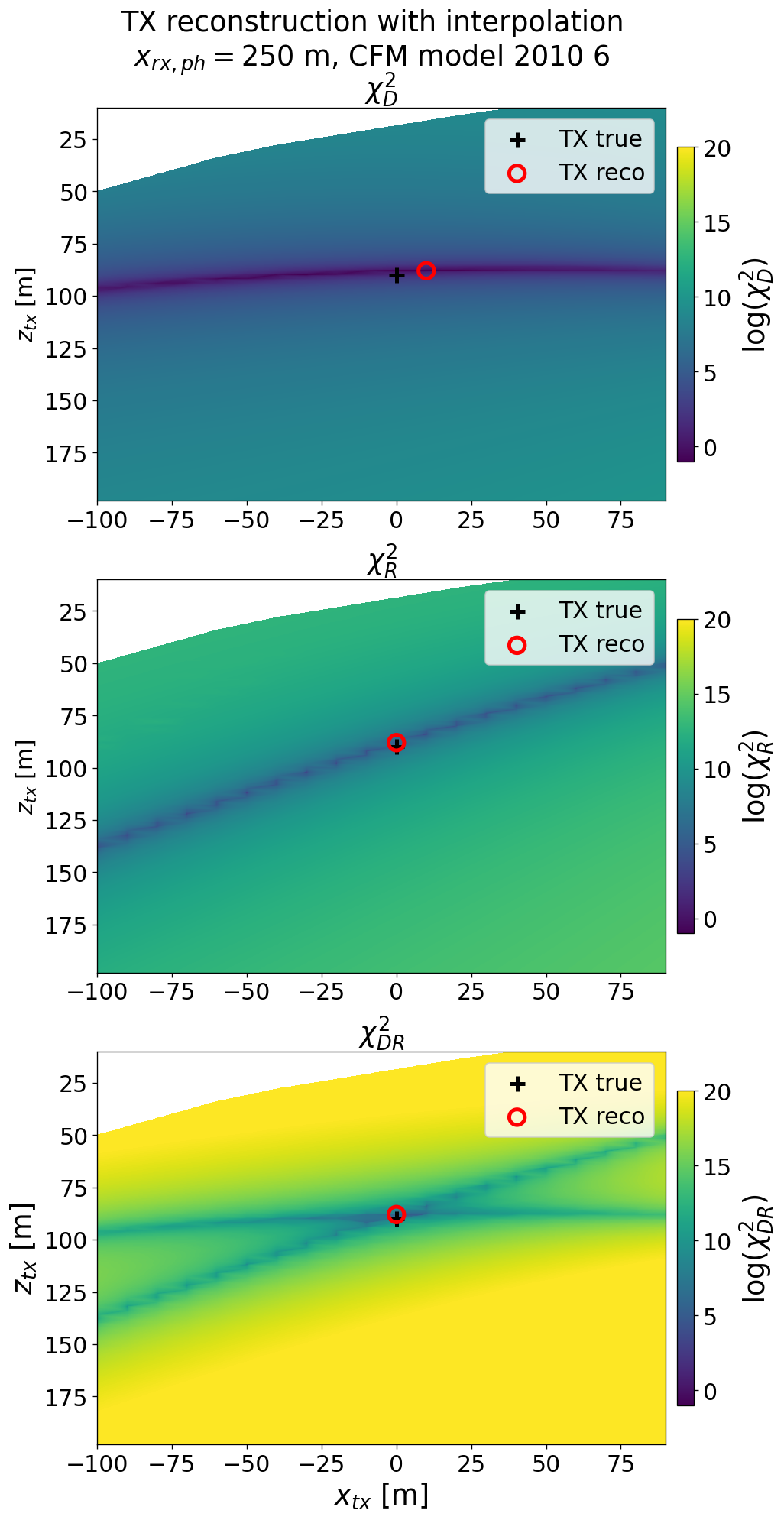}
    \caption{Maps of the penalty score $\chi^{2}_{D}$, $\chi^{2}_{R}$, and $\chi^{2}_{DR}$ for a phased array located at $x_{rx,ph} = 250 \, \mathrm{m}$ and the refractive index model $n^{CFM}_{2010,6}(z)$. The best fit value is indicated with a red circle, where as the true TX position used to initialize the \texttt{MEEP} simulation is indicated with a black cross.}
    \label{fig:tx_reco_map}
\end{subfigure}
\begin{subfigure}{\columnwidth}
    \raggedright
    \centering
    \includegraphics[width=0.9\linewidth]{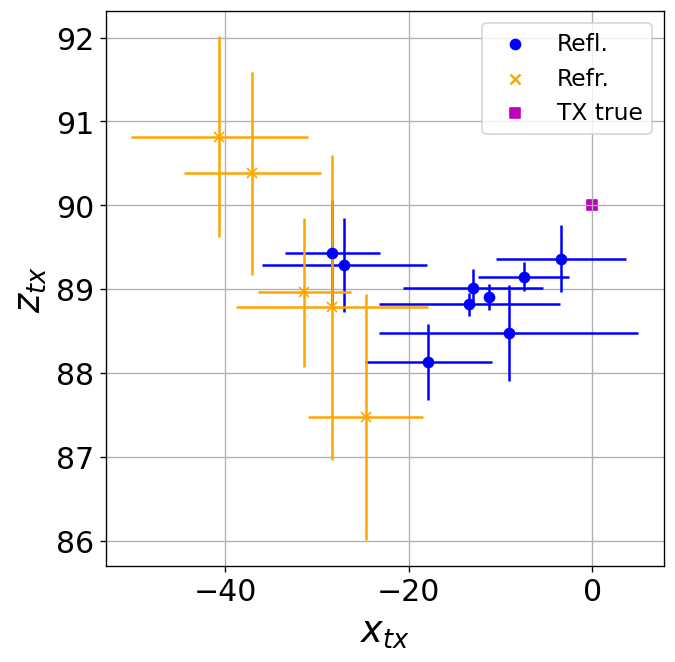}
    \caption{Reconstructed TX positions for phased arrays located at ranges $ 150 \, \mathrm{m} \leq x_{rx,ph} \leq 550 \, \mathrm{m}$. The points correspond to the mean value of $(x_{tx,reco},z_{tx,reco})$ among the ice models, while the error-bars show the standard deviation of the range and depth among the ice models.}
    \label{fig:tx_best_fit}
\end{subfigure}
\caption{Left (\ref{fig:tx_best_fit}): penalty score map over TX range. Right (\ref{fig:tx_best_fit}): reconstructed TX positions.}
\label{tx-reco-plots}
\end{figure*}

The Figures presented here provide supplementary visualization of the vertex reconstruction method outlined in section \ref{vertex-reco}. We calculated penalty scores as a function of transmitter (TX) position for the primary and secondary signals using the \texttt{RadioPropa} ray tracer in \texttt{NuRadioMC} and the \texttt{MEEP} simulated waveforms for each ice model. The ice model used for the ray tracing solution was found using a split-exponential profile $n_{nuMC}(z)$. From the linear relationship between $n(z)$ and $\rho(z)$ (expressed with Eq.\,\ref{kovacs-eq}) \cite{kovacs1995situ}, we can derive an expression equivalent to Eq.\,\ref{firn_exponential}:
\begin{equation}\label{firn_exponential-n}
        n(z) = 
        \begin{cases}
                n_{i} + (n_{s} - n_{i}) e^{-k_{0} z} & \text{if } n \leq n_{550} \\
                n_{i} + (n_{550} - n_{i}) e^{-k_{1} (z-z_{550})} & \text{if } n > n_{550}.
        \end{cases}
\end{equation}
From inspection, the exponent factors $k_{0}$ and $k_{1}$ are the same for Eq.\,\ref{firn_exponential} and Eq.\,\ref{firn_exponential-n}. The parameters were obtained from a best fit to the CFM model corresponding to January 2010: $n_{0} = 1.302$ ($\rho_{0} = 0.355 \, \mathrm{g/cm^{3}}$), $k_{0} = 0.025 \, \mathrm{m^{-1}}$, $k_{1} = 0.019 \, \mathrm{m^{-1}}$, \& $z_{550} = 16.48 \, \mathrm{m}$. The two profiles and their offsets from each other are compared in Fig. \ref{fig:n_nuMC}. For each trial position $(x_{tx}, z_{tx})$, we computed $\chi^{2}_{D}$, $\chi^{2}_{R}$ for primary and secondary signals respectively, and the combined score $\chi^{2}_{DR}$, as defined in Equation \ref{penalty-score}, and identified the optimal TX position from the minimum of $\chi^{2}_{DR}$. The optimal TX position then corresponds to an optimal propagation path $L$ and signal arrival direction at the RX for the D and R signals, respectively. The TX position is estimated independently for the different ice models used to simulate the waveforms, and the variation in TX between these models translated naturally into the seasonal variation in $L$ and $\theta_{RX}$ as shown in Sections \ref{vertex-reco} and \ref{theta-reco}, respectively.
\\~\\
In Fig. \ref{fig:tx_reco_map} we show the penalty score map calculated for a phased array located at $x_{rx,ph} = 250 \, \mathrm{m}$ and the refractive index model $n^{CFM}_{2010,6}(z)$, with the scores for the primary $\chi^{2}_{D}$ and secondary pulses shown in the top and middle subplots, and the combined score shown in the lower subplot. The $\chi^{2}_{D}$ and $\chi^{2}_{R}$ maps each show a strip of low amplitudes (low penalty scores) for various values of range; this strip corresponds to positions where a ray propagating from the source to the receiver arrive at similar angles and are launched at similar angles. Since the penalty score is calculated from the relative arrival times at the phased array, they are naturally sensitive to the angle of the wavefront with respect to the simulated phased array. The difference in arrival angles for the primary and secondary (reflected or refracted) signals leads to the bands being inclined with respect to one another, and their product creates a cross-like shape in the $\chi^{2}_{DR}$ map, whose crossing point aligns with the minimum $\chi^{2}_{DR}$ value.
\\~\\
In Fig. \ref{fig:tx_best_fit}, we show the best-fit TX positions calculated for phased arrays at different ranges, with the variation between the ice models indicated with error-bars; the central point shows the mean TX position among the ice models and the error-bars represent the standard deviation over the ice models. Here, we can associate a systematic error with the systematic offset between the refractive index model used to generate the ray tracing solutions, for all of the CFM ice models used in this study. This manifests as the offset between the mean TX reconstruction values and the true TX. These are distinct from the variation between the ice models illustrated with the error-bars. The systematic offset in range in particular is due to the offset $\Delta n = n_{nuMC}(z) - n_{CFM}(z) $ of the \texttt{NuRadioMC} model relative to the CFM models. This causes an increase in the absolute delay time of the ray arriving at the receivers, and an increase in the delay relative to the first arriving signal, leading to an overestimate in the distance to the source that increases as the phased array moves down-range. 
\\~\\
The model to model variation was generally larger for the reconstructed range as compared to the depth, and this held true for when the phased array was located in the reflection zone and refraction zones. For the range direction, this variation was $\frac{\delta x_{tx,reco}}{x_{tx,reco}} \approx 8 \, \mathrm{m}$ in both zones, whereas the depth variation was larger in the refraction zone $\frac{\delta z_{tx,reco}}{z_{tx,reco}} \approx 1.35 \, \mathrm{m}$ compared to the reflection zone $\frac{\delta z_{tx,reco}}{z_{tx,reco}} \approx 0.4 \, \mathrm{m}$. The systematic offset of the vertex reconstruction is generally larger than the seasonal variation, with offsets in range increasing from $x_{tx,reco} - x_{tx,true} \approx -5 \, \mathrm{m}$ at $x_{rx,ph} = 160 \, \mathrm{m}$ to $x_{tx,reco} - x_{tx,true} \approx -30 \, \mathrm{m}$ at $x_{rx,ph} = 320 \, \mathrm{m}$. Within the refraction zone, the offset in the depth direction varied in both directions relative to the $z_{tx,true}$ to an extent of $87 \, \mathrm{m} \lesssim |z_{tx,reco}-z_{tx,true}| \lesssim 91 \, \mathrm{m}$, slightly larger than that for the reflection zone $88 \, \mathrm{m} \lesssim |z_{tx,reco}-z_{tx,true}| \lesssim 89 \, \mathrm{m}$.
\end{document}